\documentclass[fleqn,11pt,letterpaper]{article}

\usepackage{graphicx}

\usepackage{longtable}
\usepackage{tabularx}
\usepackage{booktabs}
\usepackage{multirow}
\usepackage{tikz}
\usepackage{color, colortbl}
\definecolor{Gray}{gray}{0.9}
\usepackage{tablefootnote}
\usepackage{caption}
\usepackage{subcaption}
\usepackage{placeins}
\usepackage{amsmath}
\usepackage{amsfonts}
\usepackage{amssymb}
\usepackage{amsmath}
\usepackage{amsthm}
\newcommand{\N}{\mathbb{N}}

\newcommand{\R}{\mathbb{R}}

\newcommand{\E}{\mathbb{E}}

\usepackage{tocvsec2}
\usepackage{appendix}
\usepackage[margin=2.57cm]{geometry}
\usepackage{hyperref} 
\usepackage{fancyhdr}
\usepackage[round]{natbib}
\usepackage{authblk}
\usepackage{lipsum}
\usepackage{lscape}
\usepackage{gensymb}
\usepackage{textcomp}
\usepackage{titletoc}

\begin{document}
\title{Neural networks for nonlinear regression with serially correlated disturbances: Evidence from cloud cover} 
\date{}

\author{Sebastian Jensen\textsuperscript{a,*}}
\author{Siem Jan Koopman\textsuperscript{b}}

\affil[]{\textsuperscript{a}Aarhus University and Center for Research in Energy (CoRE), Denmark}

\affil[]{\textsuperscript{b}Vrije Universiteit Amsterdam, The Netherlands}

\maketitle

\begingroup
\renewcommand{\thefootnote}{\fnsymbol{footnote}}

\footnotetext[1]{%
	Corresponding author:
	\href{mailto:smjensen@econ.au.dk}{%
	\nolinkurl{smjensen@econ.au.dk}}%
}
\endgroup

{\let\newpage\relax\maketitle}

\begin{abstract}
We propose a new treatment of nonlinear regression with serially correlated disturbances that incorporates autoregressive moving average structures into feedforward neural networks. The resulting model provides an alternative to modeling temporal dependence using lagged variables. In simulations, the proposed method accurately recovers regression functions of varying complexity and the underlying error dynamics across a range of time-series lengths and signal-to-noise ratios. Finite-sample properties and out-of-sample predictive performances are shown to be robust to model misspecification induced by omitted lagged variables and incorrect specification of the error dynamics. Cloud cover is an important factor in climate projections. In an empirical study of cloud cover prediction for a grid of locations within and around the Mediterranean Sea, our proposed model yields more accurate predictions than existing methods, including long short-term memory networks. Serially correlated disturbances in place of lagged variables improve predictive accuracy across a range of land and ocean environments. Improvements over linear models with serially correlated disturbances are particularly pronounced in mountain areas, consistent with the presence of stronger nonlinear effects in cloud formation in such regions.

\end{abstract}


\newpage

\section{Introduction}

Machine learning methods, including neural networks, are increasingly used in data science, econometrics and statistics \citep{hastie2009elements,Athey2019,Efron2020}. These methods offer considerable flexibility for modeling nonlinear relationships in regressions, even in high-dimensional settings. Temporal dependence is often accommodated by including a set of lagged dependent (output) and lagged regressor (input) variables in the regression component. The disturbances are typically left unmodeled or treated as coming from a white noise process, under the implicit assumption that all predictable temporal dependence is absorbed by a sufficiently rich set of lagged variables. Any remaining serial correlation in the disturbance process therefore reflects misspecification. This approach of expanding the predictor space may, however, unnecessarily increase the number of predictors, particularly in a setting that is already high-dimensional. Furthermore, in a machine learning context, it is difficult to use statistical inference to guide the choice of variables and their lag lengths. An alternative approach to modeling temporal dependence via lagged variables in the regression component is to capture it through a dynamic stochastic process for the disturbances. Although presented here as alternatives, the two approaches can be combined. A classical time series model, such as the autoregressive moving average (ARMA) model, provides a parsimonious and interpretable representation of temporal dependence. In the linear framework, the use of such structured stochastic models for the disturbance process is well established in the literature. However, integrating structured stochastic models for the disturbances into flexible machine learning frameworks remains largely unexplored and is the focus of this paper.

In this paper, we develop a methodology for flexible nonlinear regression and prediction with serially correlated disturbances by integrating ARMA error structures into feedforward neural networks. The resulting model is denoted as NNARMA. We focus on ARMA dynamics for the disturbances as they provide a parsimonious representation of a wide variety of serial correlation patterns, see \cite{boxjenkins2015} for a textbook treatment. We focus on feedforward neural networks for a number of reasons. They are universal approximators \citep{HORNIK1989, Cybenko1989, LESHNO1993} and are known to perform well in practice on a number of regression and predictions tasks \citep{Kelly2020,Jensen2023}. Theoretical justification is starting to be developed \citep{Farrell2021}. \cite{Schmidt-Hieber2020} shows that feedforward neural networks are superior to classical nonparametric methods, such as kernel-based estimators and splines, if the unknown target function is a composition of simpler functions. To reduce overfitting, neural networks incorporate various regularization techniques \citep{Goodfellow-et-al-2016}. \cite{Bach2017} and \cite{Bauer2019} show that feedforward neural networks solve the curse of dimensionality that affects classical nonparametric methods \citep{Stone1982}.

The proposed estimation procedure is a conceptually straightforward generalization of well-known techniques for linear regression with serially correlated disturbances. We compute a Cholesky decomposition of the disturbance covariance structure using the Durbin-Levinson algorithm \citep{Levinson1946,Durbin1960isr} and pre-whiten the model to obtain serially uncorrelated disturbances. Estimation proceeds by minimizing the residual sum of squares of the pre-whitened model, corresponding to what is also known as feasible generalized least squares. Complications arise from the nonlinear regression function. Our main contribution is the development of a neural network framework that is able to jointly estimate the unknown regression function and ARMA parameters. 

We conduct an empirical study of cloud cover prediction, an important factor in climate projections. Using daily measures of cloud cover for a grid of locations within and around the Mediterranean Sea (Southern Europe and Northern Africa), the proposed NNARMA model leads to more accurate predictions than existing methods, including the long short-term memory (LSTM) network of \cite{Hochreiter1997}. The empirical results show that modeling temporal dependence through serially correlated disturbances rather than lagged variables improves predictive accuracy across a range of land and ocean locations. Improvements relative to linear models with ARMA errors are particularly pronounced in mountain areas. This finding is consistent with stronger nonlinear effects in such regions.

Simulation evidence suggests that the proposed estimation procedure accurately recovers regression functions of varying complexity and the underlying ARMA dynamics across a range of time-series lengths and signal-to-noise ratios. The NNARMA model demonstrates robustness to both ARMA misspecification of the errors and misspecification induced by omitted lagged variables. In both cases of misspecification, results remain stable except under substantial underspecification of the ARMA disturbance structure or the network architecture. 

The finite-sample properties of the estimation procedure and the out-of-sample predictive performance of the NNARMA model demonstrate robustness to both ARMA misspecification of the errors and misspecification induced by omitted lagged variables. In both cases of misspecification, results remain stable except under substantial underspecification of the ARMA disturbance structure or the network architecture. 

The remaining part of the paper is structured as follows. Section \ref{sec: literature review} discusses the literature on regression with serial correlation and our contribution. In Section \ref{sec: methodology}, we present our methodology for neural network regression with ARMA errors, including estimation and model selection. Section \ref{sec: Monte Carlo (main text)} uses Monte Carlo simulations to demonstrate the finite-sample properties and out-of-sample predictive performance of our methodology, including under various forms of misspecification. In Section \ref{sec: CFC application}, we present our empirical study of cloud cover prediction. Section \ref{sec: Conclusion} briefly summarizes and concludes.

\section{Nonlinear regression with serial correlation}\label{sec: literature review}

Consider the univariate time series process for some variable $y$ as denoted by $\{y_t\}_{t\in\mathbb{Z}}$, for time index $t$ with regular spacing. We define the nonlinear regression model with serially correlated disturbances for $y_t$ as
\begin{equation}\label{eq: model intro}
	y_t = f(x_t;\beta) + u_t, 
\end{equation}
where $f(\cdot)$ is the scalar nonlinear regression function, $x_t$ is a column vector of known regressors (explanatory variables), $\beta$ is a column vector of regression coefficients, and the disturbance sequence $\{u_t\}_{t\in\mathbb{Z}}$ is a weakly stationary process with zero mean, $\E(u_t)=0$, finite variance, $\text{Var}(u_t)=\omega ^2>0$, and serial correlation $\text{Cov}(u_t, u_{t-h}) = \gamma _h$, for $h=1,2,3,\ldots$, where $\omega ^2$ and $\gamma _h$ can be treated as unknown coefficients.

\subsection{Linear regression with ARMA disturbances}

It is common practice for the statistical treatment of model \eqref{eq: model intro} to impose parametric restrictions on both the functional form of $f(\cdot)$ and the dynamics of the serially corrrelated disturbances $u_t$. For example, one may consider model \eqref{eq: model intro} with
\begin{equation}\label{eq: model linear}
	f(x_t;\beta) = x_t\, ' \beta ,
	\hspace*{0.8cm} u_t = \phi u_{t-1} + e_t,
\end{equation}
where both $x_t$ and $\beta$ are column vectors, $\phi$ is an autoregressive coefficient, and $\{e_t\}_{t\in\mathbb{Z}}$ is a white noise sequence of innovations with zero mean, $\E(e_t)=0$, and finite variance, $\text{Var}(e_t)=\sigma ^2>0$. For this specification, model \eqref{eq: model intro} reduces to a linear regression with autocorrelated disturbances of order one, and with $\omega ^2= \sigma ^2 / (1-\phi ^2)$ and $\gamma _h = \omega ^2 \, \phi ^h$. To ensure a stationary disturbance process $\{u_t\}$, one typically imposes the restriction $|\, \phi \, | < 1$ such that $\omega ^2, \gamma _h<\infty$, for all $h$. The resulting linear regression model with serially correlated errors is the framework from which the Durbin-Watson test is developed for the null hypothesis $H_0 : \phi = 0$ \citep{Durbin50,Durbin51}. The best linear unbiased estimate (BLUE) of $\beta$, for a given and known value of $\phi$, can be obtained by least squares after pre-whitening the disturbances using the data transformations
\[
y_t^\ast = y_t - \phi y_{t-1},\hspace*{0.8cm}
x_t^\ast = x_t - \phi x_{t-1},\hspace*{0.8cm}
t=2,3,4,\ldots .
\]
The least squares method applied to $y_t^\ast=x_t^\ast \beta + e_t$ delivers the BLUE of $\beta$.

In this paper, we consider the autoregressive moving average process, denoted by  ARMA$(p,q)$, for the disturbance sequence $\{u_t\}_{t\in \mathbb{Z}}$. It is specified as
\begin{equation}\label{eq: model ARMA}
	u_t = \phi _1 u_{t-1} + \ldots + \phi _p u_{t-p} +  e _t + \omega _1 e_{t-1} + \ldots + \omega _q e_{t-q},
\end{equation}
with autoregressive coefficients $\phi _1, \ldots , \phi _p$ and moving average coefficients $\omega _1,\ldots,\omega _q$. Particular stability conditions can be imposed on the autoregressive coefficients to ensure a stationary process $\{u_t\}$ \citep{hamilton1994series}. The least squares method applied to the regression model \eqref{eq: model linear} with ARMA disturbances from \eqref{eq: model ARMA} is originally treated by \cite{Pierce1971}. The exact maximum likelihood approach of \cite{HarveyPhillips1979} entails the joint estimation of $\beta$ and the ARMA coefficients by adopting data transformations and using the Kalman filter to numerically compute the likelihood function. We adopt a similar approach, but rely on the pre-whitening transformations carried out by the Durbin--Levinson algorithm, which is tailored for the case of ARMA disturbances, and apply it to nonlinear neural network regression.

\subsection{Parametric and nonparametric nonlinear regression}

The nonlinear regression model \eqref{eq: model intro} with a known parametric functional form for $f(\cdot)$ and serially uncorrelated disturbances $\{u_t\}$ is treated in textbooks such as \cite{Gallant1987} and \cite{BatesWatts1988}, which establish the asymptotic theory of the nonlinear least squares (NLS) estimator. When disturbances follow an ARMA process \eqref{eq: model ARMA}, NLS remains consistent but is inefficient. This inefficiency has motivated likelihood-based methods that jointly estimate the nonlinear regression function $f(\cdot)$ and the ARMA parameters \citep{harvey1990, hamilton1994series} using the pre-whitening ideas in \cite{HarveyPhillips1979}, thereby yielding efficiency gains and more reliable inference.

In practice, it can be challenging to justify a known parametric functional form for $f(\cdot)$, motivating the use of nonparametric and semiparametric regression methods. Kernel-based estimators such as the Nadaraya–Watson estimator \citep{Nadaraya1964, Watson1964} and local polynomial estimators, together with spline methods, have been shown to be consistent under mixing or other weak‑dependence conditions \citep{Robinson1983, Robinson1988, HardleTsybakov1997}; see also \cite{PaganUllah1999} and \cite{LiRacine2007}. However, as in the parametric case, ignoring serial correlation in the disturbances may lead to inefficiency. Accordingly, methods based on pre-whitening transformations have been proposed in classical nonparametric regression settings \citep{Xiao2003, Su2006, LIU2010}.

\subsection{Neural network regression}

Neural networks have emerged as powerful tools for nonlinear regression in data rich environments. They can handle high-dimensional input vectors $x_t$, whereas classical nonparametric methods suffer from the curse of dimensionality. Early econometric work of \cite{WhiteDomowitz1984} established consistency of feedforward networks as flexible nonlinear regressors, while the universal approximation results of \cite{HORNIK1989} provided the theoretical foundation for their use in regression; see also \cite{White1989}. Extensions to time series settings have been considered, for example, by \cite{Chen1999}, who establish results under mixing conditions.

In the context of time series forecasting, \cite{Zhang2003} considers hybrid neural network methods for ARMA models, while \cite{Terasvirta2006} show how neural network regressions can be embedded in variable selection methods. Finally, the research of \cite{sun2021} is most closely related to our study. It considers a neural network regression model with serially correlated disturbances. The key differences are that \cite{sun2021} focus on a system of multiple time series variables where single regressions take place on the other variables, assume autoregressive disturbances of order one,  $u_t = \phi u_{t-1}+e_t$, for each regression,  and rely on approximate methods for pre-whitening the data. We treat the single equation \eqref{eq: model intro} with output variable $y_t$ and input vector $x_t$, assume ARMA disturbances $u_t$ as in \eqref{eq: model ARMA}, integrate the Durbin-Levinson pre-whitening within the neural network method, and jointly estimate the nonlinear function and ARMA parameters.

As in other nonparametric techniques, to control for the bias-variance tradeoff, estimation relies on a set of smoothing parameters. For example, \cite{Altman, Hart1991, HERRMANN1992,RayTsay2007, Krivobokova2007,BRABANTER2018} discuss the choice of smoothing parameters in nonparametric regression with serially correlated disturbances. In our neural network approach, the bias-variance tradeoff is controlled by both the number of layers through which the data are transformed and the size of each layer. We find that early stopping \cite{Prechelt2012}, a regularization technique, can alleviate the need for extensive model selection, making it practical to use a sufficiently flexible network architecture.

Finally, an alternative to the explicit modeling of serial correlation in the disturbances of model \eqref{eq: model intro} is to assume that disturbances are uncorrelated after lagged values of the dependent variable and the regressors are included in the regression function $f(\cdot)$. In the linear framework, this strategy corresponds to autoregressive distributed lag (ADL) models \citep{HendryNielsen2007}. Within nonlinear and nonparametric settings, this strategy is standard in machine learning methods \citep{ChenTsay1993NAAR,Medeiros2021}. Modern machine learning methods such as recurrent neural networks and their specialized form, long short-term memory (LSTM) networks, are designed to exploit lagged information to capture complex nonlinear dynamics \citep{rumelhart1986,Hochreiter1997}. The LSTM variant has been shown to achieve state-of-the-art performance for a wide range of problems \citep{Jozefowicz2015, Schmidhuber2017} and thus serves as a natural benchmark in our empirical analysis.

\section{Neural networks with ARMA disturbances}\label{sec: methodology}

We consider an observed univariate time series $y_1,\ldots,y_T$ of length $T$ (output), an observed vector time series $x_1,\ldots, x_T$ comprising $K$ regressors (input), and a univariate series of disturbances $u_1,\ldots,u_T$. In matrix notation, our proposed NNARMA model is given by
\begin{equation}\label{eq: NNARMA vector}
	y=f(X;\theta _2) + u,
\end{equation}
where
\[
y := (y_1,\ldots,y_T)', \quad X := [x_1,\ldots,x_T]', \quad f(X) := (f(x_1),\ldots,f(x_T))', \quad u := (u_1,\ldots,u_T)',
\]
such that $y,f(X),u\in \R^T$ and $X\in \R^{T\times K}$, and where $f(\cdot;\theta _2)$ represents a feedforward neural network with parameter set $\theta_2$. We assume that the disturbances $u_1,\ldots,u_T$ are a sample from a weakly stationary ARMA$(p,q)$ process as specified in \ref{eq: model ARMA}, with parameter set $ \theta_1= \{ \phi_1,\ldots,\phi_p, \omega_1,\ldots,\omega_q \} $ and innovation variance $\sigma^2$. It follows that the vector $u$ has zero mean and covariance matrix $\text{Var}(u) = \Omega(\theta_1, \sigma^2) = \sigma^2 \Psi(\theta_1) $, where $ \Psi(\theta_1) $ is symmetric and has a Toeplitz (or band) structure, with elements that are functions of $\theta_1$. We compute a Cholesky decomposition of the scaled covariance matrix $ \Psi(\theta_1) = C(\theta_1)^{-1} [C(\theta_1)^{-1}]' $, where $ C(\theta_1) $ is a $ T \times T $ lower triangular matrix computed via the Durbin--Levinson algorithm; details are provided in Appendix \ref{sec: cholesky}. The (inverse) Cholesky factor $C(\theta_1)$ is used for pre-whitening the data. Pre-multiplying \eqref{eq: NNARMA vector} by $ C(\theta_1) $ yields a regression model with serially uncorrelated disturbances with mean zero and variance $\sigma^2$:
\begin{equation}
	y^\ast = C(\theta_1)f(X;\theta_2) + u^\ast, \label{eq: C formulation}
\end{equation}
where $ y^\ast = C(\theta_1)y $ and $ u^\ast = C(\theta_1) u $, with mean vector $\E(u^\ast)=0$ and covariance matrix $\text{Var}(u^\ast)=\sigma^2 I_T$. If $f(\cdot)$ is linear, its parameter set is simply $\theta_2 = \{\beta\}$, where $\beta$ is a column vector of regression coefficients, and the matrix factor $C(\theta_1)$ can be pre-multiplied directly onto $X$ to obtain $X^\ast=C(\theta _1) X$. The least squares estimator is then obtained by minimizing the residual sum of squares in the pre-whitened regression model $ \lVert y^\ast - X^\ast \beta \rVert_2^2 $ with respect to $\beta$, subject to a given set of values for $\theta_1$. The joint estimation of $\theta_1$ and $\beta$ can be obtained by NLS, that is, by numerically minimizing the least squares objective with respect to $\{\theta _1,\beta\}$. Next, we extend this approach to the nonlinear case where $f(\cdot)$ is represented by a neural network. 

For the transformed NNARMA model \eqref{eq: C formulation} where $ f(\cdot) $ is represented by a neural network, the complication arises that $f(\cdot)$ does not commute. 
However, the compositional structure induced by the layered network architecture allows us to implement the regression component in \eqref{eq: C formulation} through the scheme
\begin{equation} \label{eq: output NN}
	C(\theta_1)f(X;\theta_2) = Z^{\ast (h)} \beta, \qquad Z^{\ast (h)} = C(\theta_1) Z^{(h)} ,
\end{equation}
where $\beta$ remains a regression coefficient vector and $h\in\mathbb{N}$ hidden layers jointly construct the recursive system of matrix equations
\begin{align}
	Z^{(h)} &= g \left( \kappa^{(h)} + Z^{(h-1)} \Gamma^{(h)} \right), \nonumber \\
	&\vdots  \label{eq: hidden last} \\
	Z^{(1)} &= g \left( \kappa^{(1)} + X \Gamma^{(1)} \right),\nonumber 
\end{align}
with element-wise nonlinear activation function $g(\cdot)$, intercept vector $\kappa ^{(\ell)}$, and coefficient matrix $\Gamma ^{(\ell)}$, for $\ell=1,\ldots,h$, and parameter set $\theta_2 = \{\beta, \kappa^{(1)},\ldots,\kappa^{(h)}, \Gamma^{(1)},\ldots, \Gamma^{(h)} \}$. The pre-whitening mechanism in \eqref{eq: output NN} is novel in a neural network context. Instead of applying the pre-whitening mechanism directly to $X$ as in the linear model, it is applied to the variables derived by the network $Z^{(h)}$ to obtain $ Z^{\ast (h)} $. The output of the network in \eqref{eq: output NN} is linear in $ Z^{\ast (h)}$, such that the vector $Z^{\ast (h)} \beta \in \mathbb{R}^T$. The variables derived by the network constitute nonlinear transformations of the regressors $ X $, learned through the sequence of feedforward hidden layers in \eqref{eq: hidden last}. The output of each layer is a $ T \times m_\ell $ matrix $ Z^{(\ell)} $, $ \ell = 1,\ldots,h $, obtained from an affine transformation of its inputs followed by elementwise application of the nonlinear (activation) function $ g(\cdot) $. The function $ g(\cdot) $ induces nonlinear input transformations in each layer, introducing flexibility into the regression function and allowing the network to represent nonlinear relationships between $X$ and $y$. We use the Swish function, defined as $ g(z) = z \left(1 + \exp(-z) \right)^{-1} $ for $z\in \mathbb{R}$, which is a generalization of the widely used rectified linear unit (ReLU; \citealp{NairHinton2010}). It has been shown to perform well in comparison with ReLU for a range of problems \citep{ramach2018}. In each hidden layer, $ \kappa^{(\ell)} $ is a $ m_\ell $-dimensional vector of intercepts that is added row-wise, and $ \Gamma^{(\ell)} $ is a $ m_{\ell-1} \times m_\ell $ coefficients matrix. We follow the convention to set $ m_0 \equiv K $. The hyperparameter $ h\geq 1 $ denotes the number of hidden layers (the network depth), and $ m_\ell $ denotes the number of columns in $ Z^{(\ell)} $ (the size of the $ \ell $th hidden layer). The choice of network architecture, as characterized by its depth and layer sizes, is a standard model selection problem subject to a bias-variance tradeoff. Model selection is discussed in Section \ref{sec: Model selection (methodology)}.

To compute predictions $\hat y = f(X; \hat \theta_2) + \hat u $ based on the NNARMA model in \ref{eq: NNARMA vector}, neural network predictions of the regression component (without the pre-whitening mechanism) $f(X; \hat \theta_2) $ are combined with ARMA predictions of the disturbance component $\hat u$. The latter is conveniently obtained from the Durbin--Levinson algorithm applied to the neural network residuals.

\subsection{Parameter estimation}\label{sec: estimation}

To jointly estimate the set of ARMA parameters $ \theta_1= \{ \phi_1,\ldots,\phi_p, \omega_1,\ldots,\omega_q \} $ and the set of neural network parameters $ \theta_2 = \{\beta, \kappa^{(1)},\ldots,\kappa^{(h)}, \Gamma^{(1)},\ldots, \Gamma^{(h)} \} $, we minimize the residual sum of squares from the pre-whitened model in \eqref{eq: C formulation},
\begin{equation}\label{eq: loss function}
	\min_{\theta_1} \min_{\theta_2} \lVert y^\ast - C(\theta_1)f(X; \theta_2) \rVert_2^2 ,
\end{equation}
subject to two sets of constraints within $\theta _1$ given by
\begin{equation} \label{eq: stationarity condition}
	1 - \phi_1 z - \ldots - \phi_p z^p \neq 0,  \qquad 1 + \omega_1 z + \ldots + \omega_q z^q \neq 0, \qquad z \in \mathbb{C}, \qquad \lvert z \rvert \leq 1, 
\end{equation}
which are imposed to enforce stability, for a stationary process $u_t$, and invertibility, for unique identification of $\omega_1,\ldots,\omega _q$, respectively \citep{hamilton1994series}. We solve the joint minimization problem by profiling out $ \theta_2 $ from the sum of squares. For a given $ \theta_1 $, an estimate of $ \theta_2 $ is obtained by numerically solving the inner minimization problem in \eqref{eq: loss function} using standard techniques from the neural networks literature:
\begin{equation}\label{eq: theta_2}
	\hat{\theta}_2(\theta_1) = \arg \underset{\theta_2}{\min} \lVert y^\ast - C(\theta_1)f(X; \theta_2) \rVert_2^2.
\end{equation}
We rely on the popular Adam variant of gradient descent \citep{kingma2014adam} to obtain estimates from \eqref{eq: theta_2}. We use all data available in the estimation sample when evaluating gradients, sometimes referred to as batch learning. An estimate of $ \theta_1 $ is obtained by solving the outer minimization problem in \eqref{eq: loss function}, for the given choice of $ \theta_2 =\hat{\theta}_2(\theta_1)$ and subject to the constraints in \eqref{eq: stationarity condition}:
\begin{equation}\label{eq: theta_1}
	\hat{\theta}_1 = \arg \underset{\theta_1}{\min} \lVert y^\ast - C(\theta_1)f(X; \hat{\theta}_2(\theta_1)) \rVert_2^2 .
\end{equation}
We use Jones' reparametrization of the parameters in $\theta _1$ to impose the stability and invertibility constraints in \eqref{eq: stationarity condition}, allowing \eqref{eq: theta_1} to be solved via unconstrained optimization; see Appendix \ref{sec: Jones} for details. We therefore rely simply on traditional (gradient free) numerical optimization techniques of \cite{Powell1964} to obtain \eqref{eq: theta_1} subject to the constraints implied by \eqref{eq: stationarity condition}. Finally, the estimate of $ \sigma^2 $ is obtained from
\begin{equation}
	\hat \sigma^2 = \frac{\lVert y^\ast - C(\hat \theta_1)f(X; \hat \theta_2(\hat \theta_1)) \rVert_2^2 }{T}.
\end{equation}
We emphasize that the estimation procedure above solves the minimization problem in \eqref{eq: loss function} with respect to $\theta_1$ and $\theta_2$ subject to constraints \eqref{eq: stationarity condition}. It is \emph{not} an iterative step-by-step method such as the Cochrane--Orcutt procedure, where the estimate of $\theta_1$ is fixed when minimizing over $\theta_2$, after which the estimate of $\theta_2$ is fixed when minimizing over $\theta_1$.

The Adam algorithm used to solve \eqref{eq: theta_2} and the Powell algorithm used to solve \eqref{eq: theta_1} are both local optimization routines. For the Adam algorithm, we distinguish between cold starts, in which the optimization with respect to $\theta_2$ is initialized from scratch, and warm starts, in which it is initialized from the solution obtained for the previous candidate value of $\theta_1$. Cold starts may lead to more thorough exploration of the $\theta_2$ parameter space, whereas warm starts may lead to faster convergence. We use cold starts throughout, except for the Monte Carlo experiments involving repeated estimation across many ARMA specifications (Section \ref{sec: ARMA misspecification main text}), where warm starts are used to reduce computational time. Appendix \ref{sec:Initialization MC Appendix} provides further details on the initialization of both algorithms and presents a simulation-based comparison of the two initialization schemes for the Adam algorithm. We find little difference between them.

\subsection{Early stopping}\label{sec: Early stopping}

We employ early stopping \citep{Prechelt2012} to mitigate overfitting and reduce the need for extensive model selection. In practice, early stopping reduces sensitivity to the choice of network architecture; see Section \ref{sec: Model selection (methodology)}.

The observation sample is split into an estimation and a validation sample. Let $T$ and $T_\text{val}$ denote the lengths of the estimation and the validation sample, with $T_{\text{tot}} = T + T_\text{val}$ denoting the length of the joint estimation and validation sample. The validation sample consists of observations $(X_t,y_t)$ for $t=T+1,\ldots,T_{\text{tot}}$, and follows the estimation sample in time to avoid information spillover due to serial correlation (look-ahead bias). We define
\begin{align*}
	y_{1:T_{\text{tot}}} := (y_1,\ldots,y_{T_{\text{tot}}})', \qquad X_{1:T_{\text{tot}}}  := [x_1,\ldots,x_{T_{\text{tot}}}]', \\
	f(X_{1:T_{\text{tot}}}) := (f(x_1),\ldots,f(x_{T_{\text{tot}}}))', \qquad u_{1:T_{\text{tot}}} := (u_1,\ldots,u_{T_{\text{tot}}})'.
\end{align*}
Let $ C(\hat \theta_1)_{1:T_{\text{tot}}} $ denote the Cholesky factor of $ \text{Var}(u_{1:T_{\text{tot}}}) $. Denote by $ C(\hat \theta_1)_{T+1:T_{\text{tot}}} $ the submatrix formed by the last $T_\text{val}$ rows of $ C(\hat \theta_1)_{1:T_{\text{tot}}} $. After each iteration of the Adam algorithm used to estimate the parameters of the neural network $ \theta_2 $, early stopping monitors the loss on the validation sample based on candidate choices $\hat \theta_1, \hat \theta_2$,
\begin{equation}\label{eq: val loss}
	L_{\text{val}} = \lVert C(\hat \theta_1)_{T+1:T_{\text{tot}}} ( y_{1:T_{\text{tot}}} - f(X_{1:T_{\text{tot}}}; \hat \theta_2) ) \rVert_2^2.
\end{equation} 
Pre-multiplication by $ C(\hat \theta_1)_{T+1:T_{\text{tot}}} $ in \eqref{eq: val loss} restricts attention to the validation sample. The Adam algorithm is stopped when the validation loss $L_{\text{val}} $ fails to decrease by more than a prescribed tolerance over a pre-specified number of consecutive iterations, defined by a patience parameter. We retain the estimate of $ \theta_2 $ associated with the lowest validation loss across iterations (not necessarily the last iteration). We fix the patience parameter at 50 iterations throughout. The stopping tolerance is calibrated relative to the scale of the validation loss. Specifically, if $L_{\text{val},0} $ denotes the validation loss after a short preliminary optimization of 50 iterations, early stopping is based on an absolute tolerance equal to $\epsilon L_{\text{val},0} $, where $\epsilon$ is a fixed relative tolerance. This yields a stopping rule that is approximately invariant to rescaling of the validation loss and aligns it more closely with the relative objective function tolerance employed by the Powell algorithm. Accordingly, we use the same relative tolerance, $\epsilon=10^{-4}$, to specify the stopping rule in the Adam algorithm and the relative objective function tolerance in the Powell algorithm.

\subsection{Model selection}\label{sec: Model selection (methodology)}

Our proposed NNARMA model requires choices for the number of hidden layers in the neural network $h$, the size of each layer $m_1,\ldots,m_h$, and the ARMA orders $p$ and $q$. Model selection can be performed based on in-sample criteria or out-of-sample predictive performances using time series cross-validation. However, repeated re-estimation across candidate network architectures and ARMA specifications quickly becomes computationally prohibitive. We therefore adopt a practical strategy that avoids such repeated estimation and is motivated by simulation evidence. Specifically, we fix a sufficiently flexible network architecture combined with early stopping and select the ARMA orders $p$ and $q$ by minimizing the Bayesian information criterion (BIC) applied to residuals from a preliminary neural network regression that assumes serially uncorrelated errors. 

Simulation evidence demonstrates largely stable validation loss across network architectures, with mostly very narrow networks (each layer size is small) tending to produce higher losses (Appendix \ref{sec:Model selection MC Appendix}). With early stopping, performance is therefore largely insensitive to the precise choice of architecture once the network is sufficiently flexible. Further, it is shown that the ARMA specification selected by our practical strategy generally converges to the true specification as the sample size increases.

\section{Monte Carlo experiments}\label{sec: Monte Carlo (main text)}

In this section, we summarize the results from a comprehensive suite of Monte Carlo experiments. We examine the finite-sample properties of the estimation procedure in Section \ref{sec: estimation}, with early stopping as described in Section \ref{sec: Early stopping}, and evaluate the out-of-sample predictive performance of the NNARMA model relative to existing feedforward neural networks with lagged variables. We consider both correct specification of the NNARMA model and misspecification arising from the ARMA component as well as from omitted lagged variables (dynamic misspecification). Additional details on the Monte Carlo experiments, including the data-generating processes and supplementary results, are provided in Appendix \ref{sec: Monte Carlo (appendix)}.

\subsection{Monte Carlo setup and benchmark models}\label{sec: MC setup Main text}

For all experiments except those involving dynamic misspecification (Section \ref{sec:dynammic misspecification maint text}), we generate a univariate response variable $y_t$ as the sum of a single regression function $f(x_{1t},x_{2t})$ (a hump-shaped or a sinusoidal regression function, respectively) and a noise term. The regression input variables $x_{1t}$ and $x_{2t}$ are generated by stationary $\text{AR}(1)$ zero-mean processes, with autoregressive coefficients $0.8$ and $0.7$, respectively. The noise term is generated by a stationary $\text{ARMA}(1,2)$ zero-mean process with autoregressive coefficient $\phi _1 = 0.9$ and moving average coefficients $\omega _1 = -0.5$ and $\omega _2 = 0.2$. 

We split each Monte Carlo sample into three consecutive parts: (1) an estimation sample (first $ 60\% $), (2) a validation sample used for early stopping (next $ 20\% $), and (3) a test sample reserved for out-of-sample evaluation (final $ 20 \% $). All models are estimated once, and we compute out-of-sample predictions conditional on the regressors (and lagged response variables) in the test sample. This setup mimics the empirical setting in Section \ref{sec: CFC application}.

For convenience, we denote by $\text{NNARMA}(p,q)$ the model proposed in Section \ref{sec: methodology} with $\text{ARMA}(p,q)$ disturbances. We denote by $\text{NN}(\ell_y,\ell_x)$ the feedforward neural network benchmark models with lagged variables, where $\ell_y,\ell_x \in \N $ denote the number of lagged dependent variables and regressors, respectively, in addition to the contemporaneous regressors. Under this notation, $\text{NNARMA}(0,0)$ and $\text{NN}(0,0)$ are equivalent. We consider 8 benchmark models: $\text{NN}(0,0)$, $\text{NN}(1,0)$, $\text{NN}(0,1)$, $\text{NN}(1,1)$, $\text{NN}(2,2)$, $\text{NN}(3,3)$, $\text{NN}(5,5)$, $\text{NN}(10,10)$.
We fix the network architecture to two hidden layers with $ 32 $ units in the first and $ 16 $ in the second, both for the NNARMA model and the benchmark models. This architecture is not intended to be optimal, but rather serves as a representative, sufficiently flexible choice for illustrating model performance without extensive tuning; see the discussion in Section \ref{sec: Model selection (methodology)}. We use the same network architecture in the NNARMA model and the benchmark models to ensure that differences in predictive performance reflect differences in model specification rather than architecture tuning. 

We denote the signal-to-noise ratio by $r$ and the sample size by $T$. Our results below are aggregated for both regression functions and all combinations of $ r \in \{0.05, 1.0\} $ and $ T \in \{250, 1000, 5000\} $ using 100 Monte Carlo replications. Appendix \ref{sec: Monte Carlo (appendix)} extends the analysis to $ r \in \{0.05, 0.1, 0.2, 1.0\} $ and $ T \in \{250, 500, 1000, 5000\} $ for a more comprehensive evaluation. These additional results do not alter the conclusions of this section.

\begin{figure}[t!]
	\centering
	\begin{subfigure}[c]{0.47\textwidth}
		\centering
		\includegraphics[width=\textwidth]{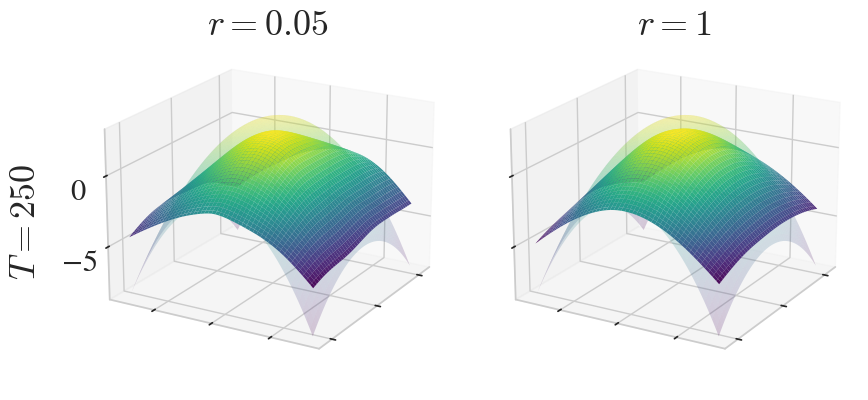}
	\end{subfigure}
	\hfill
	\begin{subfigure}[c]{0.47\textwidth}
		\centering
		\includegraphics[width=\textwidth]{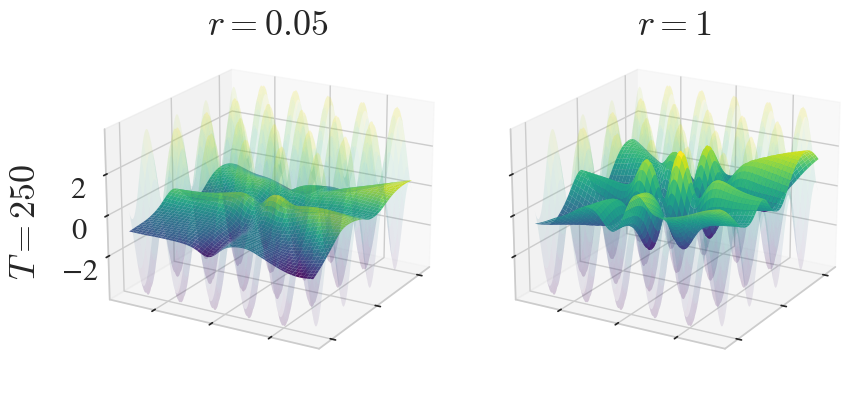}
	\end{subfigure} \\ \vspace*{-0.45cm}
	\begin{subfigure}[c]{0.47\textwidth}
		\centering
		\includegraphics[width=\textwidth]{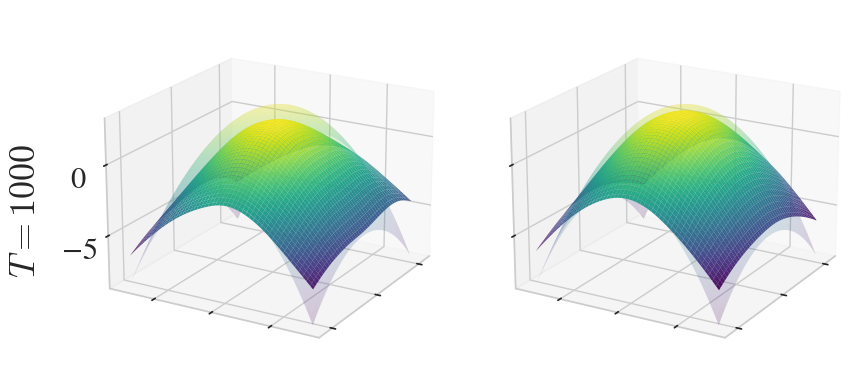}
	\end{subfigure}
	\hfill
	\begin{subfigure}[c]{0.47\textwidth}
		\centering
		\includegraphics[width=\textwidth]{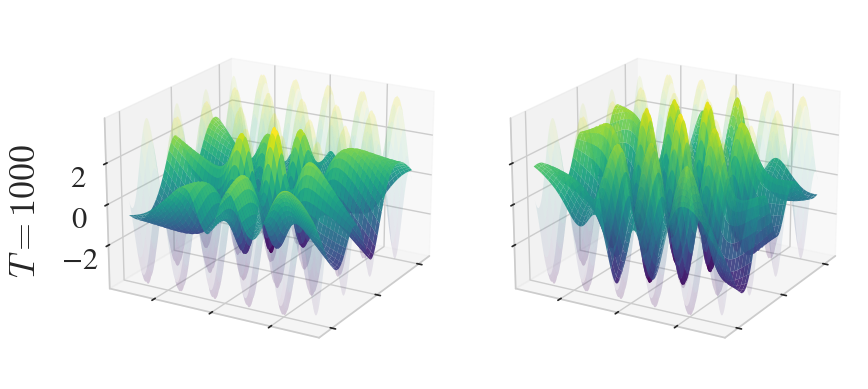}
	\end{subfigure} \\ \vspace*{-0.45cm}
	\begin{subfigure}[c]{0.47\textwidth}
		\centering
		\includegraphics[width=\textwidth]{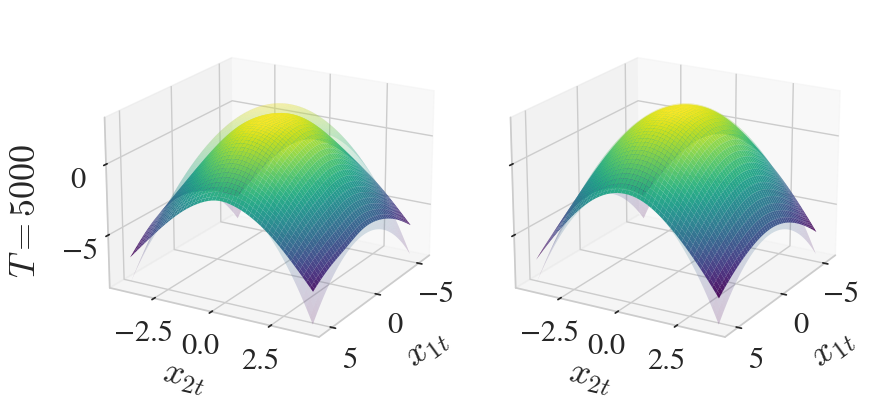}
		\subcaption{Hump-shaped regression function}
	\end{subfigure}
	\hfill
	\begin{subfigure}[c]{0.47\textwidth}
		\centering
		\includegraphics[width=\textwidth]{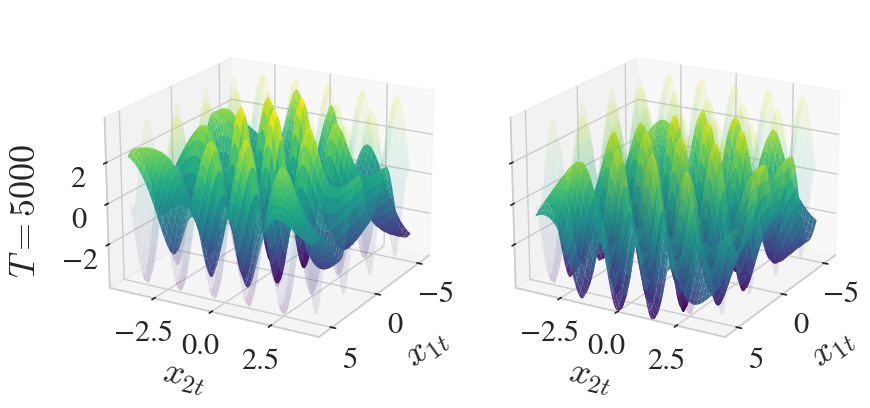}
		\subcaption{Sinusoidal regression function}
	\end{subfigure} 
	\caption{Average estimate of the hump-shaped (panel a) and the sinusoidal (panel b) regression functions. Subplots are arranged by $r$ (signal-to-noise ratio; columns) and $T$ (sample size; rows). The true function is shown transparently in the background.}
	\label{fig: estimated reg func hump sine}
\end{figure}

\begin{figure}[t!]
	\centering
	\includegraphics[width=\textwidth]{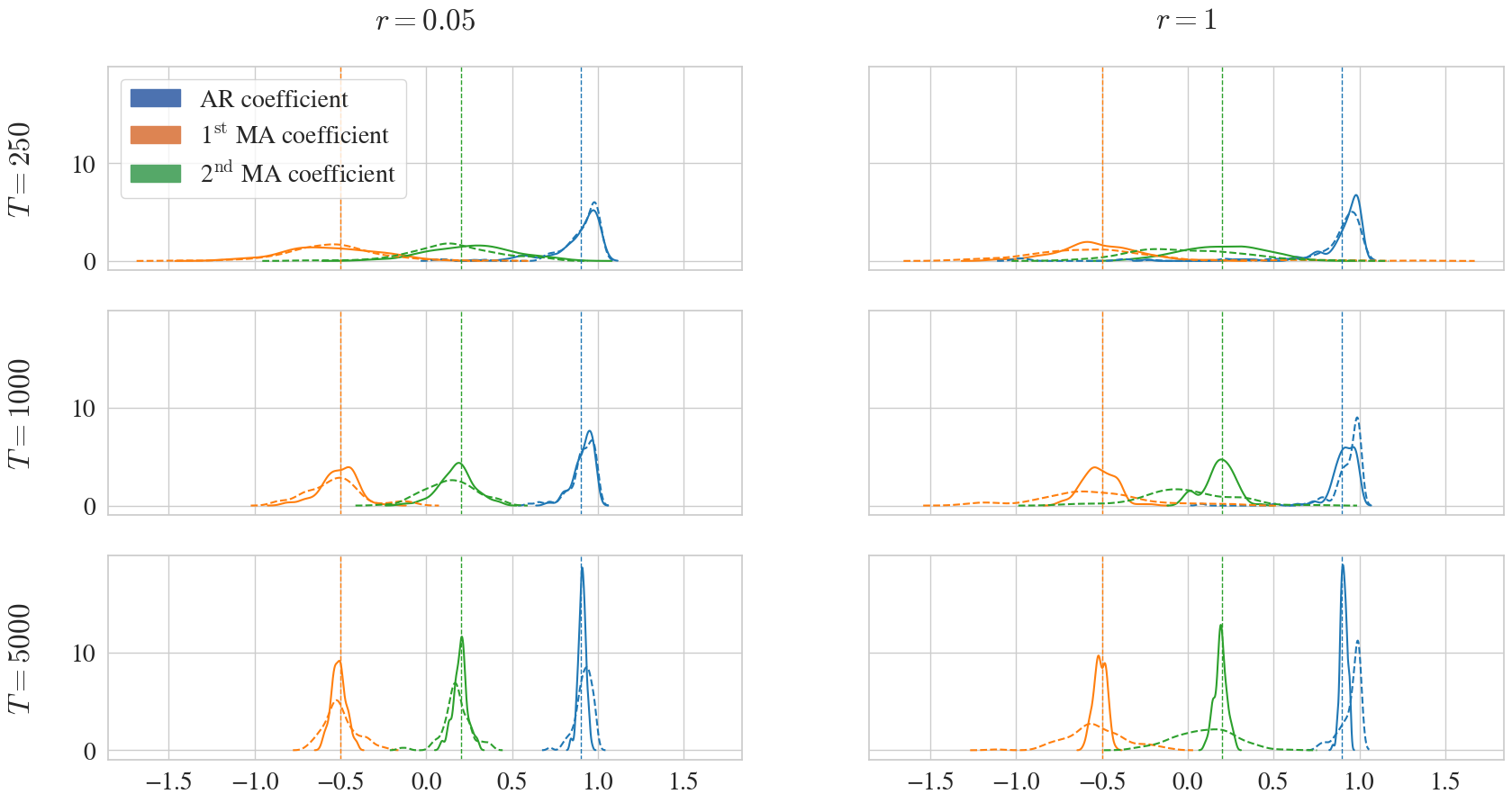}
	\caption{Sampling distribution of the ARMA coefficient estimates for the hump-shaped (solid) and sinusoidal (dashed) regression functions. Subplots are arranged by $r$ (columns) and $T$ (rows). Vertical dashed lines indicate the true coefficient values.}
	\label{fig: ARMA coeffs main}
\end{figure}

\subsection{Finite sample properties}\label{sec: Finite sample properties main text}

We start by examining the finite sample properties of the proposed estimation procedure, including early stopping, under correct specification of the $\text{NNARMA(1,2)}$ model. 

Figure \ref{fig: estimated reg func hump sine} displays the NNARMA average estimate of the hump-shaped and sinusoidal regression function. For both regression functions, accuracy of the estimated surface is increasing with $r$ and $T$ as expected. For high $r$ (strong signal) and large $T$ (large sample), both functions are estimated with a high degree of accuracy. The hump-shaped function is estimated with reasonable accuracy even for low $r$ (noisy signal) and small $T$ (small sample), especially in the interior of the input region. All estimated surfaces exhibit reduced accuracy near the boundaries of the input region, a pattern that is particularly pronounced for the sinusoidal function and is well known in nonparametric settings \citep{Cattaneo2020} and neural network applications \citep{Jensen2023}. At low $r$, the sinusoidal function is estimated fairly well provided the sample size is large enough. It generally requires a larger sample size than for the hump-shaped function. 

Corresponding to the surface estimates in Figure \ref{fig: estimated reg func hump sine}, Figure \ref{fig: ARMA coeffs main} displays the sampling distribution of the ARMA coefficient estimates. Throughout, sampling distributions are estimated by kernel density estimation using a Gaussian kernel and the bandwidth suggested by \cite{Silverman}. The behavior of the coefficient estimates closely mirrors that of the surface estimates: estimation accuracy decreases with $r$ and increases with $T$. For low $r$ and large $T$, all coefficients are accurately estimated for both regression functions. For the hump-shaped function, estimation accuracy is high even for high $r$ and small $T$. For the sinusoidal function, the AR coefficient is estimated accurately for high $r$ and small $T$, although the sampling distribution peaks slightly above the true value. For the MA coefficients, the distributions are more dispersed than for the hump-shaped function, especially for high $r$, and concentrate more slowly around their true values as $T$ increases.

\begin{figure}[t!]
	\centering
	\includegraphics[width=\textwidth]{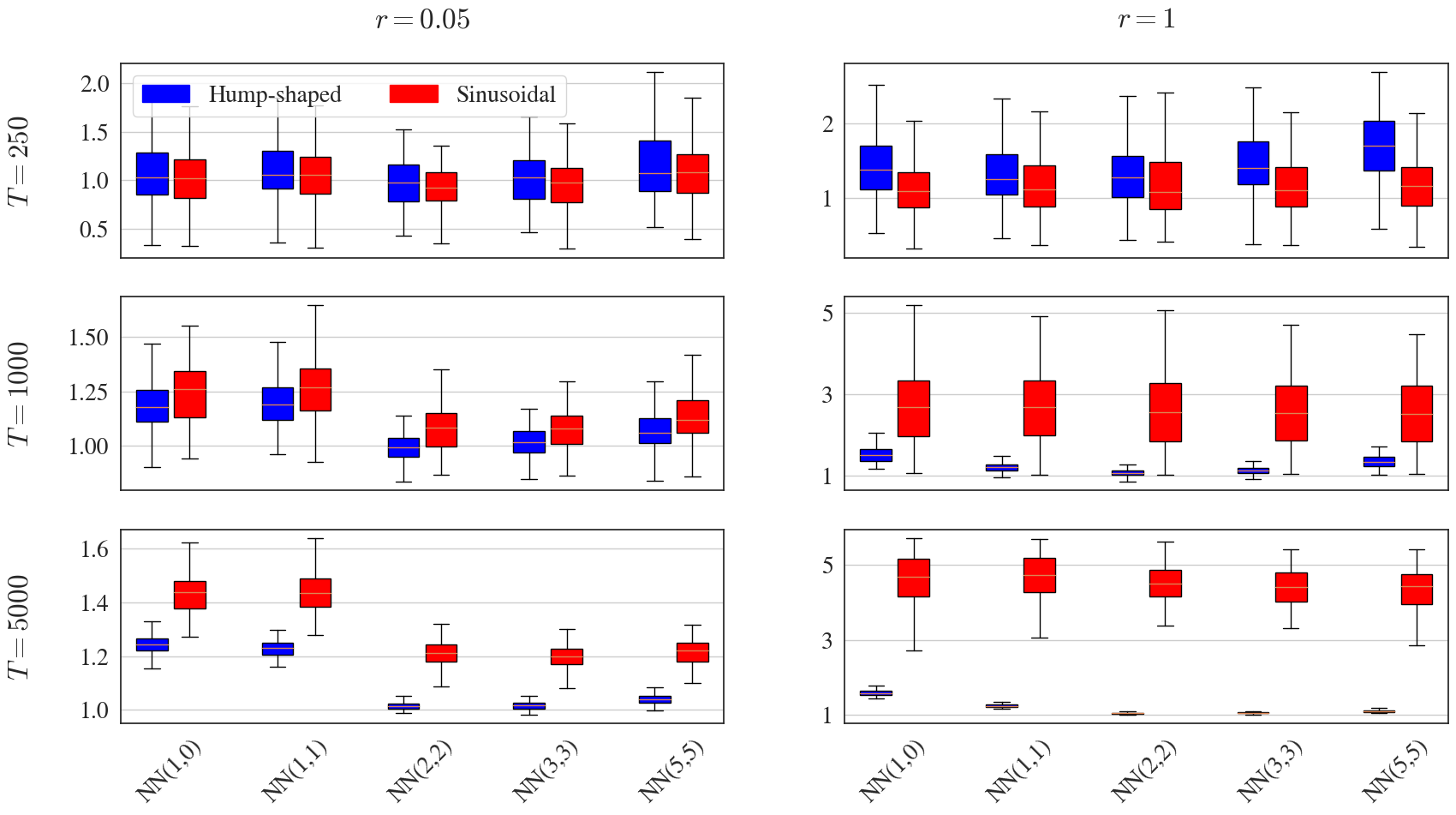}
	\caption{Out-of-sample results under correct specification of NNARMA. Box plots of relative MSEs normalized against NNARMA for the hump-shaped (blue) and sinusoidal (red) regression functions. Values above one indicate lower MSE for NNARMA. Subplots are arranged by $r$ (columns) and $T$ (rows).}
	\label{fig: OOS main}
\end{figure}

\begin{figure}[t!]
	\centering
	\includegraphics[width=\textwidth]{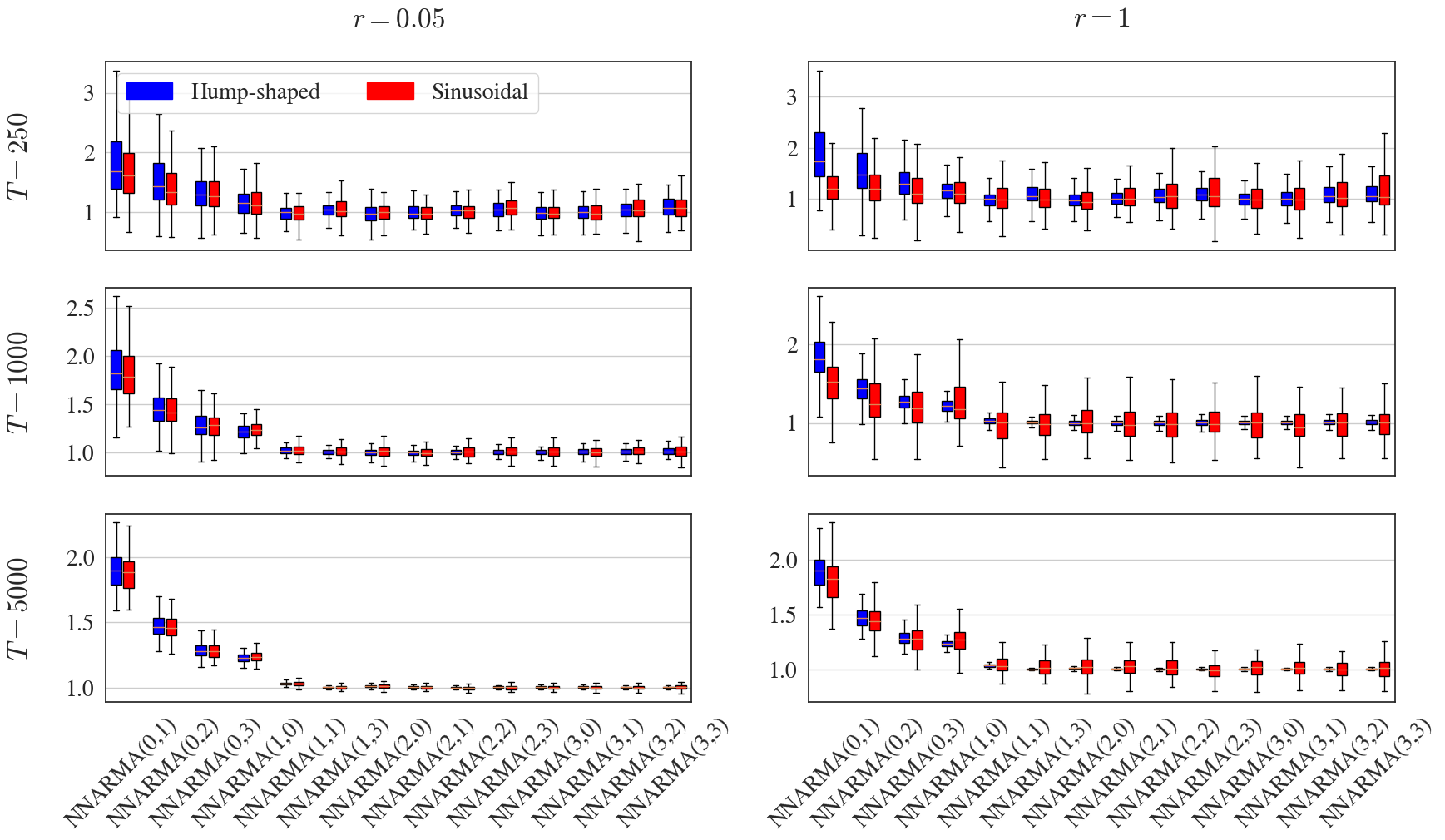}
	\caption{Out-of-sample results under ARMA misspecification, with $\text{NNARMA}(1,2)$ as the correctly specified model. Box plots of relative MSEs normalized against $\text{NNARMA}(1,2)$ for the hump-shaped (blue) and sinusoidal (red) regression functions. Values above one indicate lower MSE for $\text{NNARMA}(1,2)$. Subplots are arranged by $r$ (columns) and $T$ (rows).}
	\label{fig: OOS main ARMA mis}
\end{figure}

\begin{figure}[t!]
	\centering
	\includegraphics[width=\textwidth]{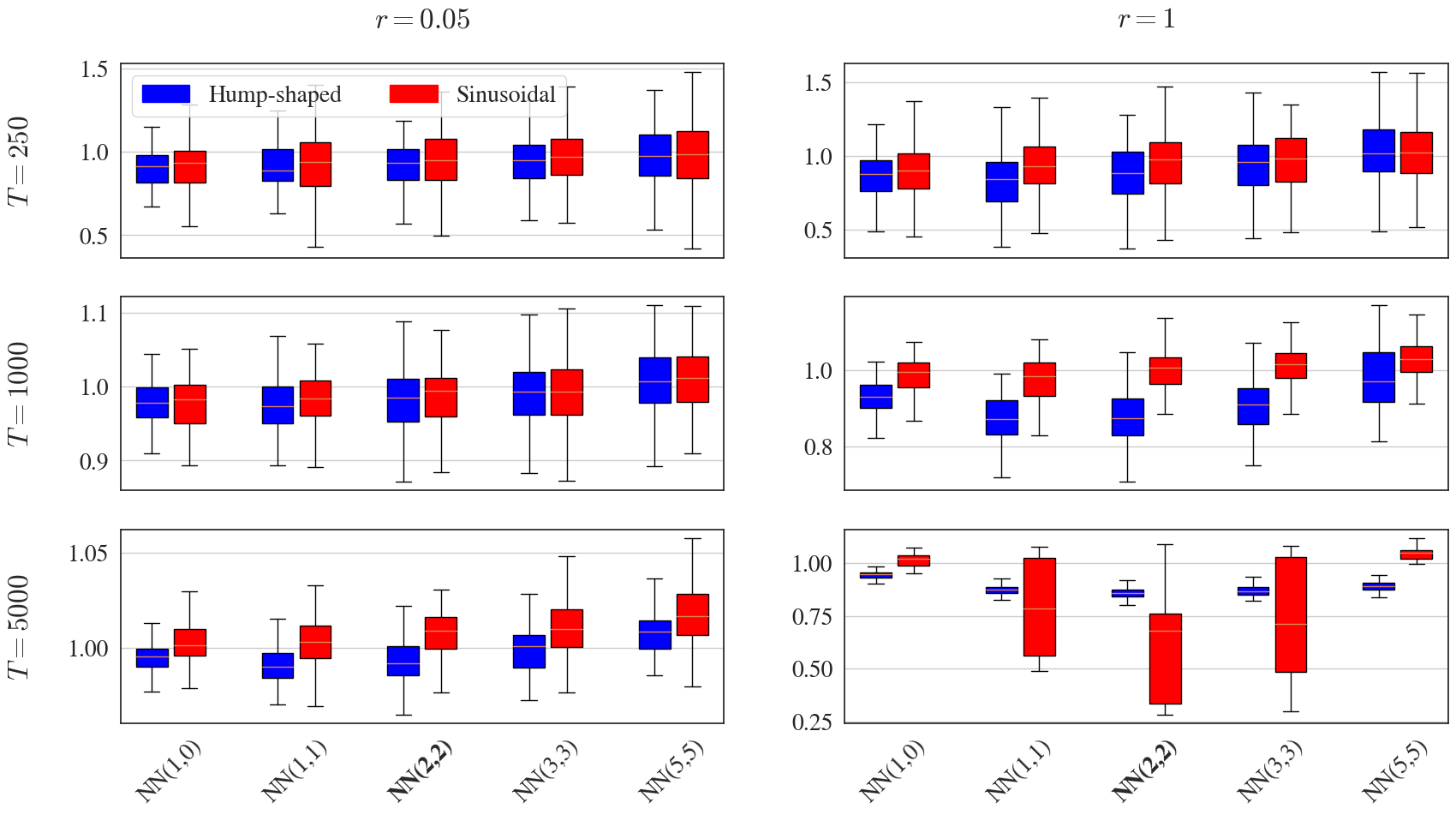}
	\caption{Out-of-sample results under dynamic misspecification of NNARMA, with $\text{NN}(2,2)$ (bold) as the correctly specified model. Box plots of relative MSEs normalized against NNARMA for the hump-shaped (blue) and sinusoidal (red) regression functions. Values above one indicate lower MSE for NNARMA. Subplots are arranged by $r$ (columns) and $T$ (rows). The NNARMA model selects $p$ and $q$ automatically for each replication using the practical model selection strategy described in Section \ref{sec: Model selection (methodology)}.}
	\label{fig: OOS main dyn mis}
\end{figure}

\subsection{Out-of-sample prediction performance}\label{sec: Out-of-sample main text}

We next extend the in-sample analysis using the same simulated data, and evaluate the out-of-sample prediction accuracy of the correctly specified $\text{NNARMA(1,2)}$ model relative to the benchmark models with lagged variables. 

Figure \ref{fig: OOS main} presents Box plots of the relative out-of-sample mean squared error (MSE) for the benchmark models, normalized against the NNARMA model. To avoid distorting Figure \ref{fig: OOS main}, we omit the results for $\text{NN}(0,0)$ and $\text{NN}(0,1)$, whose MSEs are substantially higher, and for $\text{NN}(10,10)$, whose performance is worse than that of specifications with fewer lags. The omitted results are reported in Appendix \ref{sec: Finite sample properties benchmarks}. The vast majority of relative MSEs exceed one, indicating that the NNARMA model yields more accurate predictions than the benchmarks in most replications, with particularly pronounced gains for the sinusoidal function. The improvements in accuracy of NNARMA relative to the benchmarks models are increasing with $r$ and $T$.

\subsection{ARMA misspecification}\label{sec: ARMA misspecification main text} 

We examine the effect of ARMA misspecification on the finite-sample properties and out-of-sample predictive performances of the NNARMA models using the same simulated data set as above. Specifically, we consider all $16$ $\text{NNARMA}(p,q)$ model specifications with $p,q\in\{0,1,2,3\}$. Here, we focus on the out-of-sample performance. Appendix \ref{sec: MC Misspecification} presents a detailed discussion of both the finite-sample properties and out-of-sample performance.

Figure \ref{fig: OOS main ARMA mis} presents Box plots of the relative out-of-sample MSE for all NNARMA misspecified models, except $\text{NNARMA}(0,0)$, normalized against the correct specification $\text{NNARMA}(1,2)$. The $\text{NNARMA}(0,0)$ specification yields substantially higher MSE values, especially for the sinusoidal function, and is therefore omitted to avoid distorting the graphs; the omitted results are reported in Appendix \ref{sec: MC Misspecification}. Prediction accuracy is lower than that of the correct specification when the ARMA structure is substantially underspecified. In particular, the specifications $\text{NNARMA}(0,j)$, for $j=1,2,3$, and $\text{NNARMA}(1,0)$  produce less accurate predictions than $\text{NNARMA(1,2)}$. The accuracy of $\text{NNARMA(1,1)}$ and all specifications where the ARMA structure is overspecified fluctuates around the same level as $\text{NNARMA(1,2)}$, for all combinations of $T$ and $r$.

\subsection{Dynamic misspecification}\label{sec:dynammic misspecification maint text}

In our final simulation experiment, we analyze the out-of-sample predictive performance of the NNARMA model when it is misspecified and the benchmark model $\text{NN}(2,2)$ is correctly specified. For this experiment, we generate a new data set for $y_t$ from the following data generation process,
\[
y_t = f(0.4x_{t}+0.3x_{t-1}+0.2x_{t-2}) + 0.5 y_{t-1} + 0.4 y_{t-2} + e_t, \qquad e_t \sim NID (0, 1/r),
\]
with $2\times 1$ regressor vector $x_t=(x_{1t},x_{2t})'$, for $t=1,\ldots,T$. We consider the same hump-shaped and sinusoidal regression functions for $f(\cdot)$ and the same values for $T$ and $r$. The simulated data set for $x_t$ remains the same as above. To analyze the effect of omitting lagged variables from the $\text{NN}(2,2)$ specification, we compare its out-of-sample MSEs to the model specifications $\text{NN}(1,0)$, $\text{NN}(1,1)$, and NNARMA. In the latter case, the NNARMA orders $p$ and $q$ are selected automatically for each replication using the practical strategy discussed in Section \ref{sec: Model selection (methodology)}. For completeness, we also include the overspecified models $\text{NN}(3,3)$ and $\text{NN}(5,5)$. A detailed discussion of this simulation exercise is provided in Appendix \ref{sec: MC Misspecification Dynamics}, together with results for $\text{NN}(0,0)$, $\text{NN}(0,1)$, and $\text{NN}(10,10)$. The Appendix also reports results for a similar experiment in which $\text{NN}(0,2)$ is the correct specification.

Box plots of the relative out-of-sample MSEs, normalized against the NNARMA model, are presented in Figure \ref{fig: OOS main dyn mis}. Differences in predictive accuracy between the correct specification $\text{NN}(2,2)$ and the NNARMA model are generally small, but tend to become more pronounced in favor of $\text{NN}(2,2)$ for high $r$ and large $T$, with the largest difference observed for the sinusoidal function with $r=1$ and $T=5000$. In this scenario, the correct specification $\text{NN}(2,2) $ is also considerably more accurate than most of the remaining benchmark models. Appendix \ref{sec: MC Misspecification Dynamics} reports the model rankings by MSEs for each replication. For the sinusoidal function with $r \in \{0.05, 0.1, 0.2\}$ and $T=5000$, the NNARMA model ranks above the correctly specified $\text{NN}(2,2)$ model in most replications.

\section{Cloud cover prediction}\label{sec: CFC application}

Clouds play a crucial role in the climate system \citep{Hughes1984}. They are central to the global water cycle \citep{Bengtsson2010}, particularly through preciptation formation and distribution \citep{Stephens2005}. They also act as energy gatekeepers for the climate system by reflecting incoming solar radiation (cooling) and blocking outgoing terrestrial radiation (warming; \citealp{StephensETAl2012,LiuEtAl2023}). Future changes in cloud fractional cover (the proportion of the sky covered by clouds) are expected to affect global warming and, in turn, both climate and society \citep{Svennevik2024}. However, projections of future cloud cover are subject to substantial uncertainty \citep{Zelinka2020}, largely because the accurate representation of clouds remains a major bottleneck in numerical climate models, limiting the reliability of climate projections \citep{StevensBony2013,BonyEtAl2015,GrundnerEtAl2025,ThuyEtAl2025}. A more accurate representation of clouds in numerical climate models is therefore required. One solution is a statistical downscaling procedure based on regression as discussed by \cite{Svennevik2024}. We propose NNARMA as a candidate regression model, and show that it is useful for representing the nonlinear statistical relations between cloud cover and other climate variables and for improving cloud cover predictions.

\subsection{Data}\label{sec: empirical data}

We consider the European Cloud Cover data set from \cite{Svennevik2024}, available through the Open Science Framework \citep{DataOSF}. The data set comprises satellite observations of cloud fractional cover together with reanalysis fields for $(i)$ air temperature in Kelvin (K), $(ii)$ specific humidity expressed in kilograms of water vapor per kilogram of air (kg/kg), $(iii)$ relative humidity, measured by the amount of water vapor in the air as a percentage of the amount the air could potentially hold at that temperature and pressure, and $(iv)$ surface pressure in Pascals (Pa). Cloud fractional cover is the response variable in our analysis, and is sourced from the METeosat Second Generation (MSG) cloud mask from the European Organisation for the Exploitation of Meteorological Satellites (EUMETSAT; \citealp{Schmetz2002}). Sensor failures introduce missing values that are random in time but perfectly synchronous across all spatial locations. The remaining four variables serve as regressors and are sourced from the 5th Generation Reanalysis data (ERA5) from the European Centre for Medium-Range Weather Forecasts \citep{Hersbach2020}. They are retrieved from the surface or the closest pressure level (1000hPa). No missing values occur for the four regressor variables. 

Cloud formation occurs when air becomes saturated with water vapor. Saturation can arise either from decreases in temperature (cooling) or from increases in water vapor. These mechanisms are reflected in the four regressors: $(i)$ temperature determines the saturation level; $(ii)$ specific humidity measures the amount of water vapor present; $(iii)$ relative humidity defines proximity to saturation; $(iv)$ surface pressure captures additional dynamical effects through its link to vertical air movement, with low surface pressure typically associated with rising air that cools and promotes condensation. Relative humidity is derived as a nonlinear transformation of temperature and specific humidity, and therefore may imply potential redundancy in the flexible neural network models. Since relative humidity encodes key thermodynamic relationships governing saturation, this variable helps in representing the resulting nonlinearities in the data \citep{Romps2014}. At the same time, relative humidity reflects only proximity to saturation, and cloud formation is influenced by thermodynamic structure in addition to relative humidity \citep{Slingo1987}.

\begin{figure}[t!]
	\centering
	\includegraphics[width=\textwidth]{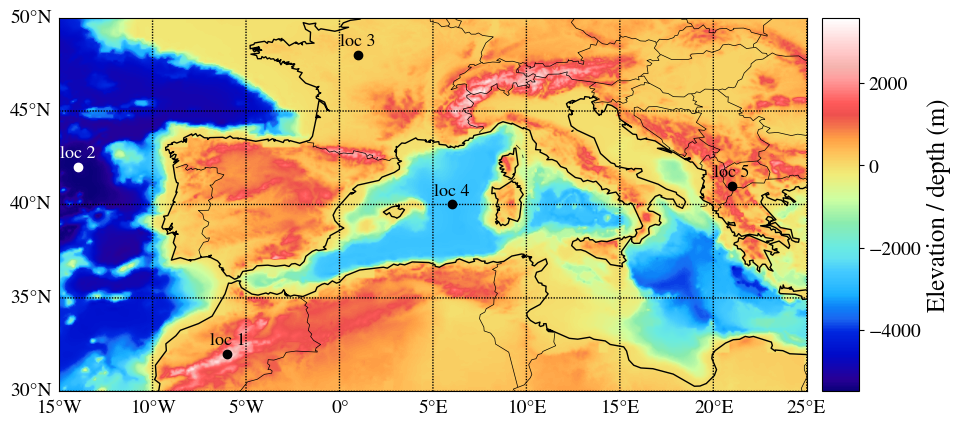}
	\caption{Topography map. Figure \ref{fig: TS plot 1} shows time series plots for the example locations labeled 1--5. Elevation and depth are obtained from the 5-minute Gridded Global Relief Data (ETOPO5; \citealp{NGDC1993})}
	\label{fig: topography map}
\end{figure}

The geographical domain of the data set is between 30\degree \,and 50\degree \,latitude (degrees north) and -15\degree \,and 25\degree \,longitude (degrees east), covering Southern Europe and Northern Africa as shown in the topography map in Figure \ref{fig: topography map}. The data are provided at a 0.25\degree horizontal resolution with hourly sampling from April 2004 to December 2018. We average the data to daily frequency and sample these at a 1.0\degree horizontal resolution. The temporal aggregation eliminates most missing values, as days with at least one valid observation yield a complete daily record. In total, we consider $861$ geographical locations. At each location, the effective sample size is $T=5370$ observations. A total of $11$ observations are omitted due to missing values, and $7$ initial observations are reserved as initial conditions for benchmark models with lagged dependent variables, see the discussion below. The initial observations are also omitted in models without lagged dependent variables to ensure a common sample across specifications.

Appendix \ref{sec: Appendix supplementary figures}, Figure \ref{fig: Sumary plot appendix} summarizes the spatial distribution of the sample mean and standard deviation of cloud fractional cover and the regressors, computed over time at each location. Mean cloud fractional cover is higher over the eastern North Atlantic Ocean, particularly between 40\degree \, and 50\degree \, latitude, than over land and the Mediterranean Sea. Variability follows the opposite pattern. Over the ocean, a steady supply of water vapor and limited temperature variation maintain consistently high and stable relative humidity and cloud cover. Over land, a limited supply of water vapor reduces mean cloud cover, while strong day--night temperature fluctuations increase the variability of relative humidity and cloud cover. The Mediterranean Sea exhibits large variability in specific humidity due to alternating dry continental and moist marine air masses, with cloud cover showing intermediate mean and variability.

\subsection{Empirical design and benchmark methods}

For each of the $861$ geographical locations, we separately predict cloud fractional cover from the four regressors, which are treated as given and known. We compare the prediction accuracy of the NNARMA model to that of the following benchmarks: a linear regression model with and without a lagged dependent variable $y_{t-1}$, as adopted in \cite{Svennevik2024}, a linear regression model with ARMA disturbances, and an LSTM neural network with and without a lagged dependent variable $y_{t-1}$. The recurrent structure of the LSTM network incorporates all prior input information when predicting cloud cover. Appendix \ref{sec: Appendix expanded set of benchmarks} reproduces the key results from this section for an expanded set of benchmarks, which includes the same models but with seven lagged dependent variables $y_{t-1},\ldots,y_{t-7}$. In all specifications, an overall intercept term is included to align with the linear benchmark models.

We adopt the same two-layer network architecture (with $ 32 $ and $ 16 $ units) as in the Monte Carlo study for both the NNARMA model and the LSTM network. We select the ARMA specifications for the NNARMA model and the linear regression with ARMA disturbances using the practical strategy discussed in Section \ref{sec: Model selection (methodology)}. We consider 36 $\text{ARMA}(p,q)$ specifications, for $p,q \in \{0,1,2,3,4,5\}$, as candidates.  

We split the daily sample into three consecutive parts: an estimation sample (2004--2012), a validation sample used for early stopping (2013--2015), and a test sample reserved for out-of-sample evaluation (2016--2018). Only the NNARMA model and the LSTM network rely on early stopping. The linear benchmarks do not, and for these models we collapse the period from 2004--2015 into an extended estimation sample. All models are estimated once. No re-estimation is performed, and predictions are conditional on the regressors (and possibly lagged dependent variables) in the test sample. This design is motivated by \cite{Svennevik2024} and mimics the Monte Carlo design in Section \ref{sec: Monte Carlo (main text)}. We measure prediction accuracy by out-of-sample mean squared errors and use the \cite{Diebold1995} test to determine statistically significant differences across models.\footnote{Diebold--Mariano tests are computed using a Bartlett lag window and a truncation lag of $7$. Similar results are obtained with truncation lags in $\{1,2,3,4,5,6,14,21\}$ and are available upon request.}

\begin{figure}[t!]
	\centering
	\includegraphics[width=\textwidth]{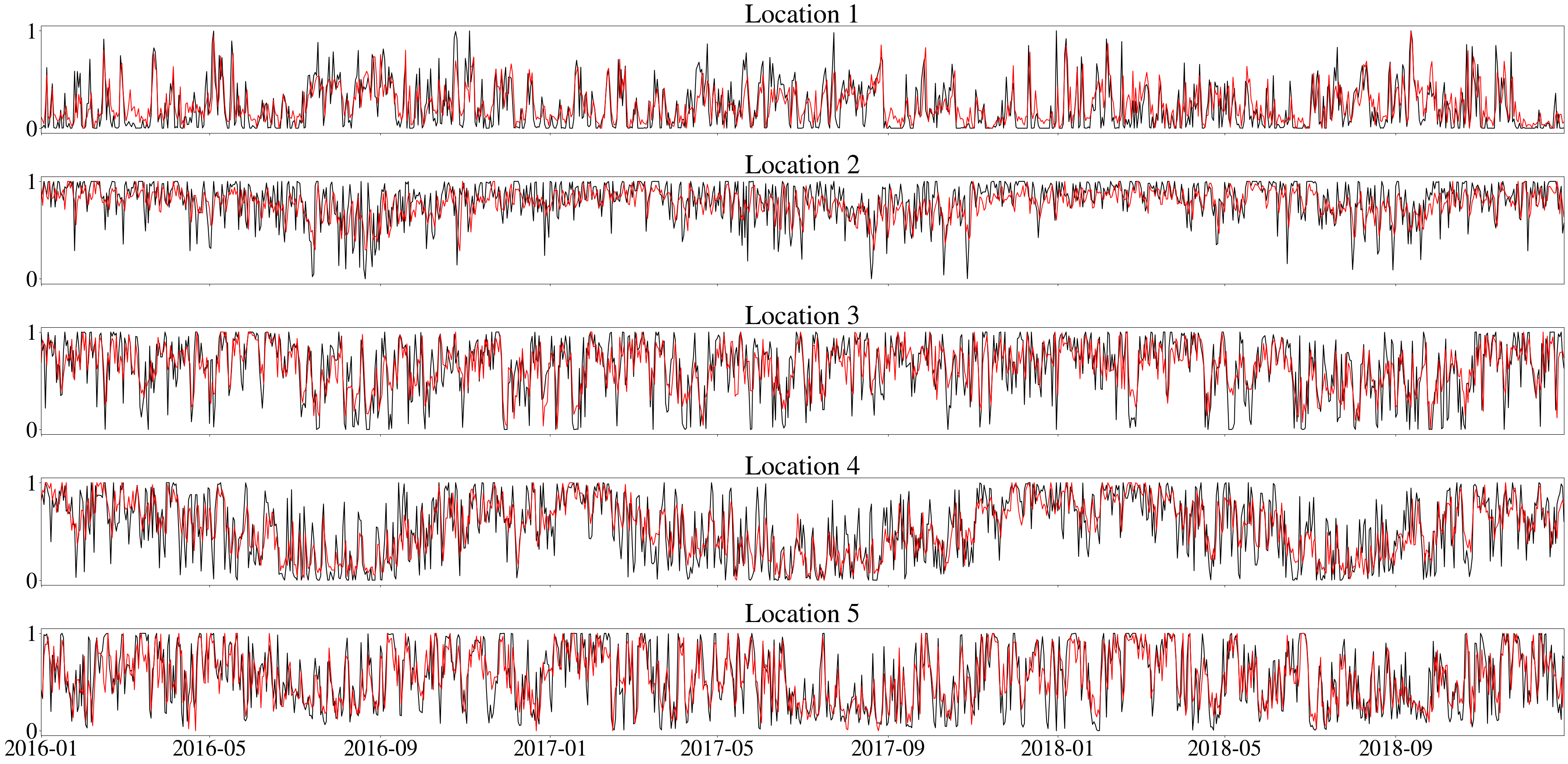}
	\caption{NNARMA predictions over the tests sample (red) together with observed cloud fractional cover (black) for the locations indicated in Figure \ref{fig: topography map}.}
	\label{fig: TS plot 1}
\end{figure}

The response variable $y_t$ lies in the unit interval $ [0,1]$. To avoid introducing additional model complexity, we follow \cite{Svennevik2024} and do not explicitly enforce this constraint through the loss function nor through structural constraints such as bounded link functions. All models are estimated with regular squared error loss. Regressors are standardized to have zero mean and unit variance.  Ex post, predictions below 0 are set to 0 and predictions above 1 are set to 1. Appendix \ref{sec: Appendix supplementary figures}, Figure \ref{fig: Truncations box plot} shows box plots, across geographical locations, of the percentage of unrestricted out-of-sample predictions falling below 0, above 1, and outside the admissible range $[0,1]$. The median percentage of unrestricted predictions outside the admissible range is 1.5\% for NNARMA and 1.8\% or below for the remaining models. Thus, violations of the range constraint are rare in practice.

\subsection{Results}

Figure \ref{fig: TS plot 1} presents the NNARMA out-of-sample predictions for the five illustrative locations indicated in Figure \ref{fig: topography map} together with the observed cloud fractional cover time series. These five locations span a representative range of land and ocean environments that are present among the total $861$ locations in our study. Consistent with the data discussion above, these time series exhibit distinct temporal dynamics. The NNARMA model appears to capture much of this variation across locations.

\begin{figure}[t!]
	\centering
	\begin{subfigure}[c]{\textwidth}
		\centering
		\includegraphics[width=\textwidth]{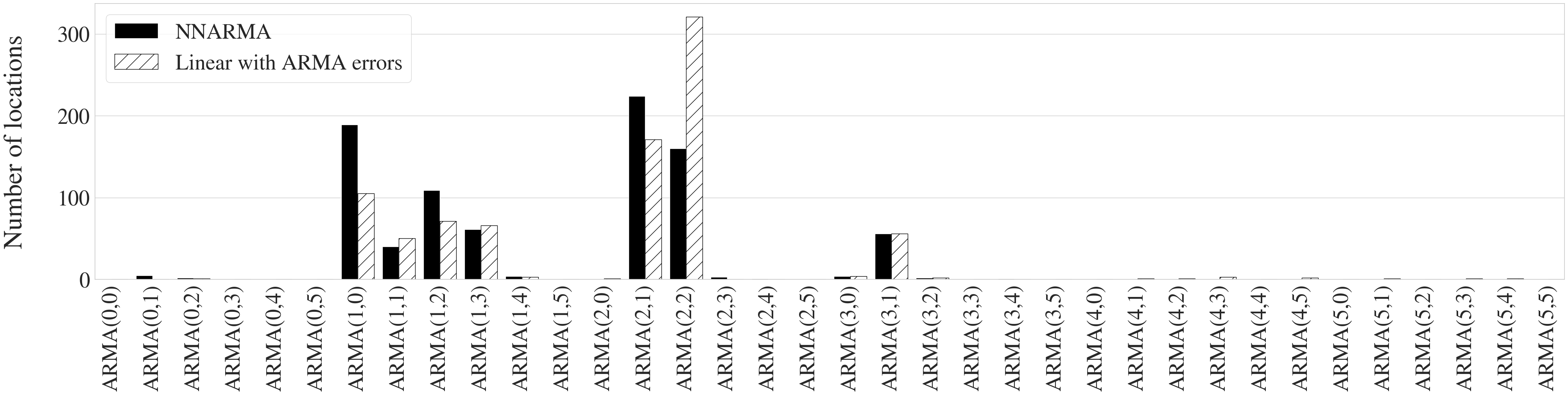}
		\subcaption{Empirical distribution of ARMA specifications.}
	\end{subfigure} \\ \vspace*{0.2cm}
	\begin{subfigure}[c]{\textwidth}
		\centering
		\includegraphics[width=\textwidth]{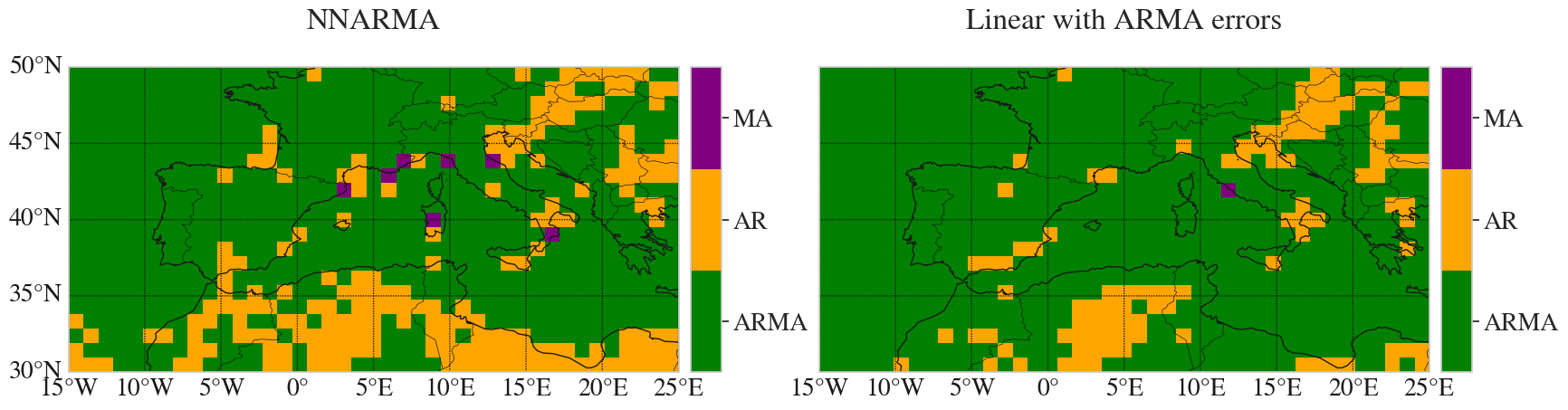}
		\subcaption{Spatial distribution of model classes.} 
	\end{subfigure} 
	\caption{Empirical distribution of ARMA specifications across the 861 geographical locations (panel a) and spatial distribution of model classes (AR, MA, or ARMA; panel b) selected by NNARMA and the linear model with ARMA errors.}
	\label{fig: ARMA selections across locations}
\end{figure}

\begin{figure}[t!]
	\centering
	\includegraphics[width=\textwidth]{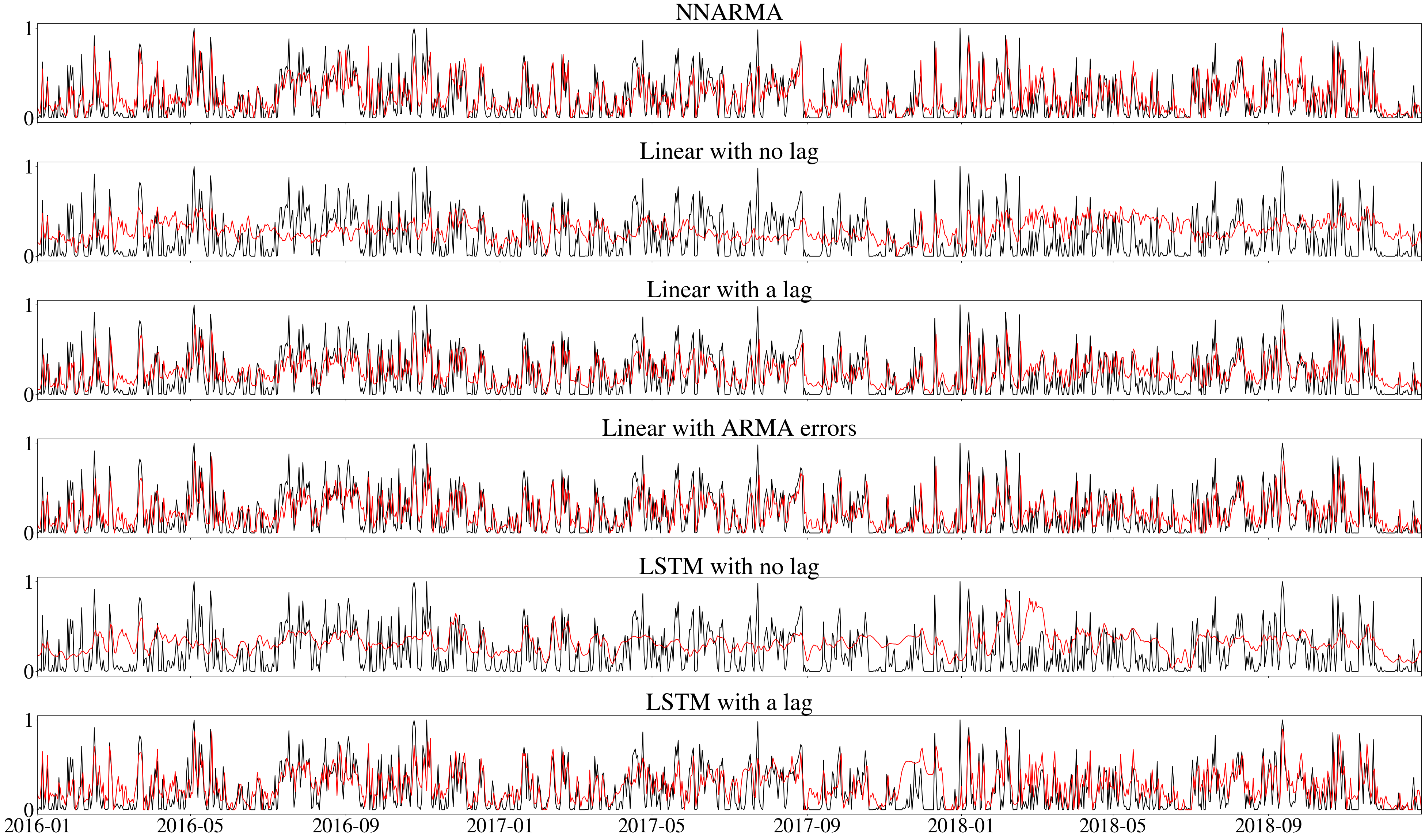}
	\caption{NNARMA and benchmark predictions over the tests sample (red) together observed cloud fractional cover (black) for the location labeled 1 in Figure \ref{fig: topography map}.}
	\label{fig: TS plot many 1}
\end{figure}

\begin{figure}[t!]
	\centering
	\includegraphics[width=\textwidth]{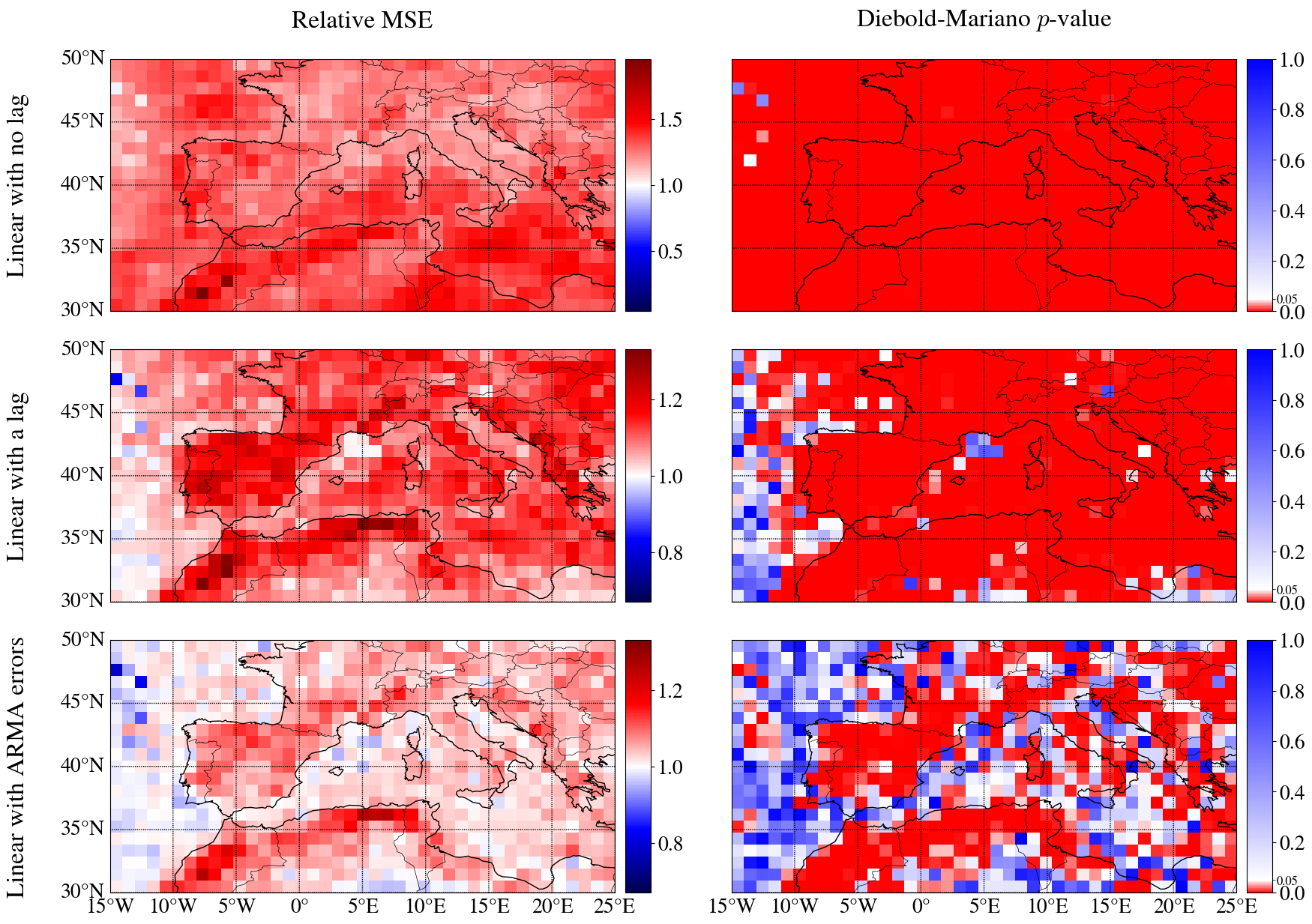}
	\caption{NNARMA and the linear benchmarks. Relative MSEs over the test sample normalized to NNARMA (left), where values above one (red) indicate lower error for NNARMA. $p$-values from a two-sided Diebold-Mariano test of equal predictive accuracy to NNARMA (right), where values below $0.05$ (red) indicate significant differences in predictive accuracy at that level.}
	\label{fig: Linear MSEDM}
\end{figure}

\begin{figure}[t!]
	\centering
	\includegraphics[width=\textwidth]{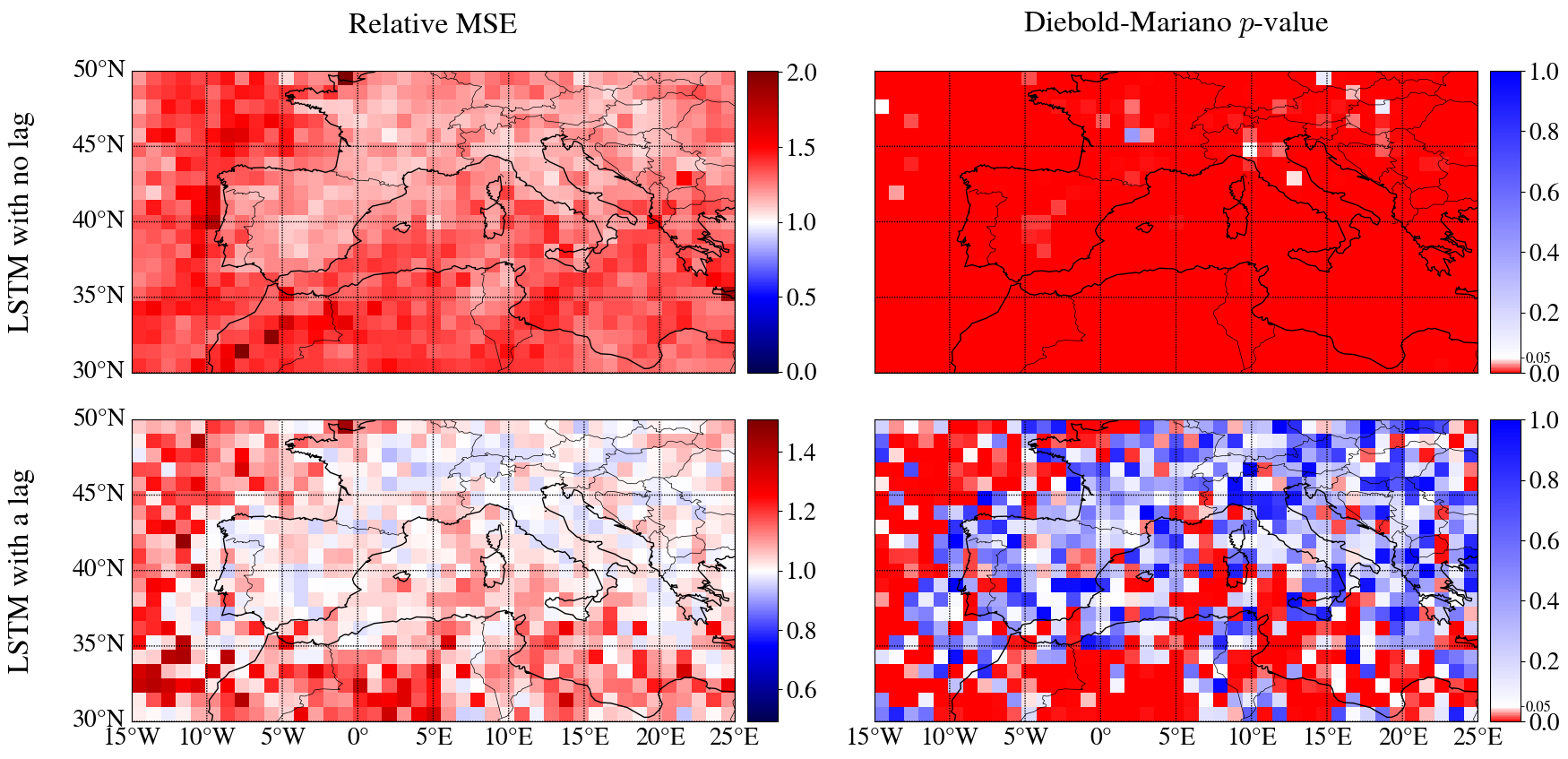}
	\caption{NNARMA and the LSTM network. Relative MSEs over the test sample normalized to NNARMA (left), where values above one (red) indicate lower error for NNARMA. $p$-values from a two-sided Diebold-Mariano test of equal predictive accuracy to NNARMA (right), where values below $0.05$ (red) indicate significant differences in predictive accuracy at that level.}
	\label{fig: LSTM MSEDM}
\end{figure}

The variation in temporal dynamics across locations is reflected in the heterogeneity of ARMA specifications selected for the NNARMA model and its linear counterpart; see Figure \ref{fig: ARMA selections across locations}. Across locations, the most frequent selections are $\text{ARMA}(2,1)$ and $\text{ARMA}(2,2)$ for the NNARMA model and the linear model, respectively, while the extreme candidate specifications $\text{ARMA}(0,0)$ and $\text{ARMA}(5,5)$ are never selected (panel a). The linear model more often includes moving-average dynamics than the NNARMA model, most notably over the Atlas mountains in Northern Africa, consistent with a higher degree of model misspecification in the linear model (panel b).

Figure \ref{fig: TS plot many 1} illustrates the NNARMA predictions in comparison with those of the benchmarks models for location 1. Most visibly, the versions of the linear model and the LSTM network without lagged dependent variables capture the least amount of variation in the observed series. Their accuracy increases substantially with the inclusion of a lagged dependent variable $y_{t-1}$. Including higher-order lags of the dependent variable has a negligible effect (Appendix \ref{sec: Appendix expanded set of benchmarks}, Figure \ref{fig: Appendix TS plot many 1}). Similar patterns are observed across the remaining locations.

Figures \ref{fig: Linear MSEDM} and \ref{fig: LSTM MSEDM} summarize prediction accuracy across the entire geographical domain of the Mediterranean by comparing the NNARMA model with the linear and the LSTM benchmark models, respectively. The figures report relative MSEs normalized against NNARMA, together with the $p$-values from two-sided Diebold-Mariano tests. 

The NNARMA model provides significantly more accurate predictions than the linear model, with or without a lagged dependent variable, across most locations in the geographical domain (Figure \ref{fig: Linear MSEDM}). This pattern is consistent with cloud formation being inherently threshold-based and nonlinear with respect to thermodynamic variables (e.g., \citealp{SundqvistEtAl1989, XuRandall1996, Romps2014}). 

The same figure shows that the use of ARMA disturbances in place of lagged dependent variables significantly improves accuracy in the linear model. The relative advantage of ARMA disturbances is consistent with the ARMA selections for the linear model, which often include moving-average dynamics that may be important for accurate predictions but difficult to capture using a finite number of lagged dependent variables (Figure \ref{fig: ARMA selections across locations}). The predictive gains from ARMA disturbances are further characterized in Appendix \ref{sec: Appendix supplementary figures}, Figure \ref{fig: lin vs lin}, which also shows that adding ARMA disturbances to the linear model yields the largest improvements across the Mediterranean Sea, Northern Africa, and the eastern North Atlantic Ocean.

The NNARMA model often performs significantly better than the linear model with ARMA errors in mountain regions (compare Figures \ref{fig: Linear MSEDM} and \ref{fig: topography map}), characterized by low surface pressure and large variability in relative humidity (compare Figure \ref{fig: Linear MSEDM} and Appendix \ref{sec: Appendix supplementary figures}, Figure \ref{fig: Sumary plot appendix}). This pattern is consistent with how mountains modify airflow and temperature through terrain-induced vertical motion and associated thermodynamic changes (e.g., \citealp{SMITH1979,Houze2012}), which can amplify nonlinear effects, including threshold behavior, regime switching, and interactions. Mountain terrain forces air to rise and sink over slopes, with rapid cooling during ascent and rapid warming during descent. These processes generate large variability in relative humidity and repeatedly push air above and below the saturation threshold, creating abrupt, value-dependent changes in cloud-response sensitivity. Related to this, mountain regions exhibits strong regime switching because the terrain makes vertical wind motion highly sensitive to wind direction and atmospheric stability. Small changes in regional wind patterns or atmospheric stability can trigger a switch between cloud-forming and cloud-suppressing states, which may be better captured by the flexibility of the NNARMA model. In addition, mountains increase the importance of interaction effects. One example is the coupling of temperature and specific humidity. As air is forced to rise or sink over slopes, cooling or warming occurs without an immediate change in specific humidity, bringing the air closer to or farther away from saturation. As a result, temperature directly modulates the effect of specific humidity on cloud formation. Vertical motion also promotes mixing between air masses with different temperature and specific humidity. The NNARMA model naturally captures these interaction effects. In contrast, oceans typically experience smoother and more coherent atmospheric variability, leading to more gradual cloud responses and allowing the linear models to perform comparatively well.

Throughout most of the geographical domain, the NNARMA model achieves significantly higher predictive accuracy than the LSTM network without a lagged dependent variable (Figure \ref{fig: LSTM MSEDM}). Including a lagged dependent variable in the LSTM network leads to statistically indistinguishable accuracy relative to NNARMA across most of mainland Europe, with each model occasionally outperforming the other. Across most of the remaining domain, including large portions of the Mediterranean Sea, Northern Africa, and the eastern North Atlantic Ocean, the NNARMA model continues to significantly outperform the LSTM network. These areas largely coincide with those where ARMA disturbances also yield the largest predictive gains in the linear model, as discussed above.

The patterns in Figures \ref{fig: Linear MSEDM} and \ref{fig: LSTM MSEDM} are essentially unchanged when higher-order lags are included in the benchmark models (Appendix \ref{sec: Appendix expanded set of benchmarks}, Figures \ref{fig: Appendix linear MSEDM} and \ref{fig: Appendix LSTM MSEDM}). In conclusion, incorporating ARMA structures can improve predictive accuracy in both linear and neural network models in this application of cloud cover prediction.

\FloatBarrier
\section{Conclusion}\label{sec: Conclusion}

In this paper, we have introduced the NNARMA model, which integrates autoregressive moving average (ARMA) error structures into feedforward neural networks for flexible nonlinear regression and prediction with serially correlated disturbances.
We have developed a method for the joint estimation of the neural network and ARMA coefficients.
Simulation experiments have demonstrated accurate estimation of unknown regression functions and the underlying dynamics in the disturbances across a range of time series lengths and signal-to-noise ratios. An empirical study of cloud cover has shown improved prediction accuracy of NNARMA relative to existing methods, including LSTM networks. ARMA errors yield improvements over lagged variables across a range of locations; gains relative to linear models with ARMA errors are particularly pronounced in mountain regions.

Our findings underscore the complementary roles of flexible nonlinear regression and structured stochastic modeling of the disturbances within the proposed framework. Explicitly representing temporal dependence through ARMA disturbances may provide advantages relative to the inclusion of lagged variables in the regression component in settings characterized by structured temporal dynamics. The simulations further indicate that ARMA disturbances can partially mitigate the impact of omitted lags in the regression equation, suggesting that the disturbance-based representation of temporal dependence can accommodate certain forms of dynamic misspecification. The neural network regression component enables the model to accommodate complex relationships that may not be well captured by linear specifications. The pronounced differences observed in mountain regions provide a concrete illustration of this mechanism. Together, these elements illustrate the potential of integrating structured stochastic modeling into flexible machine learning frameworks such as neural networks. From a practical perspective, the NNARMA model shows robustness to the choices of ARMA specification and network architecture, with noticeable deterioration only under substantial underspecification. 

While the proposed NNARMA model emphasizes flexible nonlinear regression and structured stochastic modeling of the disturbances, several aspects remain open for further development. In particular, a rigorous theoretical analysis of the statistical properties of the proposed estimation procedure under serial dependence has not yet been established, formal uncertainty quantification is not yet available, and interpretability of the estimated nonlinear relationships may warrant further investigation. These considerations reflect broader methodological challenges in reconciling flexible neural network methods with formal statistical inference, and with particular attention to dynamic stochastic systems.
Further, as alternatives to ARMA structures for the disturbances, stochastic processes implied by hidden Markov models and spatial-temporal models can, for example, be considered. Continued efforts to strengthen the theoretical foundation and inferential tools for such advanced methods may further enhance their applicability in time series analysis and forecasting.


\clearpage
\bibliographystyle{chicago}
\bibliography{References}

\begin{thebibliography}{}

\bibitem[\protect\citeauthoryear{Altman}{Altman}{1990}]{Altman}
Altman, N.~S. (1990).
\newblock Kernel smoothing of data with correlated errors.
\newblock {\em Journal of the American Statistical Association\/}~{\em
  85\/}(411), 749--759.

\bibitem[\protect\citeauthoryear{Athey and Imbens}{Athey and
  Imbens}{2019}]{Athey2019}
Athey, S. and G.~W. Imbens (2019).
\newblock Machine learning methods that economists should know about.
\newblock {\em Annual Review of Economics\/}~{\em 11\/}(Volume 11, 2019),
  685--725.

\bibitem[\protect\citeauthoryear{Bach}{Bach}{2017}]{Bach2017}
Bach, F. (2017).
\newblock Breaking the curse of dimensionality with convex neural networks.
\newblock {\em Journal of Machine Learning Research\/}~{\em 18\/}(19), 1--53.

\bibitem[\protect\citeauthoryear{Bates and Watts}{Bates and
  Watts}{1988}]{BatesWatts1988}
Bates, D.~M. and D.~G. Watts (1988).
\newblock {\em Nonlinear Regression Analysis and Its Applications}.
\newblock New York: Wiley.

\bibitem[\protect\citeauthoryear{Bauer and Kohler}{Bauer and
  Kohler}{2019}]{Bauer2019}
Bauer, B. and M.~Kohler (2019).
\newblock {On deep learning as a remedy for the curse of dimensionality in
  nonparametric regression}.
\newblock {\em The Annals of Statistics\/}~{\em 47\/}(4), 2261 -- 2285.

\bibitem[\protect\citeauthoryear{Bengtsson}{Bengtsson}{2010}]{Bengtsson2010}
Bengtsson, L. (2010).
\newblock The global atmospheric water cycle.
\newblock {\em Environmental Research Letters\/}~{\em 5\/}(2), 025202.

\bibitem[\protect\citeauthoryear{Bennedsen, Hillebrand, and Jensen}{Bennedsen
  et~al.}{2023}]{Jensen2023}
Bennedsen, M., E.~Hillebrand, and S.~Jensen (2023).
\newblock A neural network approach to the environmental {K}uznets curve.
\newblock {\em Energy Economics\/}~{\em 126}, 106985.

\bibitem[\protect\citeauthoryear{Bony, Stevens, Frierson, Jakob, Kageyama,
  Pincus, Shepherd, Sherwood, Siebesma, Watanabe, and Webb}{Bony
  et~al.}{2015}]{BonyEtAl2015}
Bony, S., B.~Stevens, D.~Frierson, C.~Jakob, M.~Kageyama, R.~Pincus,
  T.~Shepherd, S.~Sherwood, A.~Siebesma, M.~Watanabe, and M.~Webb (2015).
\newblock Clouds, circulation and climate sensitivity.
\newblock {\em Nature Geoscience\/}~{\em 8}, 261--268.

\bibitem[\protect\citeauthoryear{Box, Jenkins, Reinsel, and Ljung}{Box
  et~al.}{2015}]{boxjenkins2015}
Box, G., G.~Jenkins, G.~Reinsel, and G.~Ljung (2015).
\newblock {\em Time Series Analysis: Forecasting and Control}.
\newblock Wiley Series in Probability and Statistics. Wiley.

\bibitem[\protect\citeauthoryear{Brabanter, Cao, Gijbels, and
  Opsomer}{Brabanter et~al.}{2018}]{BRABANTER2018}
Brabanter, K.~D., F.~Cao, I.~Gijbels, and J.~Opsomer (2018).
\newblock Local polynomial regression with correlated errors in random design
  and unknown correlation structure.
\newblock {\em Biometrika\/}~{\em 105\/}(3), pp. 681--690.

\bibitem[\protect\citeauthoryear{Cattaneo, Jansson, and Ma}{Cattaneo
  et~al.}{2020}]{Cattaneo2020}
Cattaneo, M.~D., M.~Jansson, and X.~Ma (2020).
\newblock Simple local polynomial density estimators.
\newblock {\em Journal of the American Statistical Association\/}~{\em
  115\/}(531), 1449--1455.

\bibitem[\protect\citeauthoryear{Chen and Tsay}{Chen and
  Tsay}{1993}]{ChenTsay1993NAAR}
Chen, R. and R.~S. Tsay (1993).
\newblock Nonlinear additive {ARX} models.
\newblock {\em Journal of the American Statistical Association\/}~{\em
  88\/}(423), 955--967.

\bibitem[\protect\citeauthoryear{Chen and White}{Chen and
  White}{1999}]{Chen1999}
Chen, X. and H.~White (1999).
\newblock Improved rates and asymptotic normality for nonparametric neural
  network estimators.
\newblock {\em IEEE Transactions on Information Theory\/}~{\em 45\/}(2),
  682--691.

\bibitem[\protect\citeauthoryear{Cybenko}{Cybenko}{1989}]{Cybenko1989}
Cybenko, G. (1989).
\newblock Approximation by superpositions of a sigmoidal function.
\newblock {\em Mathematics of Control, Signals and Systems\/}~{\em 2\/}(4),
  303--314.

\bibitem[\protect\citeauthoryear{Diebold and Mariano}{Diebold and
  Mariano}{1995}]{Diebold1995}
Diebold, F.~X. and R.~S. Mariano (1995).
\newblock Comparing predictive accuracy.
\newblock {\em Journal of Business \& Economic Statistics\/}~{\em 13\/}(3),
  253--263.

\bibitem[\protect\citeauthoryear{Durbin}{Durbin}{1960}]{Durbin1960isr}
Durbin, J. (1960).
\newblock The fitting of time-series models.
\newblock {\em Review of the International Statistical Institute\/}~{\em
  28\/}(3), 233--244.

\bibitem[\protect\citeauthoryear{Durbin and Watson}{Durbin and
  Watson}{1950}]{Durbin50}
Durbin, J. and G.~S. Watson (1950).
\newblock Testing for serial correlation in least squares regression, i.
\newblock {\em Biometrika\/}~{\em 37\/}(3–4), 409--428.

\bibitem[\protect\citeauthoryear{Durbin and Watson}{Durbin and
  Watson}{1951}]{Durbin51}
Durbin, J. and G.~S. Watson (1951).
\newblock Testing for serial correlation in least squares regression, ii.
\newblock {\em Biometrika\/}~{\em 38\/}(1–2), 159--179.

\bibitem[\protect\citeauthoryear{Efron}{Efron}{2020}]{Efron2020}
Efron, B. (2020).
\newblock Prediction, estimation, and attribution.
\newblock {\em Journal of the American Statistical Association\/}~{\em
  115\/}(530), 636--655.

\bibitem[\protect\citeauthoryear{Farrell, Liang, and Misra}{Farrell
  et~al.}{2021}]{Farrell2021}
Farrell, M.~H., T.~Liang, and S.~Misra (2021).
\newblock Deep neural networks for estimation and inference.
\newblock {\em Econometrica\/}~{\em 89\/}(1), 181--213.

\bibitem[\protect\citeauthoryear{Gallant}{Gallant}{1987}]{Gallant1987}
Gallant, A.~R. (1987).
\newblock {\em Nonlinear Statistical Models}.
\newblock New York: Wiley.

\bibitem[\protect\citeauthoryear{Goodfellow, Bengio, and Courville}{Goodfellow
  et~al.}{2016}]{Goodfellow-et-al-2016}
Goodfellow, I., Y.~Bengio, and A.~Courville (2016).
\newblock {\em {Deep Learning}}.
\newblock MIT Press.
\newblock \url{http://www.deeplearningbook.org}.

\bibitem[\protect\citeauthoryear{Greff, Srivastava, Koutník, Steunebrink, and
  Schmidhuber}{Greff et~al.}{2017}]{Schmidhuber2017}
Greff, K., R.~K. Srivastava, J.~Koutník, B.~R. Steunebrink, and J.~Schmidhuber
  (2017).
\newblock Lstm: A search space odyssey.
\newblock {\em IEEE Transactions on Neural Networks and Learning
  Systems\/}~{\em 28\/}(10), 2222--2232.

\bibitem[\protect\citeauthoryear{Grundner, Beucler, Savre, Lauer, Schlund, and
  Eyring}{Grundner et~al.}{2025}]{GrundnerEtAl2025}
Grundner, A., T.~Beucler, J.~Savre, A.~Lauer, M.~Schlund, and V.~Eyring (2025).
\newblock Reduced cloud cover errors in a hybrid ai-climate model through
  equation discovery and automatic tuning.
\newblock {\em Scientific Reports\/}~{\em 15\/}(1), 43836.

\bibitem[\protect\citeauthoryear{Gu, Kelly, and Xiu}{Gu
  et~al.}{2020}]{Kelly2020}
Gu, S., B.~Kelly, and D.~Xiu (2020).
\newblock {Empirical Asset Pricing via Machine Learning}.
\newblock {\em The Review of Financial Studies\/}~{\em 33\/}(5), 2223--2273.

\bibitem[\protect\citeauthoryear{Hamilton}{Hamilton}{1994}]{hamilton1994series}
Hamilton, J.~D. (1994).
\newblock {\em Time Series Analysis}.
\newblock Princeton University Press.

\bibitem[\protect\citeauthoryear{H\"{a}rdle and Tsybakov}{H\"{a}rdle and
  Tsybakov}{1997}]{HardleTsybakov1997}
H\"{a}rdle, W. and A.~Tsybakov (1997).
\newblock Local polynomial estimators under dependence.
\newblock {\em Annals of Statistics\/}~{\em 25}, 118--148.

\bibitem[\protect\citeauthoryear{Hart}{Hart}{1991}]{Hart1991}
Hart, J.~D. (1991).
\newblock Kernel regression estimation with time series errors.
\newblock {\em Journal of the Royal Statistical Society. Series B
  (Methodological)\/}~{\em 53\/}(1), 173--187.

\bibitem[\protect\citeauthoryear{Harvey}{Harvey}{1990}]{harvey1990}
Harvey, A.~C. (1990).
\newblock {\em {The Econometric Analysis of Time Series}\/} (2nd ed.).
\newblock The MIT Press.

\bibitem[\protect\citeauthoryear{Harvey and Phillips}{Harvey and
  Phillips}{1979}]{HarveyPhillips1979}
Harvey, A.~C. and G.~D.~A. Phillips (1979).
\newblock Maximum likelihood estimation of regression models with
  autoregressive- moving average disturbances.
\newblock {\em Biometrika\/}~{\em 66\/}(1), 49--58.

\bibitem[\protect\citeauthoryear{Hastie, Tibshirani, and Friedman}{Hastie
  et~al.}{2009}]{hastie2009elements}
Hastie, T., R.~Tibshirani, and J.~H. Friedman (2009).
\newblock {\em The Elements of Statistical Learning: Data Mining, Inference,
  and Prediction\/} (2 ed.).
\newblock Springer Series in Statistics. New York, NY: Springer.

\bibitem[\protect\citeauthoryear{He, Zhang, Ren, and Sun}{He
  et~al.}{2015}]{He_2015}
He, K., X.~Zhang, S.~Ren, and J.~Sun (2015).
\newblock {Delving Deep into Rectifiers: Surpassing Human-Level Performance on
  ImageNet Classification}.
\newblock In {\em The IEEE International Conference on Computer Vision (ICCV)}.

\bibitem[\protect\citeauthoryear{Hendry and Nielsen}{Hendry and
  Nielsen}{2007}]{HendryNielsen2007}
Hendry, D.~F. and B.~Nielsen (2007).
\newblock {\em Econometric Modeling: A Likelihood Approach}.
\newblock Princeton, NJ: Princeton University Press.

\bibitem[\protect\citeauthoryear{Herrmann, Gasser, and Kneip}{Herrmann
  et~al.}{1992}]{HERRMANN1992}
Herrmann, E., T.~Gasser, and A.~Kneip (1992).
\newblock Choice of bandwidth for kernel regression when residuals are
  correlated.
\newblock {\em Biometrika\/}~{\em 79\/}(4), 783--795.

\bibitem[\protect\citeauthoryear{Hersbach, Bell, Berrisford, Hirahara,
  Horányi, Muñoz-Sabater, Nicolas, Peubey, Radu, Schepers, Simmons, Soci,
  Abdalla, Abellan, Balsamo, Bechtold, Biavati, Bidlot, Bonavita, De~Chiara,
  Dahlgren, Dee, Diamantakis, Dragani, Flemming, Forbes, Fuentes, Geer,
  Haimberger, Healy, Hogan, Hólm, Janisková, Keeley, Laloyaux, Lopez, Lupu,
  Radnoti, de~Rosnay, Rozum, Vamborg, Villaume, and Thépaut}{Hersbach
  et~al.}{2020}]{Hersbach2020}
Hersbach, H., B.~Bell, P.~Berrisford, S.~Hirahara, A.~Horányi,
  J.~Muñoz-Sabater, J.~Nicolas, C.~Peubey, R.~Radu, D.~Schepers, A.~Simmons,
  C.~Soci, S.~Abdalla, X.~Abellan, G.~Balsamo, P.~Bechtold, G.~Biavati,
  J.~Bidlot, M.~Bonavita, G.~De~Chiara, P.~Dahlgren, D.~Dee, M.~Diamantakis,
  R.~Dragani, J.~Flemming, R.~Forbes, M.~Fuentes, A.~Geer, L.~Haimberger,
  S.~Healy, R.~J. Hogan, E.~Hólm, M.~Janisková, S.~Keeley, P.~Laloyaux,
  P.~Lopez, C.~Lupu, G.~Radnoti, P.~de~Rosnay, I.~Rozum, F.~Vamborg,
  S.~Villaume, and J.-N. Thépaut (2020).
\newblock The era5 global reanalysis.
\newblock {\em Quarterly Journal of the Royal Meteorological Society\/}~{\em
  146\/}(730), 1999--2049.

\bibitem[\protect\citeauthoryear{Hicks, Riegler, Svennevik, and Hammer}{Hicks
  et~al.}{2023}]{DataOSF}
Hicks, S., M.~Riegler, H.~Svennevik, and H.~L. Hammer (2023).
\newblock European cloud cover.
\newblock {\em Open Science Framework\/}.
\newblock doi:10.17605/OSF.IO/KQDGX.

\bibitem[\protect\citeauthoryear{Hochreiter and Schmidhuber}{Hochreiter and
  Schmidhuber}{1997}]{Hochreiter1997}
Hochreiter, S. and J.~Schmidhuber (1997).
\newblock Long short-term memory.
\newblock {\em Neural Computation\/}~{\em 9\/}(8), 1735--1780.

\bibitem[\protect\citeauthoryear{Hornik, Stinchcombe, and White}{Hornik
  et~al.}{1989}]{HORNIK1989}
Hornik, K., M.~Stinchcombe, and H.~White (1989).
\newblock Multilayer feedforward networks are universal approximators.
\newblock {\em Neural Networks\/}~{\em 2\/}(5), 359--366.

\bibitem[\protect\citeauthoryear{Houze~Jr.}{Houze~Jr.}{2012}]{Houze2012}
Houze~Jr., R.~A. (2012).
\newblock Orographic effects on precipitating clouds.
\newblock {\em Reviews of Geophysics\/}~{\em 50\/}(1).

\bibitem[\protect\citeauthoryear{Hughes}{Hughes}{1984}]{Hughes1984}
Hughes, N.~A. (1984).
\newblock Global cloud climatologies: A historical review.
\newblock {\em Journal of Applied Meteorology and Climatology\/}~{\em 23\/}(5),
  724 -- 751.

\bibitem[\protect\citeauthoryear{Jones}{Jones}{1980}]{Jones1980}
Jones, R.~H. (1980).
\newblock Maximum likelihood fitting of arma models to time series with missing
  observations.
\newblock {\em Technometrics\/}~{\em 22}, 389--395.

\bibitem[\protect\citeauthoryear{Jozefowicz, Zaremba, and Sutskever}{Jozefowicz
  et~al.}{2015}]{Jozefowicz2015}
Jozefowicz, R., W.~Zaremba, and I.~Sutskever (2015).
\newblock An empirical exploration of recurrent network architectures.
\newblock In {\em Proceedings of the 32nd International Conference on
  International Conference on Machine Learning - Volume 37}, ICML'15, pp.\
  2342–2350. JMLR.org.

\bibitem[\protect\citeauthoryear{Kingma and Ba}{Kingma and
  Ba}{2014}]{kingma2014adam}
Kingma, D.~P. and J.~Ba (2014).
\newblock Adam: A method for stochastic optimization.
\newblock {\em arXiv preprint arXiv:1412.6980\/}.

\bibitem[\protect\citeauthoryear{Krivobokova and Kauermann}{Krivobokova and
  Kauermann}{2007}]{Krivobokova2007}
Krivobokova, T. and G.~Kauermann (2007).
\newblock A note on penalized spline smoothing with correlated errors.
\newblock {\em Journal of the American Statistical Association\/}~{\em
  102\/}(480), 1328--1337.

\bibitem[\protect\citeauthoryear{Leshno, Lin, Pinkus, and Schocken}{Leshno
  et~al.}{1993}]{LESHNO1993}
Leshno, M., V.~Y. Lin, A.~Pinkus, and S.~Schocken (1993).
\newblock Multilayer feedforward networks with a nonpolynomial activation
  function can approximate any function.
\newblock {\em Neural Networks\/}~{\em 6\/}(6), 861--867.

\bibitem[\protect\citeauthoryear{Levinson}{Levinson}{1946}]{Levinson1946}
Levinson, N. (1946).
\newblock The wiener (root mean square) error criterion in filter design and
  prediction.
\newblock {\em Journal of Mathematics and Physics\/}~{\em 25\/}(1-4), 261--278.

\bibitem[\protect\citeauthoryear{Li and Racine}{Li and
  Racine}{2007}]{LiRacine2007}
Li, Q. and J.~Racine (2007).
\newblock {\em Nonparametric Econometrics: Theory and Practice}.
\newblock Princeton, NJ: Princeton University Press.

\bibitem[\protect\citeauthoryear{Liu, Koren, Altaratz, and Chekroun}{Liu
  et~al.}{2023}]{LiuEtAl2023}
Liu, H., I.~Koren, O.~Altaratz, and M.~D. Chekroun (2023).
\newblock Opposing trends of cloud coverage over land and ocean under global
  warming.
\newblock {\em Atmospheric Chemistry and Physics\/}~{\em 23\/}(11), 6559--6569.

\bibitem[\protect\citeauthoryear{Liu, Chen, and Yao}{Liu
  et~al.}{2010}]{LIU2010}
Liu, J.~M., R.~Chen, and Q.~Yao (2010).
\newblock Nonparametric transfer function models.
\newblock {\em Journal of Econometrics\/}~{\em 157\/}(1), 151--164.
\newblock Nonlinear and Nonparametric Methods in Econometrics.

\bibitem[\protect\citeauthoryear{Masters}{Masters}{1993}]{masters1993}
Masters, T. (1993).
\newblock {\em {Practical neural network recipes in {C}++}}.
\newblock (Academic Press).

\bibitem[\protect\citeauthoryear{Medeiros, Vasconcelos, Álvaro Veiga, and
  Zilberman}{Medeiros et~al.}{2021}]{Medeiros2021}
Medeiros, M.~C., G.~F.~R. Vasconcelos, Álvaro Veiga, and E.~Zilberman (2021).
\newblock Forecasting inflation in a data-rich environment: The benefits of
  machine learning methods.
\newblock {\em Journal of Business \& Economic Statistics\/}~{\em 39\/}(1),
  98--119.

\bibitem[\protect\citeauthoryear{Nadaraya}{Nadaraya}{1964}]{Nadaraya1964}
Nadaraya, E.~A. (1964).
\newblock On estimating regression.
\newblock {\em Theory of Probability \& Its Applications\/}~{\em 9\/}(1),
  141--142.

\bibitem[\protect\citeauthoryear{Nair and Hinton}{Nair and
  Hinton}{2010}]{NairHinton2010}
Nair, V. and G.~E. Hinton (2010).
\newblock Rectified linear units improve restricted boltzmann machines.
\newblock In {\em Proceedings of the 27th International Conference on
  International Conference on Machine Learning}, ICML'10, Madison, WI, USA,
  pp.\  807–814. Omnipress.

\bibitem[\protect\citeauthoryear{{National Geophysical Data Center}}{{National
  Geophysical Data Center}}{1993}]{NGDC1993}
{National Geophysical Data Center} (1993).
\newblock 5-minute gridded global relief data ({ETOPO5}).
\newblock National Geophysical Data Center, NOAA.
\newblock Accessed: 2026-08-14.

\bibitem[\protect\citeauthoryear{Pagan and Ullah}{Pagan and
  Ullah}{1999}]{PaganUllah1999}
Pagan, A. and A.~Ullah (1999).
\newblock {\em Nonparametric Econometrics}.
\newblock Cambridge: Cambridge University Press.

\bibitem[\protect\citeauthoryear{Pierce}{Pierce}{1971}]{Pierce1971}
Pierce, D.~A. (1971).
\newblock Least squares estimation in the regression model with
  autoregressive-moving average errors.
\newblock {\em Biometrika\/}~{\em 58\/}(2), 299--312.

\bibitem[\protect\citeauthoryear{Powell}{Powell}{1964}]{Powell1964}
Powell, M. J.~D. (1964).
\newblock {An efficient method for finding the minimum of a function of several
  variables without calculating derivatives}.
\newblock {\em The Computer Journal\/}~{\em 7\/}(2), 155--162.

\bibitem[\protect\citeauthoryear{Prechelt}{Prechelt}{2012}]{Prechelt2012}
Prechelt, L. (2012).
\newblock {\em Early Stopping --- But When?}, pp.\  53--67.
\newblock Berlin, Heidelberg: Springer Berlin Heidelberg.

\bibitem[\protect\citeauthoryear{Ramachandran, Zoph, and Le}{Ramachandran
  et~al.}{2018}]{ramach2018}
Ramachandran, P., B.~Zoph, and Q.~V. Le (2018).
\newblock {Searching for Activation Functions}.
\newblock In {\em {6th International Conference on Learning Representations,
  {ICLR} 2018, Vancouver, BC, Canada, Workshop Track Proceedings}}.
  OpenReview.net.

\bibitem[\protect\citeauthoryear{Ray and Tsay}{Ray and
  Tsay}{1997}]{RayTsay2007}
Ray, B.~K. and R.~S. Tsay (1997).
\newblock Bandwidth selection for kernel regression with long-range dependent
  errors.
\newblock {\em Biometrika\/}~{\em 84\/}(4), 791--802.

\bibitem[\protect\citeauthoryear{Robinson}{Robinson}{1983}]{Robinson1983}
Robinson, P.~M. (1983).
\newblock Nonparametric estimators for time series.
\newblock {\em Journal of Time Series Analysis\/}~{\em 4\/}(3), 185--207.

\bibitem[\protect\citeauthoryear{Robinson}{Robinson}{1988}]{Robinson1988}
Robinson, P.~M. (1988).
\newblock Root-n-consistent semiparametric regression.
\newblock {\em Econometrica\/}~{\em 56}, 931--954.

\bibitem[\protect\citeauthoryear{Romps}{Romps}{2014}]{Romps2014}
Romps, D.~M. (2014).
\newblock An analytical model for tropical relative humidity.
\newblock {\em Journal of Climate\/}~{\em 27\/}(19), 7432 -- 7449.

\bibitem[\protect\citeauthoryear{Rumelhart, Hinton, and Williams}{Rumelhart
  et~al.}{1986}]{rumelhart1986}
Rumelhart, D.~E., G.~E. Hinton, and R.~J. Williams (1986).
\newblock Learning representations by back-propagating errors.
\newblock {\em Nature\/}~{\em 323}, 533--536.

\bibitem[\protect\citeauthoryear{Schmetz, Pili, Tjemkes, Just, Kerkmann, Rota,
  and Ratier}{Schmetz et~al.}{2002}]{Schmetz2002}
Schmetz, J., P.~Pili, S.~Tjemkes, D.~Just, J.~Kerkmann, S.~Rota, and A.~Ratier
  (2002).
\newblock An introduction to meteosat second generation (msg).
\newblock {\em Bulletin of the American Meteorological Society\/}~{\em
  83\/}(7), 977 -- 992.

\bibitem[\protect\citeauthoryear{Schmidt-Hieber}{Schmidt-Hieber}{2020}]{Schmidt-Hieber2020}
Schmidt-Hieber, J. (2020).
\newblock {Nonparametric regression using deep neural networks with ReLU
  activation function}.
\newblock {\em The Annals of Statistics\/}~{\em 48\/}(4), 1875 -- 1897.

\bibitem[\protect\citeauthoryear{Silverman}{Silverman}{1986}]{Silverman}
Silverman, B.~W. (1986).
\newblock {\em {Density estimation for statistics and data analysis}}.
\newblock London: Chapman and Hall.

\bibitem[\protect\citeauthoryear{Slingo}{Slingo}{1987}]{Slingo1987}
Slingo, J.~M. (1987).
\newblock The development and verification of a cloud prediction scheme for the
  ecmwf model.
\newblock {\em Quarterly Journal of the Royal Meteorological Society\/}~{\em
  113\/}(477), 899--927.

\bibitem[\protect\citeauthoryear{Smith}{Smith}{1979}]{SMITH1979}
Smith, R.~B. (1979).
\newblock The influence of mountains on the atmosphere.
\newblock Volume~21 of {\em Advances in Geophysics}, pp.\  87--230. Elsevier.

\bibitem[\protect\citeauthoryear{Stephens}{Stephens}{2005}]{Stephens2005}
Stephens, G.~L. (2005).
\newblock Cloud feedbacks in the climate system: A critical review.
\newblock {\em Journal of Climate\/}~{\em 18\/}(2), 237 -- 273.

\bibitem[\protect\citeauthoryear{Stephens, Li, Wild, Clayson, Loeb, Kato,
  L'Ecuyer, Stackhouse, Lebsock, and Andrews}{Stephens
  et~al.}{2012}]{StephensETAl2012}
Stephens, G.~L., J.~Li, M.~Wild, C.~A. Clayson, N.~Loeb, S.~Kato, T.~L'Ecuyer,
  P.~W. Stackhouse, M.~Lebsock, and T.~Andrews (2012).
\newblock An update on earth's energy balance in light of the latest global
  observations.
\newblock {\em Nature Geoscience\/}~{\em 5\/}(10), 691--696.

\bibitem[\protect\citeauthoryear{Stevens and Bony}{Stevens and
  Bony}{2013}]{StevensBony2013}
Stevens, B. and S.~Bony (2013).
\newblock What are climate models missing?
\newblock {\em Science\/}~{\em 340\/}(6136), 1053--1054.

\bibitem[\protect\citeauthoryear{Stone}{Stone}{1982}]{Stone1982}
Stone, C.~J. (1982).
\newblock Optimal global rates of convergence for nonparametric regression.
\newblock {\em The Annals of Statistics\/}~{\em 10\/}(4), 1040--1053.

\bibitem[\protect\citeauthoryear{Su and Ullah}{Su and Ullah}{2006}]{Su2006}
Su, L. and A.~Ullah (2006).
\newblock More efficient estimation in nonparametric regression with
  nonparametric autocorrelated errors.
\newblock {\em Econometric Theory\/}~{\em 22\/}(1), 98--126.

\bibitem[\protect\citeauthoryear{Sun, Lang, and Boning}{Sun
  et~al.}{2021}]{sun2021}
Sun, F.-K., C.~I. Lang, and D.~S. Boning (2021).
\newblock Adjusting for autocorrelated errors in neural networks for time
  series.

\bibitem[\protect\citeauthoryear{Sundqvist, Berge, and Kristjánsson}{Sundqvist
  et~al.}{1989}]{SundqvistEtAl1989}
Sundqvist, H., E.~Berge, and J.~E. Kristjánsson (1989).
\newblock Condensation and cloud parameterization studies with a mesoscale
  numerical weather prediction model.
\newblock {\em Monthly Weather Review\/}~{\em 117\/}(8), 1641 -- 1657.

\bibitem[\protect\citeauthoryear{Svennevik, Hicks, Riegler, Storelvmo, and
  Hammer}{Svennevik et~al.}{2024}]{Svennevik2024}
Svennevik, H., S.~Hicks, M.~Riegler, T.~Storelvmo, and H.~Hammer (2024).
\newblock A dataset for predicting cloud cover over europe.
\newblock {\em Scientific Data\/}~{\em 11\/}(1), 245.

\bibitem[\protect\citeauthoryear{Ter\"{a}svirta, Medeiros, and
  Rech}{Ter\"{a}svirta et~al.}{2006}]{Terasvirta2006}
Ter\"{a}svirta, T., M.~Medeiros, and G.~Rech (2006).
\newblock Building neural network models for time series: a statistical
  approach.
\newblock {\em Journal of Forecasting\/}~{\em 25\/}(1), 49--75.

\bibitem[\protect\citeauthoryear{Vo, Hu, Xue, and Chen}{Vo
  et~al.}{2025}]{ThuyEtAl2025}
Vo, T.~T., L.~Hu, L.~Xue, and S.~Chen (2025).
\newblock Trends in cloud covers across conus (1980–2020).
\newblock {\em Journal of Climate\/}~{\em 38\/}(19), 5371 -- 5390.

\bibitem[\protect\citeauthoryear{Watson}{Watson}{1964}]{Watson1964}
Watson, G.~S. (1964).
\newblock Smooth regression analysis.
\newblock {\em Sankhyā: The Indian Journal of Statistics, Series A
  (1961-2002)\/}~{\em 26\/}(4), 359--372.

\bibitem[\protect\citeauthoryear{White}{White}{1989}]{White1989}
White, H. (1989).
\newblock {Some Asymptotic Results for Learning in Single-Hidden-Layer
  Feedforward Networks}.
\newblock {\em Neural Networks\/}~{\em 2}, 425--431.

\bibitem[\protect\citeauthoryear{White and Domowitz}{White and
  Domowitz}{1984}]{WhiteDomowitz1984}
White, H. and I.~Domowitz (1984).
\newblock Nonlinear regression with dependent observations.
\newblock {\em Econometrica\/}~{\em 52\/}(1), 143--161.

\bibitem[\protect\citeauthoryear{Xiao, Linton, Carroll, and Mammen}{Xiao
  et~al.}{2003}]{Xiao2003}
Xiao, Z., O.~B. Linton, R.~J. Carroll, and E.~Mammen (2003).
\newblock More efficient local polynomial estimation in nonparametric
  regression with autocorrelated errors.
\newblock {\em Journal of the American Statistical Association\/}~{\em
  98\/}(464), 980--992.

\bibitem[\protect\citeauthoryear{Xu and Randall}{Xu and
  Randall}{1996}]{XuRandall1996}
Xu, K.-M. and D.~A. Randall (1996).
\newblock A semiempirical cloudiness parameterization for use in climate
  models.
\newblock {\em Journal of Atmospheric Sciences\/}~{\em 53\/}(21), 3084 -- 3102.

\bibitem[\protect\citeauthoryear{Zelinka, Myers, McCoy, Po-Chedley, Caldwell,
  Ceppi, Klein, and Taylor}{Zelinka et~al.}{2020}]{Zelinka2020}
Zelinka, M.~D., T.~A. Myers, D.~T. McCoy, S.~Po-Chedley, P.~M. Caldwell,
  P.~Ceppi, S.~A. Klein, and K.~E. Taylor (2020).
\newblock Causes of higher climate sensitivity in cmip6 models.
\newblock {\em Geophysical Research Letters\/}~{\em 47\/}(1), e2019GL085782.

\bibitem[\protect\citeauthoryear{Zhang}{Zhang}{2003}]{Zhang2003}
Zhang, G. (2003).
\newblock Time series forecasting using a hybrid arima and neural network
  model.
\newblock {\em Neurocomputing\/}~{\em 50}, 159--175.

\end{thebibliography}


\newpage
\renewcommand\appendixtocname{Appendix}
\renewcommand\appendixpagename{Appendix}
\renewcommand\appendixname{Appendix}

\appendix

\appendixpage
\addappheadtotoc

\startcontents[appendix]

\numberwithin{equation}{section}


\numberwithin{figure}{section}
\renewcommand{\thefigure}{\Alph{section}.\arabic{figure}}

\printcontents[appendix]{}{1}{}

\clearpage
\section{Cholesky decomposition of the covariance matrix}\label{sec: cholesky}

We use the Durbin--Levinson algorithm to efficiently perform a Cholesky decomposition of the scaled covariance matrix $ \Psi(\theta_1) = C(\theta_1)^{-1} [C(\theta_1)^{-1}]' $, defined in Section \ref{sec: methodology}. The algorithm is recursive and initiated from the autocovaraince function of the ARMA process. We denote by $ \gamma(h;\theta_1) $ the value of the ARMA autocovariance function at lag $h$. For simplicity, dependence on $ \theta_1 $ is suppressed in the remainder of this section, e.g. $ \Psi(\theta_1) \equiv \Psi $ and $ \gamma(h;\theta_1) \equiv \gamma(h) $.

As explained below, we use the Durbin--Levinson algorithm to perform the decomposition $ \Psi = L^{-1} D^{-1} [L^{-1}]' $, from which it can be deduced that $ C = D^{\frac{1}{2}} L $. The matrices $ L $ and $ D $ are defined as follows
\begin{align*}
	L = 
	\begin{bmatrix}
		1 & 0 & \ldots & \ldots & \ldots & 0 \\
		-\phi_{11} & 1 & 0 & \ldots & \ldots & 0 \\
		-\phi_{22} & - \phi_{21} & 1 & 0 & \ldots & 0 \\ 
		-\phi_{33} & - \phi_{32} & -\phi_{31} & 1& \ldots & 0 \\ 
		\vdots & \vdots & \ddots & \ddots & \ddots & \vdots \\
		-\phi_{T-1,T-1} & -\phi_{T-1,T-2} & -\phi_{T-1,T-3} & \ldots & -\phi_{t-1,1} & 1
	\end{bmatrix}
\end{align*}
and
\begin{align}\label{eq: D}
	D = \sigma^2 \text{diag}(v_0^{-1},v_1^{-1},\ldots,v_{T-1}^{-1}).
\end{align}
The coefficients $ \phi_{tj} $ define the orthogonal projection of $u_{t+1}$ onto the linear span of $u_1,\ldots,u_T$,
\begin{equation*}
	\hat u_{t+1} = P_{\overline{\text{sp}}\{u_1,\ldots,u_t \}} u_{t+1} = \sum_{j=1}^{t} \phi_{tj} u_{t+1-j}, \hspace*{0.2cm}  t=1,\ldots,T-1,
\end{equation*}
with mean squared errors 
\begin{equation*}
	v_t = \E \left( u_{t+1} - \hat u_{t+1} \right)^2. 
\end{equation*}
We note that $ \E (L u [Lu]') = \sigma^2 L \Psi L' = \sigma^{2} D^{-1} $. The decomposition $ \Psi = L^{-1} D^{-1} [L^{-1}]' $ follows immediately. The coefficients $ \phi_{tj} $ and mean squared errors $ v_t $ are determined from the Durbin--Levinson recursions. Set $ \phi_{11} = \gamma(1) / \gamma(0) $, $ v_0 = \gamma(0) $,
\begin{align}
	\phi_{tt} &= \frac{\gamma(t) - \sum_{j=1}^{t-1} \phi_{t-1,j} \gamma(t-j)}{v_{t-1}}, \hspace*{0.2cm} t = 1,\ldots,T-1 \label{eq: phi_tt}, \\
	\phi_{tj} &= \phi_{t-1,j} - \phi_{tt}\phi_{t-1,t-j}, \hspace*{0.2cm} j =1,\ldots,t-1,
\end{align}
and 
\begin{align*}
	v_t = v_{t-1}[1-\phi_{tt}^2].
\end{align*}
The innovation variance $\sigma^2$ cancels out in \eqref{eq: D} and \eqref{eq: phi_tt}, as the mean squared errors $ v_t $ and the autocovariances $ \gamma_t $ are both proportional to $ \sigma^2 $. Thus, $ C $ can be obtained without assumptions on $ \sigma^2 $. The coefficients $ \phi_{tt} $ equal the partial autocorrelations of the process $ u_t $ at lag $ t $; see Appendix \ref{sec: Jones}.

\section{Jones' reparametrization}\label{sec: Jones}

We use the reparametrization technique of \cite{Jones1980} to transform the constrained minimization problem in \eqref{eq: theta_1}, subject to the constraints implied by \eqref{eq: stationarity condition}, into an unconstrained problem. Here, we focus on how to impose the stability condition by reparametrization of the autoregressive coefficients $ \phi_1,\ldots,\phi_p $. The same technique applies to the invertibility condition, which can be phrased as a ``stability condition" on the negative of the moving average coefficients $ -\omega_1,\ldots,-\omega_q $.

Following \cite{Jones1980}, the stability condition in \eqref{eq: stationarity condition} is imposed by reparametrizing the autoregressive coefficients $ \phi_1,\ldots,\phi_p $ in terms of partial autocorrelations $a_1,\ldots,a_p$, constrained to the open interval $ (-1,1) $ by the transformation
\begin{equation}\
	a_t = \frac{1 - \exp(-w_t)}{1 + \exp(-w_t)}, \hspace*{0.2cm} t = 1,\ldots,p, \label{eq: a}
\end{equation}
where the coefficients $ w_1,\ldots,w_p $ are unconstrained. Based on the partial autocorrelations in \eqref{eq: a}, the autoregressive coefficients are obtained from the Durbin--Levinson recursions,
\begin{align*}
	\phi_{tt} &= a_t, \hspace*{0.2cm} t = 1,\ldots,p, \\
	\phi_{tj} &= \phi_{t-1,j} - \phi_{tt}\phi_{t-1,t-j}, \hspace*{0.2cm} j =1,\ldots,t-1,
\end{align*}
with $ \phi_j = \phi_{pj}$, $ j=1,\ldots,p $. Optimization is carried out with respect to the unconstrained coefficients $ w_1,\ldots,w_p $. 

We use the inverse of the above transformation for initialization, as it is more convenient to specify initial values in terms of the original coefficients $ \phi_1,\ldots,\phi_p $ rather than the corresponding unconstrained coefficients $ w_1,\ldots,w_p $. The inverse transformation is obtained by solving \eqref{eq: a} for the unconstrained coefficients,
\begin{align*}
	w_t = \log(1+a_t) - \log(1-a_t), \hspace*{0.2cm} t = 1,\ldots,p,
\end{align*}
where the partial autocorrelations are obtained from the Durbin--Levinson algorithm with $ \phi_{tt} = a_t $ and initialization based on the original coefficients, as explained in Appendix \ref{sec: cholesky}.

\FloatBarrier
\section{Monte Carlo experiments}\label{sec: Monte Carlo (appendix)}
\FloatBarrier

This Appendix provides a comprehensive account of the Monte Carlo experiments summarized in Section \ref{sec: Monte Carlo (main text)}, including the full data-generating processes as well as additional results and discussion. It is structured as follows. Appendix \ref{sec: MC setup Appendix} describes the Monte Carlo setup and the benchmark models used for out-of-sample evaluation. In Appendix \ref{sec: MC Finite sample properties}, we investigate the finite-sample properties of the estimation procedure in Section \ref{sec: estimation}, with early stopping as described in Section \ref{sec: Early stopping}. Appendix \ref{sec: Finite sample properties benchmarks} evaluates out-of-sample prediction accuracy. Appendices \ref{sec: MC Finite sample properties} and \ref{sec: Finite sample properties benchmarks} assume the NNARMA model is correctly specified. Appendices \ref{sec: MC Misspecification} and \ref{sec: MC Misspecification Dynamics} examine, respectively, ARMA misspecification and dynamic misspecification induced by omitted lagged variables. Appendix \ref{sec:Model selection MC Appendix} motivates a simple and practical strategy for model selection in the NNARMA model. Appendix \ref{sec:Initialization MC Appendix} provides details on the initialization of the estimation procedure, including a simulation-based comparison of cold and warm starts for the Adam algorithm.

\subsection{Monte Carlo setup and benchmark models}\label{sec: MC setup Appendix}

\begin{figure}[t!]
	\centering
	\begin{subfigure}[c]{0.47\textwidth}
		\centering
		\includegraphics[width=\textwidth]{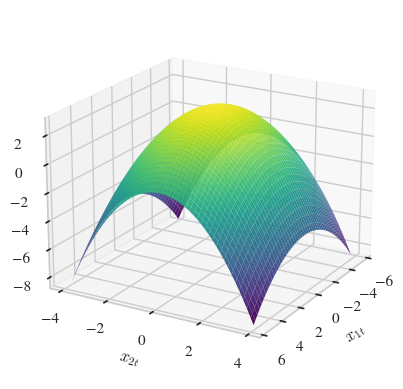}
	\end{subfigure}
	\hfill
	\begin{subfigure}[c]{0.47\textwidth}
		\centering
		\includegraphics[width=\textwidth]{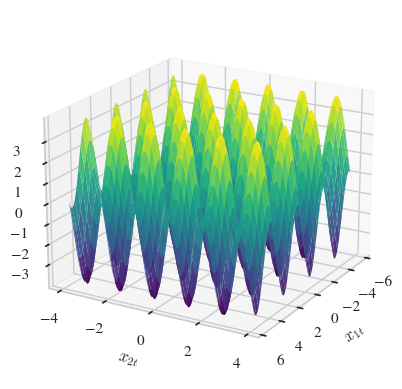}
	\end{subfigure}
	\caption{The hump-shaped (left) and sinusoidal (right) regression function.}
	\label{fig: true surfaces}
\end{figure}

For all experiments except those involving dynamic misspecification (Appendix \ref{sec: MC Misspecification Dynamics}), we generate a univariate response variable $y_t$ from two regressors $ x_{1t} $ and $ x_{2t} $, according to the data generating process
\begin{equation}\label{eq: MC dependent variable}
	y_t = f(x_{1t}, x_{2t}) + u_t, \hspace*{0.2cm} t=1,\ldots,T,
\end{equation}
where $T$ denotes the sample size. The regressors are generated as independent and stationary AR(1) processes
\begin{align}
	x_{1t} &= 0.8 x_{1t-1} + \nu_{1t}, \hspace*{0.2cm} \nu_{1t} \sim NIID(0, 1), \label{eq: MC exogenous variables 1} \\
	x_{2t} &= 0.7 x_{2t-1} + \nu_{2t}, \hspace*{0.2cm} \nu_{2t} \sim NIID(0, 1). \label{eq: MC exogenous variables 2}
\end{align}
The disturbance term in \eqref{eq: MC dependent variable} is simulated from an ARMA(1,2) process
\begin{equation}\label{eq: MC ARMA}
	u_t = 0.9 u_{t-1} + e_t - 0.5 e_{t-1} + 0.2 e_{t-2}, \hspace*{0.2cm} e_t \sim NIID(0,1/r),
\end{equation}
where $ r $ is the signal-to-noise ratio. We analyze all combinations of $ r \in \{0.05, 0.1, 0.2, 1.0\} $ and $ T \in \{250, 500, 1000, 5000\} $. Sample sizes of 250, 500, and 1000 reflect typical studies with quarterly and monthly observations, while a sample size of 5000 corresponds to the empirical study in Section \ref{sec: CFC application} (daily observations). We consider two specifications of the regression function in \eqref{eq: MC dependent variable} of varying complexity
\begin{align}
	f(x_{1t}, x_{2t}) &= 3 - 0.25 x_{1t}^2 - 0.25 x_{2t}^2, \tag{Hump-shaped} \\
	f(x_{1t}, x_{2t}) &= 2 \sin(3 x_{1t}) + 2 \sin(3 x_{2t}). \tag{Sinusoidal}
\end{align}
Both specifications are illustrated in Figure \ref{fig: true surfaces}, where each regressor is varied over the interval defined by its mean (zero) plus or minus two standard deviations. Appendix \ref{sec: MC Misspecification Dynamics} considers a modified setup in which the NNARMA model is dynamically misspecified. 

In all experiments, we split each Monte Carlo sample into three consecutive parts: (1) an estimation sample (first $ 60\% $), (2) a validation sample used for early stopping (next $ 20\% $), and (3) a test sample reserved for out-of-sample evaluation (final $ 20 \% $). The set of benchmark models used for out-of-sample evaluation comprises existing feedforward neural networks with lagged variables. Using the notation of Section \ref{sec: Monte Carlo (main text)}, they can be listed as
\begin{itemize}
	\setlength{\itemsep}{5pt}
	\item NN(0,0): $ y_t = f^\text{NN}(x_t) + u_t $
	\item NN(1,0): $ y_t = f^\text{NN}(x_t, y_{t-1}) + u_t$ 
	\item NN(0,1): $ y_t = f^\text{NN}(x_t, x_{t-1}) + u_t$
	\item NN(1,1): $ y_t = f^\text{NN}(x_t, y_{t-1}, x_{t-1}) + u_t $
	\item NN(2,2): $ y_t = f^\text{NN}(x_t, y_{t-1}, y_{t-2}, x_{t-1},x_{t-2}) + u_t $ 
	\item NN(3,3): $ y_t = f^\text{NN}(x_t, y_{t-1},\ldots,y_{t-3},x_{t-1},\ldots,x_{t-3}) + u_t $
	\item NN(5,5): $ y_t = f^\text{NN}(x_t, y_{t-1},\ldots,y_{t-5},x_{t-1},\ldots,x_{t-5}) + u_t $
	\item NN(10,10): $ y_t = f^\text{NN}(x_t, y_{t-1},\ldots,y_{t-10},x_{t-1},\ldots,x_{t-10}) + u_t $
\end{itemize}
where  $ f^\text{NN}(\cdot)$ represents a feedforward neural network and $u_t$ is left unmodeled. The benchmark models are estimated using squared error loss, as in the NNARMA model. For the latter, we use cold starts for the Adam algorithm throughout, except for the experiments involving repeated estimation across many ARMA specifications (Appendix \ref{sec: MC Misspecification}). Here, warm starts are used to reduce computational time. No overall intercept is included in the NNARMA model or the benchmark models; constant components can be represented within the network.

Throughout all experiments, we fix the network architecture to two hidden layers with $ 32 $ units in the first and $ 16 $ in the second, both for the NNARMA model and the benchmark models. This choice of network architecture is motivated in Appendix \ref{sec:Model selection MC Appendix} and serves as a representative choice for illustrating model performance without extensive tuning; see also the discussion in Section \ref{sec: Model selection (methodology)}. We use the same network architecture in the NNARMA model and the benchmark models to ensure that differences in predictive performance reflect differences in model specification rather than architecture tuning. 

Results through Appendices \ref{sec: MC Finite sample properties}--\ref{sec:Initialization MC Appendix} are, unless otherwise noted, aggregated across all combinations of $r$ and $T$ using 100 Monte Carlo replications.

\subsection{Finite sample properties}\label{sec: MC Finite sample properties}

We start by examining the finite sample properties of the proposed estimation procedure, including early stopping, under correct specification of the $\text{NNARMA(1,2)}$ model. 

\begin{figure}[t!]
	\centering
	\includegraphics[width=\textwidth]{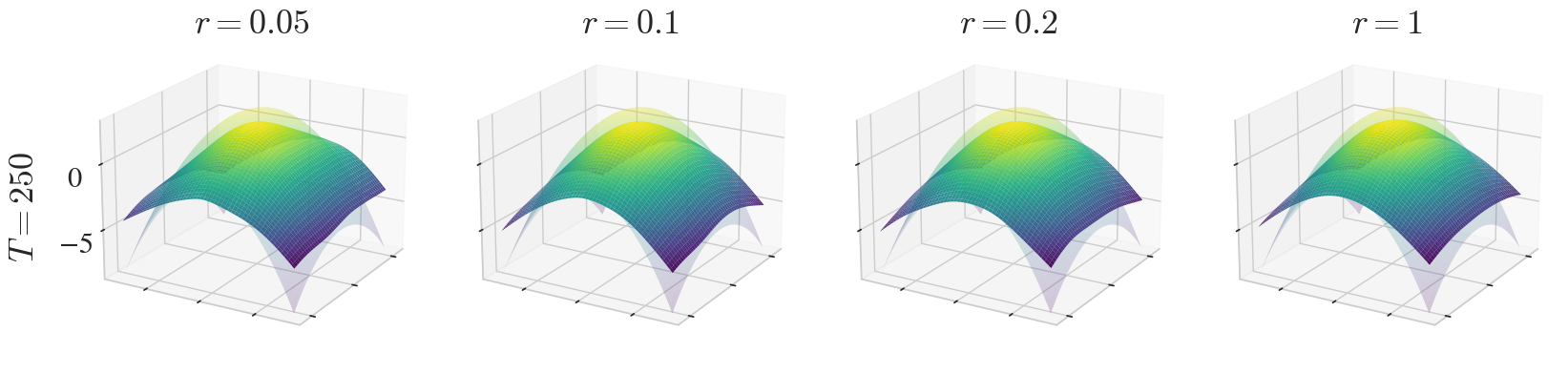} \\ \vspace*{-0.5cm}
	\includegraphics[width=\textwidth]{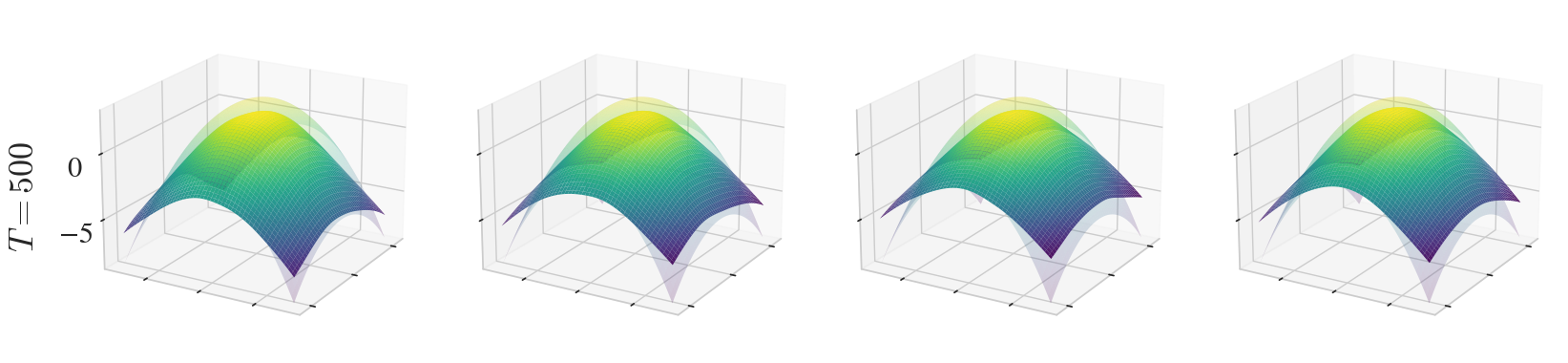} \\ \vspace*{-0.5cm}
	\includegraphics[width=\textwidth]{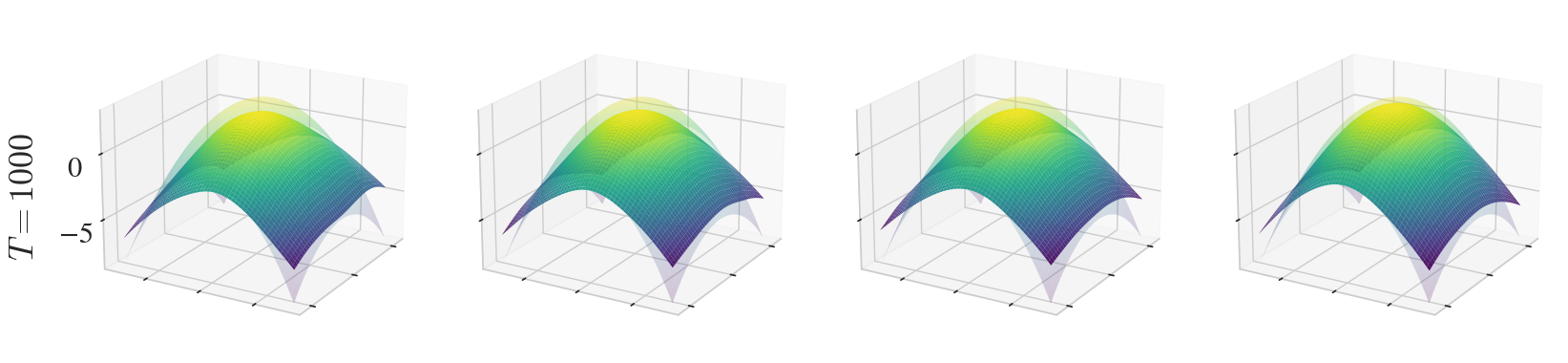} \\ \vspace*{-0.5cm}
	\includegraphics[width=\textwidth]{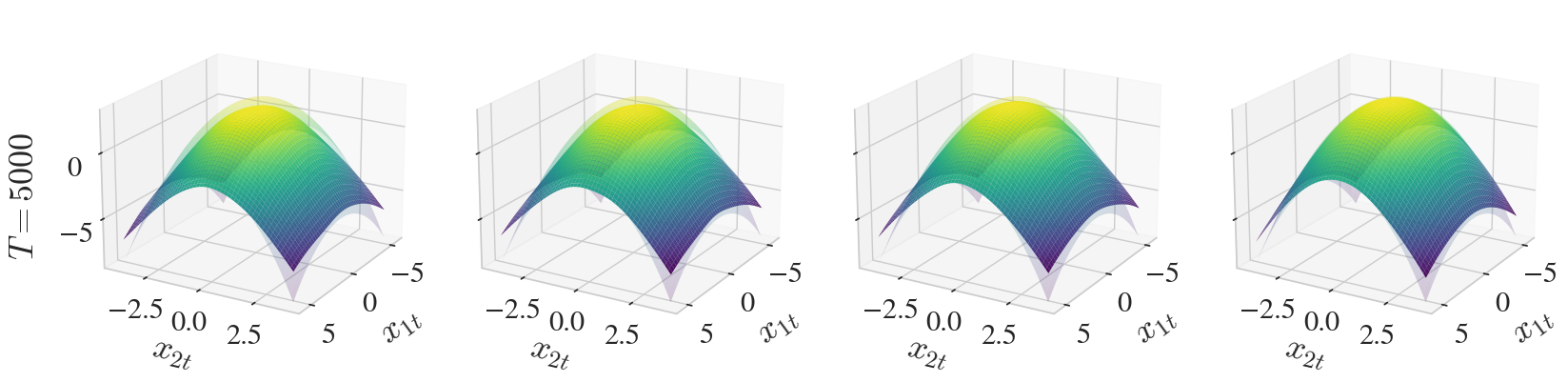} 
	\caption{Average estimate of the hump-shaped regression function. Subplots are arranged by $r$ (signal-to-noise ratio; columns) and $T$ (sample size; rows). The true function is shown transparently in the background.}
	\label{fig: estimated reg func hump}
\end{figure}

\begin{figure}[t!]
	\centering
	\includegraphics[width=\textwidth]{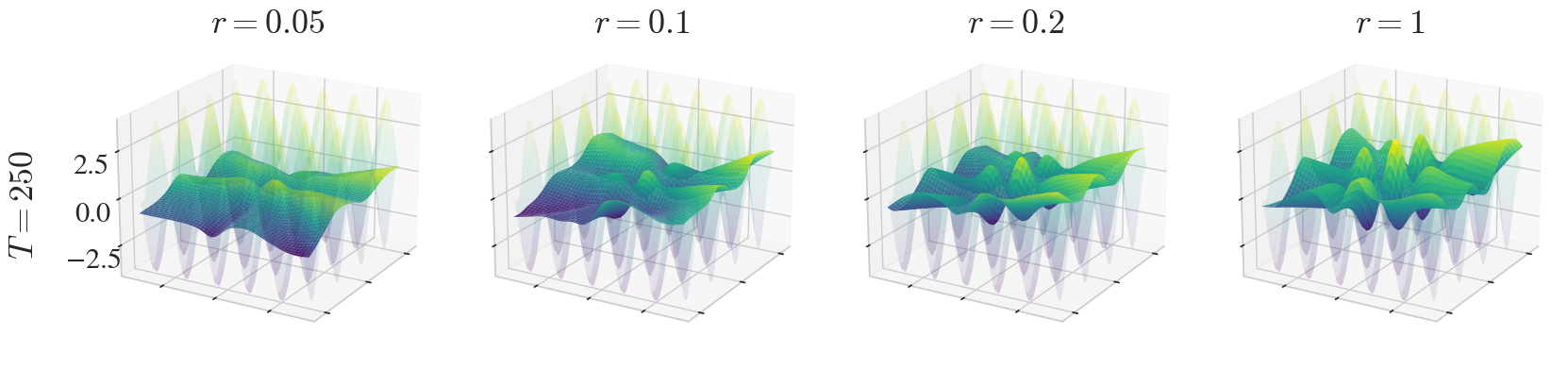} \\ \vspace*{-0.5cm}
	\includegraphics[width=\textwidth]{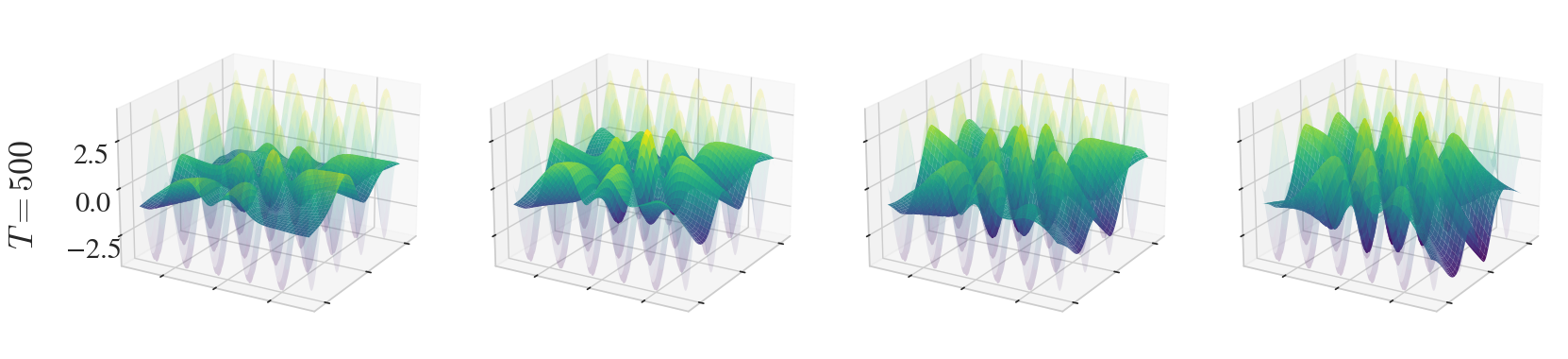} \\ \vspace*{-0.5cm}
	\includegraphics[width=\textwidth]{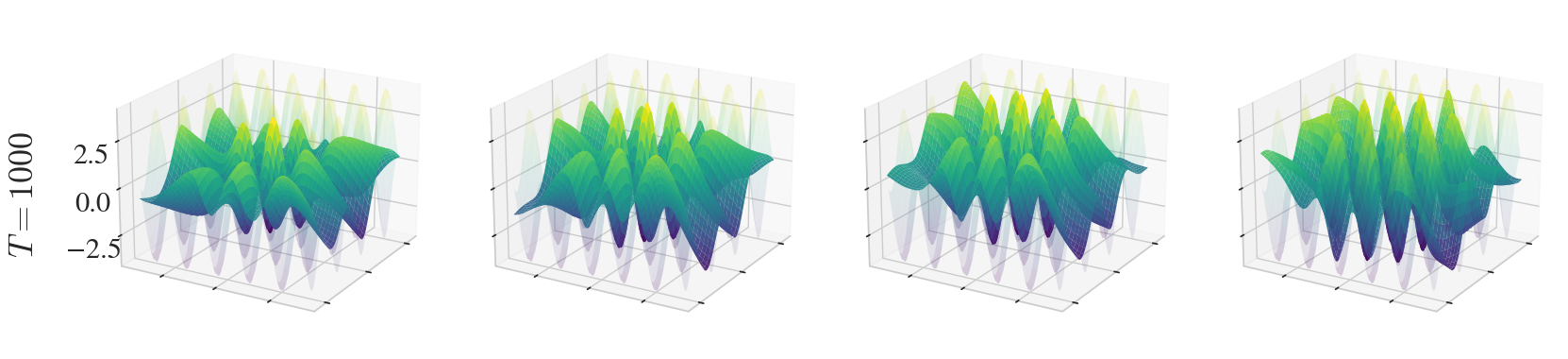} \\ \vspace*{-0.5cm}
	\includegraphics[width=\textwidth]{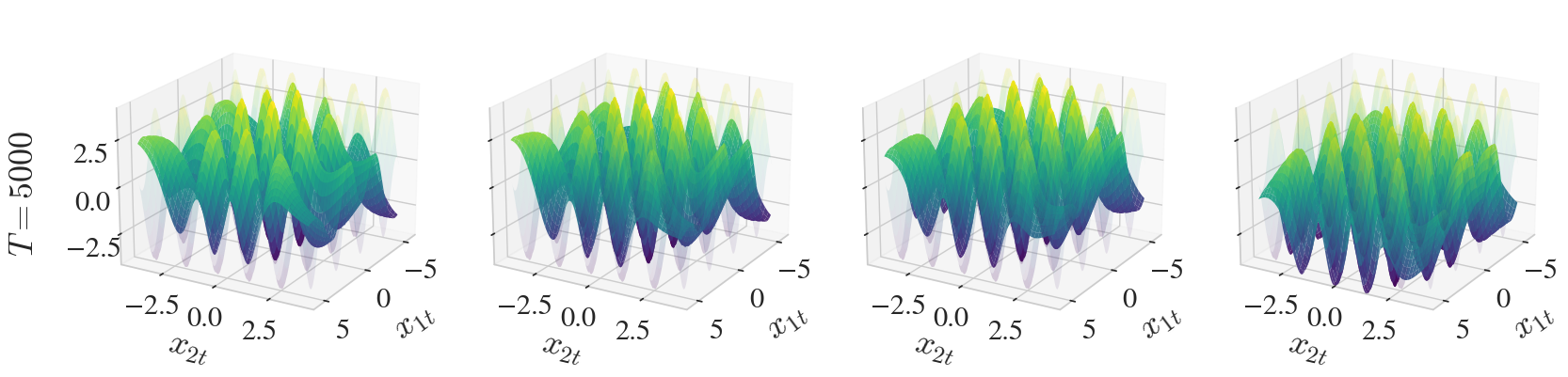} 
	\caption{Average estimate of the sinusoidal regression function. Subplots are arranged by $r$ (columns) and $T$ (rows). The true function is shown transparently in the background.}
	\label{fig: estimated reg func sine}
\end{figure}

Figures \ref{fig: estimated reg func hump} and \ref{fig: estimated reg func sine} display the average estimate of the hump-shaped and sinusoidal regression function, respectively. For both regression functions, accuracy of the estimated surface is increasing with $r$ and $T$ as expected. For high $r$ (strong signal) and large $T$ (large sample), both functions are estimated with a high degree of accuracy. The hump-shaped function is estimated with reasonable accuracy even for low $r$ (noisy signal) and small $T$ (small sample), especially in the interior of the input region. All estimated surfaces exhibit reduced accuracy near the boundaries of the input region, a pattern that is particularly pronounced for the sinusoidal function and is well known in nonparametric settings \citep{Cattaneo2020} and neural network applications \citep{Jensen2023}. At low $r$, the sinusoidal function is estimated fairly well provided the sample size is large enough. It generally requires a larger sample size than for the hump-shaped function. 

\begin{figure}[t!]
	\centering
	\begin{subfigure}[c]{\textwidth}
		\centering
		\includegraphics[width=\textwidth]{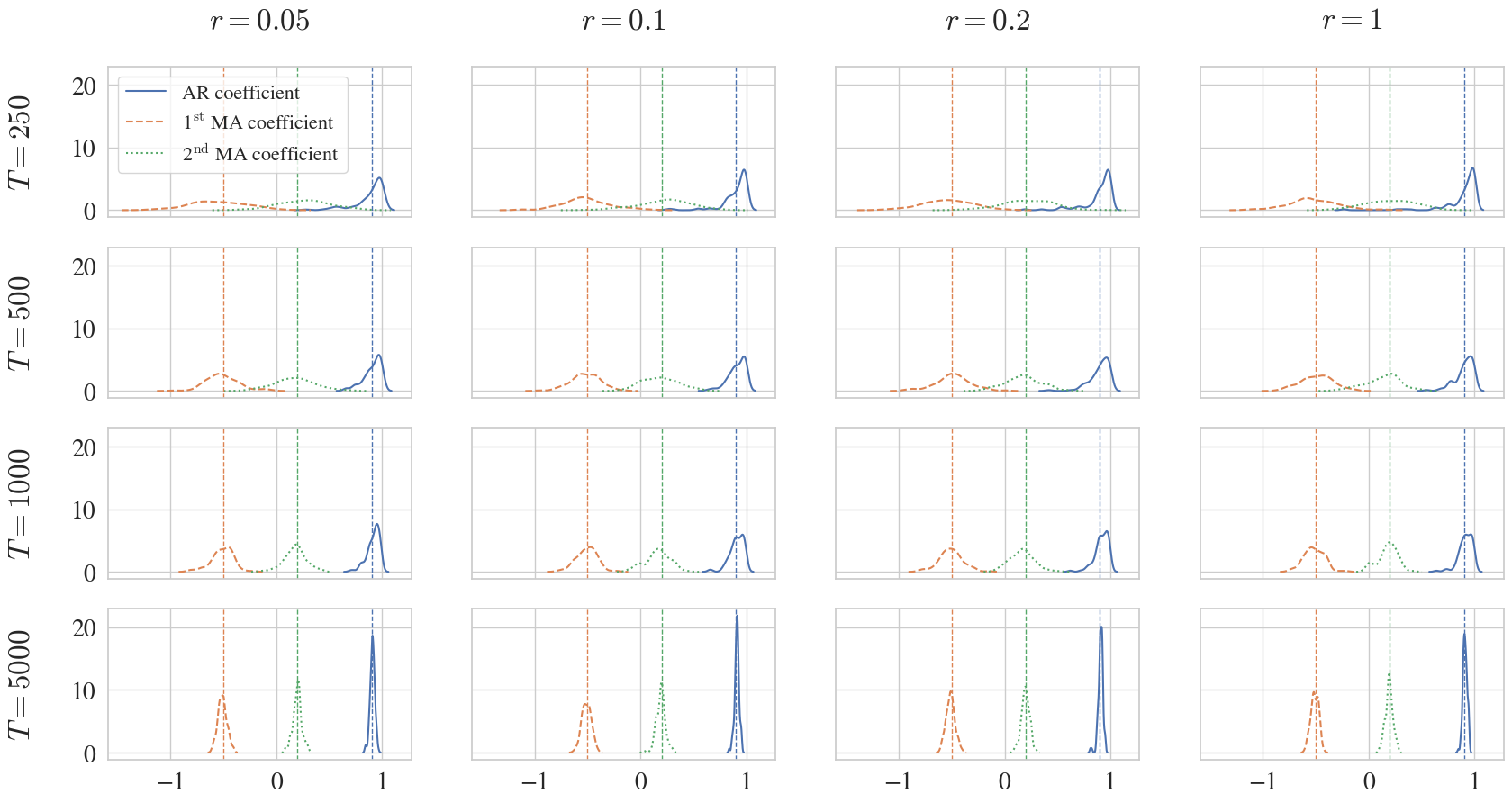}
		\subcaption{ARMA coefficient estimates}
	\end{subfigure} 
	
	\begin{subfigure}[c]{\textwidth}
		\centering
		\includegraphics[width=\textwidth]{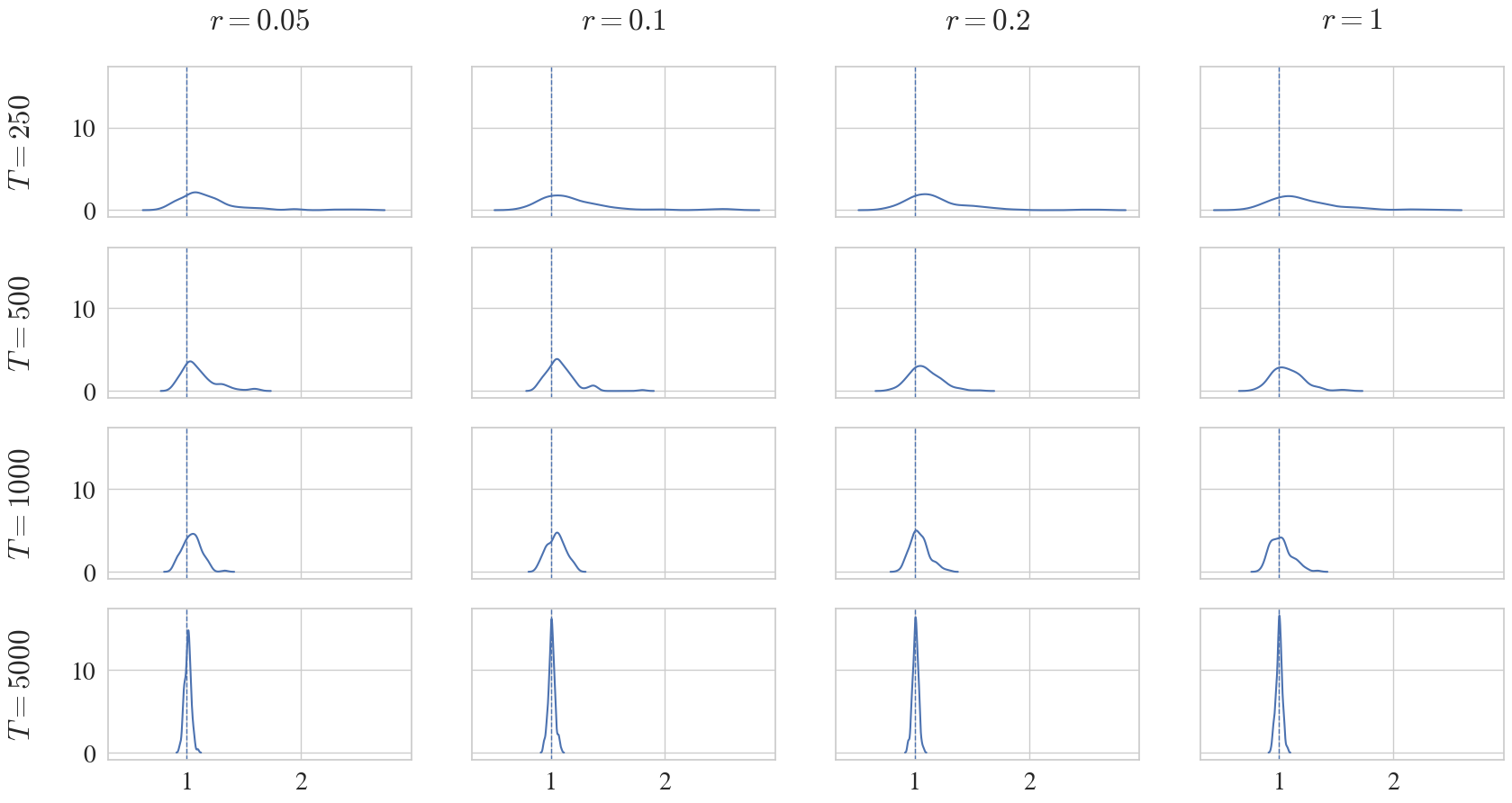}
		\subcaption{Innovation variance estimate scaled by $r$}
	\end{subfigure} 
	\caption{Hump-shaped regression function: ARMA coefficient estimates (panel a) and innovation variance estimate scaled by $r$ (panel b), both shown as sampling distributions. Subplots are arranged by $r$ (columns) and $T$ (rows). Vertical dashed lines indicate the true parameter values.}
	\label{fig: estimated arma params hump}
\end{figure}

\begin{figure}[t!]
	\centering
	\begin{subfigure}[c]{\textwidth}
		\centering
		\includegraphics[width=\textwidth]{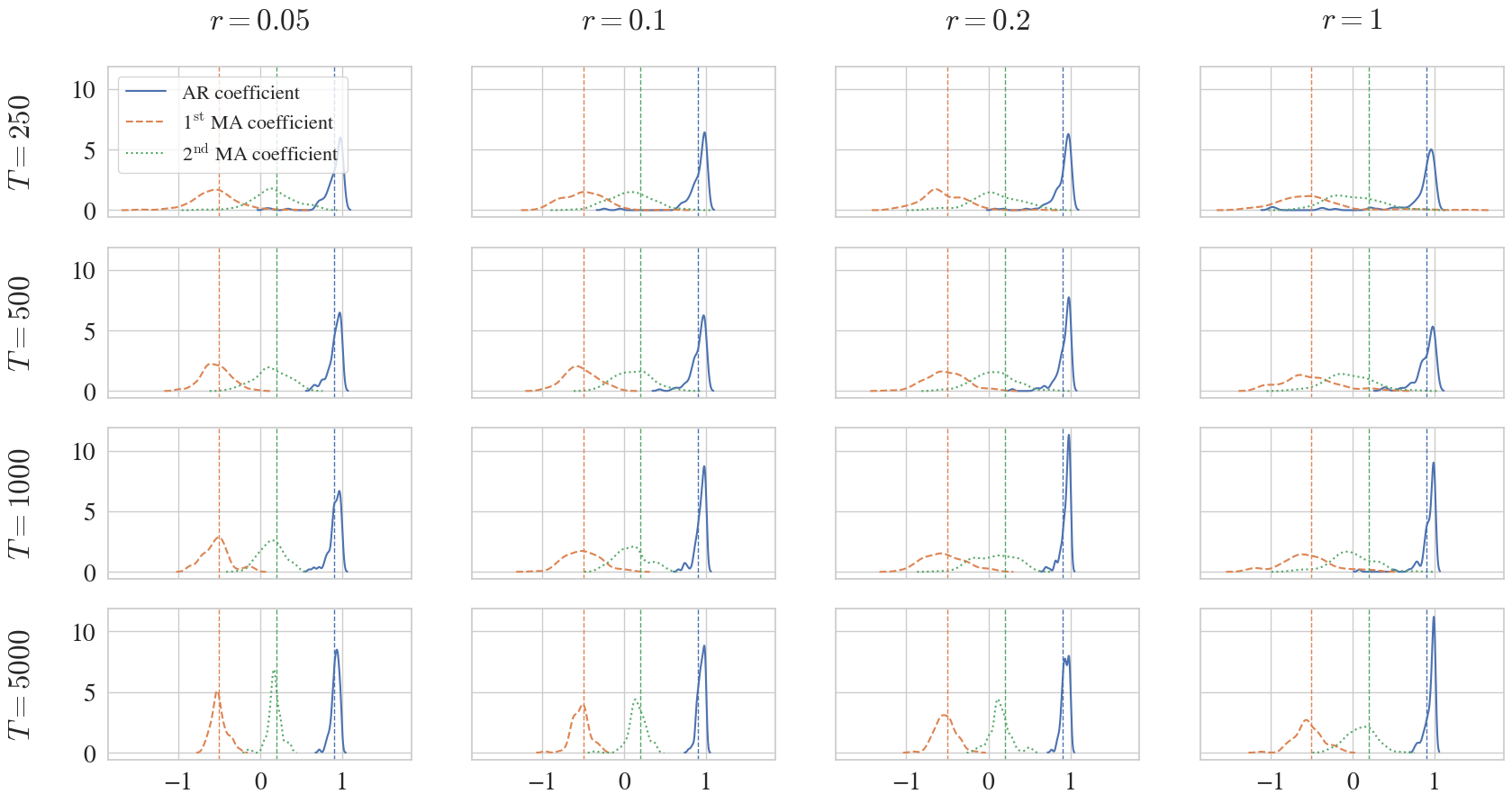}
		\subcaption{ARMA coefficient estimates}
	\end{subfigure} 
	
	\begin{subfigure}[c]{\textwidth}
		\centering
		\includegraphics[width=\textwidth]{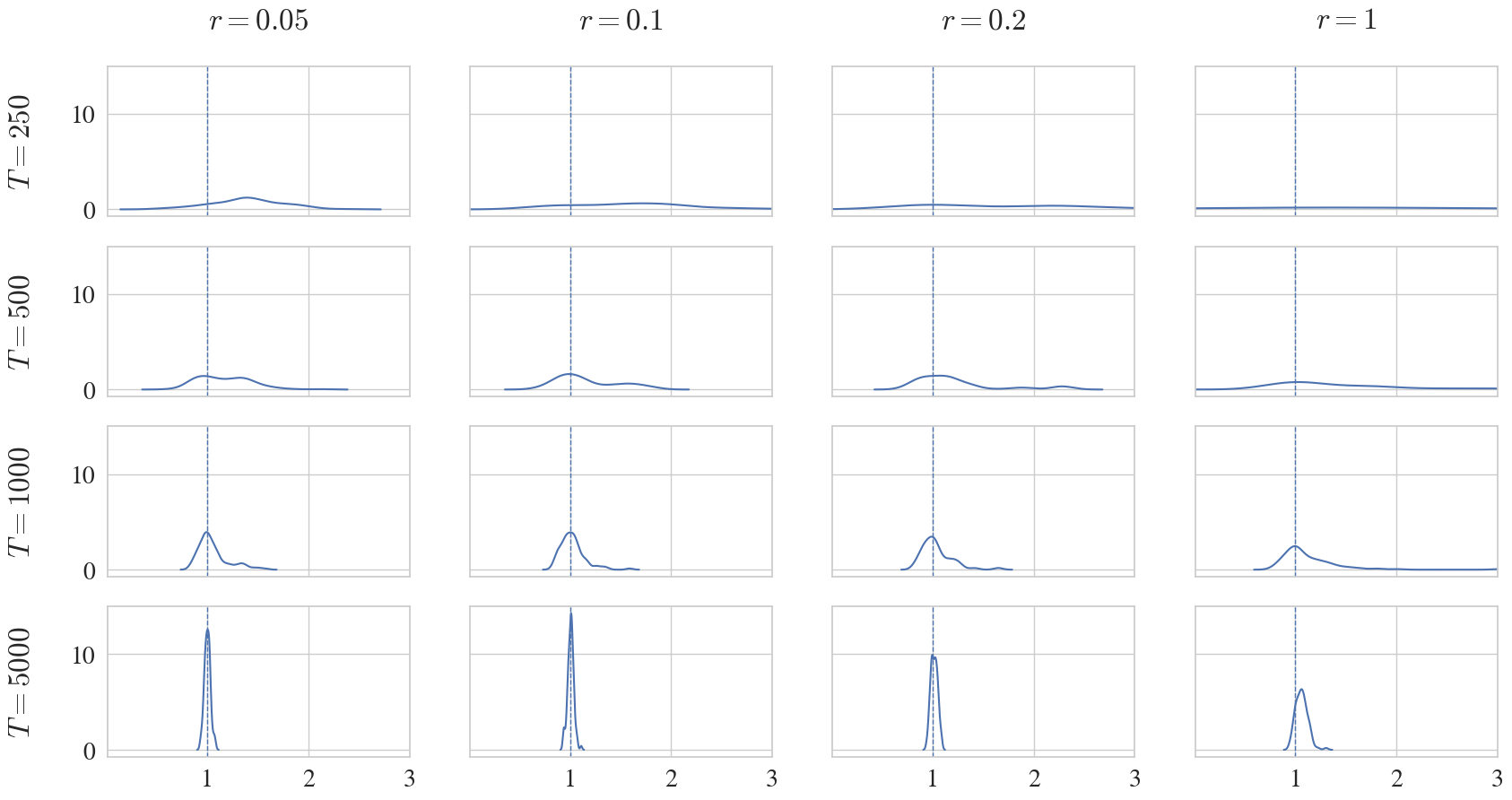}
		\subcaption{Innovation variance estimate scaled by $r$}
	\end{subfigure} 
	\caption{Sinusoidal regression function: ARMA coefficient estimates (panel a) and innovation variance estimate scaled by $r$ (panel b), both shown as sampling distributions. Subplots are arranged by $r$ (columns) and $T$ (rows). Vertical dashed lines indicate the true parameter values.}
	\label{fig: estimated arma params sine}
\end{figure}

Corresponding to the surface estimates in Figures \ref{fig: estimated reg func hump} and \ref{fig: estimated reg func sine}, Figures \ref{fig: estimated arma params hump} and \ref{fig: estimated arma params sine} display the sampling distribution of the ARMA parameter estimates. The behavior of the ARMA parameter estimates closely mirrors that of the surface estimates. Estimation accuracy is decreasing with $r$ and increasing with $T$. For low $r$ and large $T$, all parameters are accurately estimated for both regression functions. For the hump-shaped function, estimation accuracy is high even for high $r$ and small $T$. For the sinusoidal function, the AR coefficient is accurately estimated at high $r$ even for small $T$, although the sampling distribution peaks slightly above the true value. For the MA coefficients, the distributions are more dispersed than for the hump-shaped function, especially for high $r$. A larger sample size than for the hump-shaped function is required for the distributions of the MA coefficients to peak at their true values and to sharpen the distribution of the innovation variance estimate, especially when $r=1$.

\FloatBarrier
\subsection{Out-of-sample prediction performance}\label{sec: Finite sample properties benchmarks}
\FloatBarrier

This section extends the in-sample analysis using the same simulated data, and evaluates the out-of-sample prediction accuracy of the correctly specified $\text{NNARMA(1,2)}$ model relative to that of the benchmark models. 

For each Monte Carlo replication, all models are estimated once using the estimation sample, with the validation sample used for early stopping. No re-estimation is performed, and out-of-sample prediction accuracy is evaluated over the entire test sample. We compute out-of-sample predictions conditional on the regressors (and lagged response variables) in the test sample, and measure accuracy by MSE. This setup mimics the empirical setting in Section \ref{sec: CFC application}. 

\begin{figure}[t!]
	\centering
	\begin{subfigure}[c]{\textwidth}
		\centering
		\includegraphics[width=\textwidth]{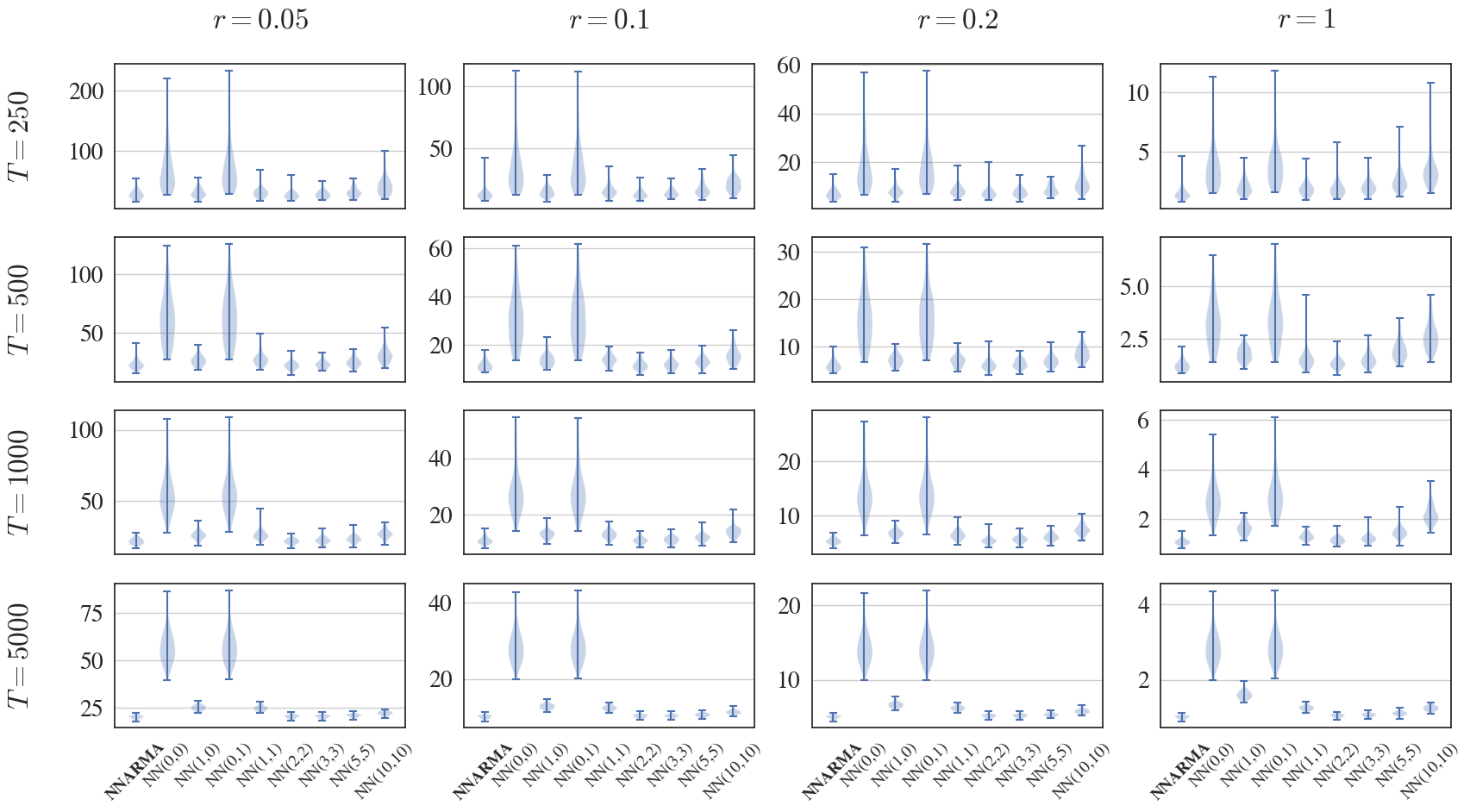}
		\subcaption{MSE sampling distribution}
	\end{subfigure} 	
	\begin{subfigure}[c]{\textwidth}
		\centering
		\includegraphics[width=\textwidth]{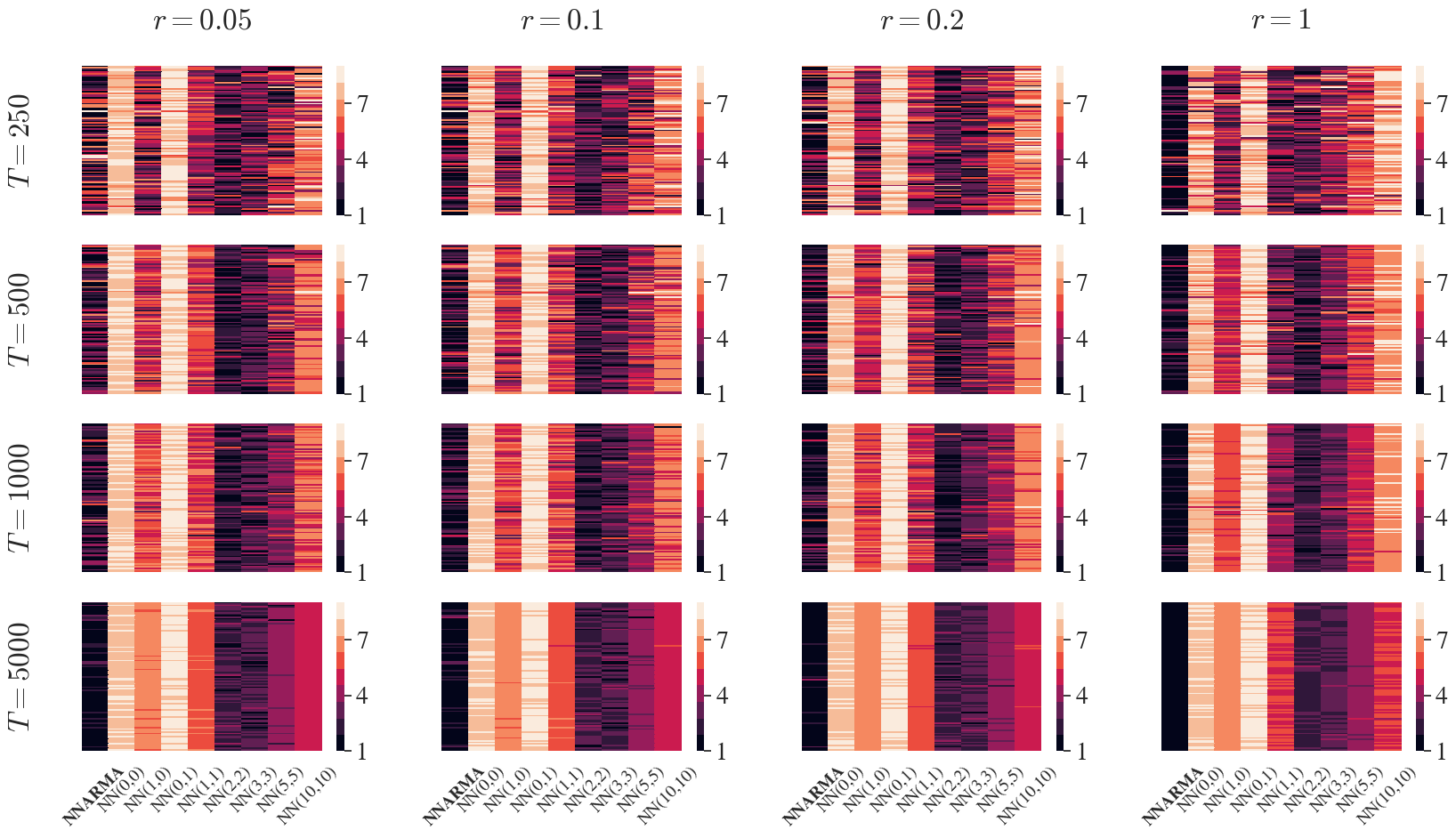}
		\subcaption{Per-sample model ranking by MSE}
	\end{subfigure} 
	\caption{Out-of-sample results under correct specification of NNARMA (bold). Hump-shaped regression function: MSE distribution (panel a) and per-sample model rankings (panel b), ordered from lowest MSE (1/dark) to highest MSE (9/light). Subplots are arranged by $r$ (columns) and $T$ (rows).}
	\label{fig: OOS hump-shaped}
\end{figure}

\begin{figure}[t!]
	\centering
	\begin{subfigure}[c]{\textwidth}
		\centering
		\includegraphics[width=\textwidth]{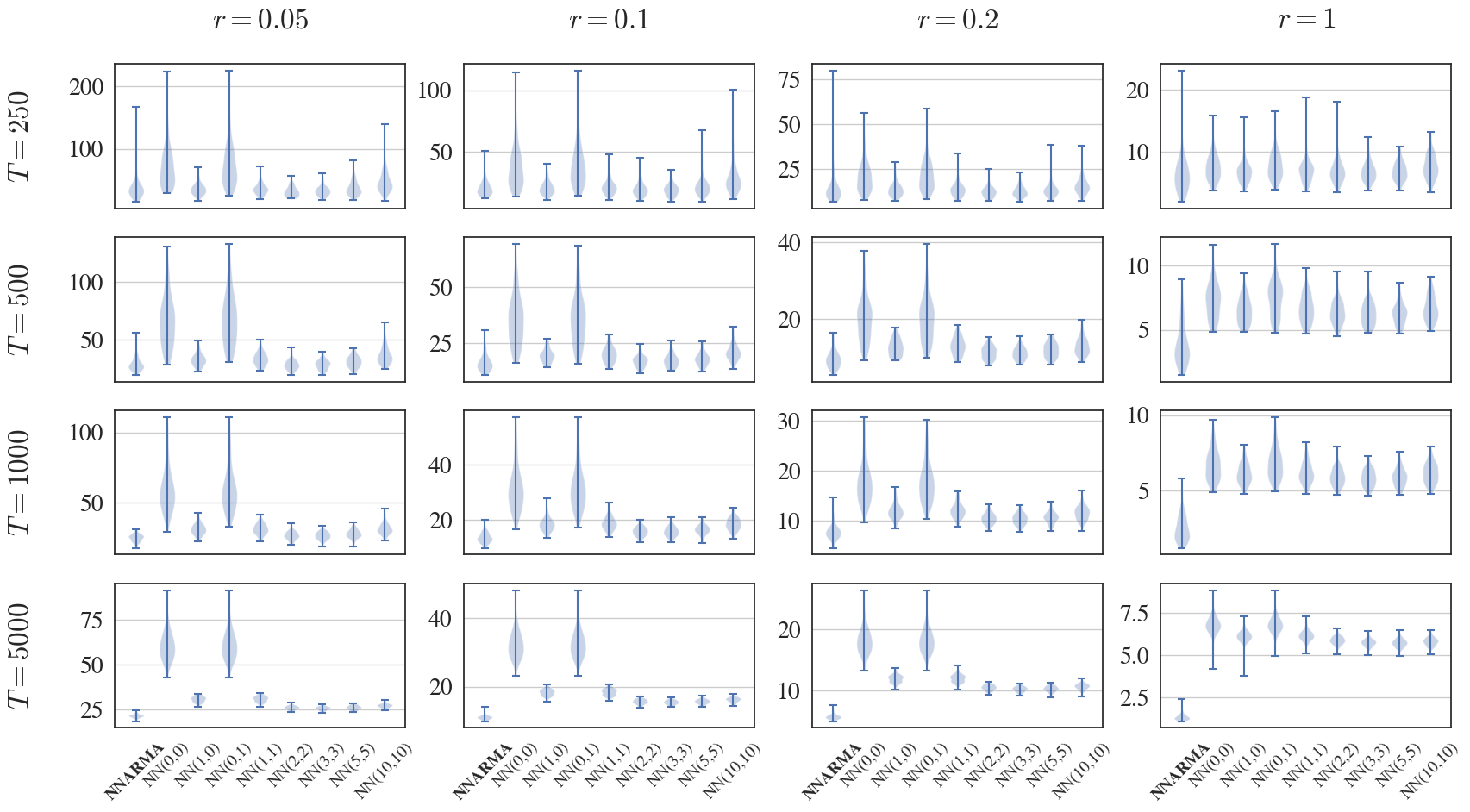}
		\subcaption{MSE sampling distribution}
	\end{subfigure} 
	\begin{subfigure}[c]{\textwidth}
		\centering
		\includegraphics[width=\textwidth]{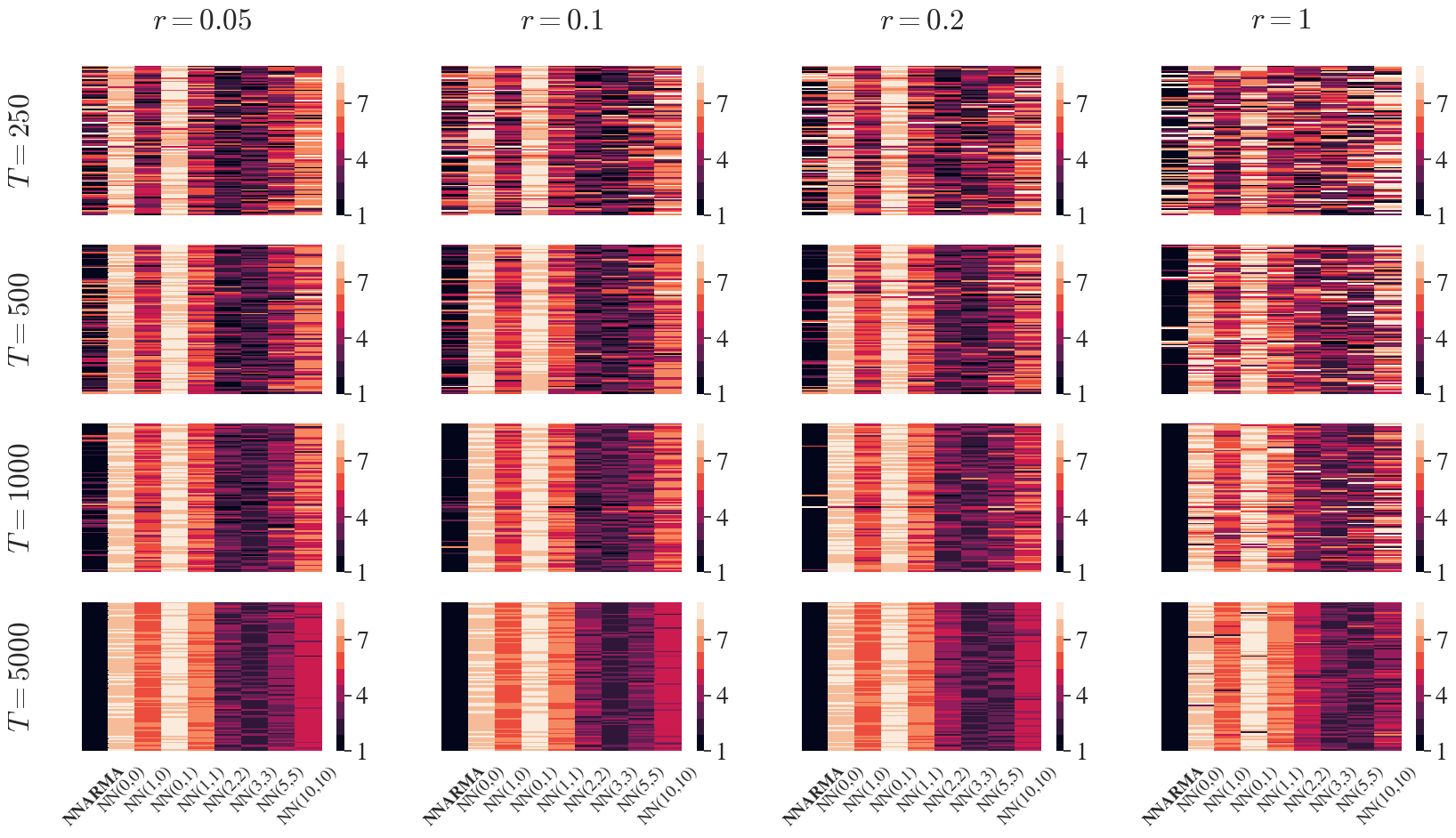}
		\subcaption{Per-sample model ranking by MSE}
	\end{subfigure} 
	\caption{Out-of-sample results under correct specification of NNARMA (bold). Sinusoidal regression function: MSE distribution (panel a) and per-sample model rankings (panel b), ordered from lowest MSE (1/dark) to highest MSE (9/light). Subplots are arranged by $r$ (columns) and $T$ (rows).}
	\label{fig: OOS sine waves}
\end{figure}

Figures \ref{fig: OOS hump-shaped} and \ref{fig: OOS sine waves} present results for the hump-shaped and sinusoidal regression functions, showing both the distribution of out-of-sample MSE values and the per-sample model ranking by MSE. The same reporting structure is adopted in all subsequent out-of-sample experiments. As expected, panel (a) shows that the NNARMA model often leads to more accurate predictions (lower MSE values) than the benchmarks, especially for the sinusoidal function. The improvements in accuracy of NNARMA relative to the benchmarks are increasing with $r$ and $T$. Accordingly, the frequency with which NNARMA is the most accurate model is increasing with $r$ and $T$, see panel (b). For both regression functions, the NNARMA model is consistently the most accurate for high $r$ and large $T$. Even for low $r$ and small $T$, the NNARMA model is typically the most accurate. The specifications $\text{NN}(0,0)$ and $\text{NN}(0,1)$ tend to yield substantially higher MSE-values than the remaining models.

\FloatBarrier
\subsection{ARMA misspecification}\label{sec: MC Misspecification}
\FloatBarrier

We examine the effect of ARMA misspecification on the finite-sample properties and out-of-sample predictive performance of the NNARMA model using the same simulated data and out-of-sample evaluation protocol as in the preceding analyses. Using the notation of Section \ref{sec: Monte Carlo (main text)}, we consider all $16$ $\text{NNARMA}(p,q)$ specifications with $p,q\in\{0,1,2,3\}$:
\begin{itemize}
	\setlength{\itemsep}{5pt}
	\item NNARMA(0,0): $ y_t = f^\text{NN}(x_t) + u_t $, $u_t \sim \text{ARMA}(0,0) $ 
	\item NNARMA(0,1): $ y_t = f^\text{NN}(x_t) + u_t $, $u_t \sim \text{ARMA}(0,1) $ 
	\item NNARMA(0,2): $ y_t = f^\text{NN}(x_t) + u_t $, $u_t \sim \text{ARMA}(0,2) $ 
	\item NNARMA(0,3): $ y_t = f^\text{NN}(x_t) + u_t $, $u_t \sim \text{ARMA}(0,3) $ 
	\item NNARMA(1,0): $ y_t = f^\text{NN}(x_t) + u_t $, $u_t \sim \text{ARMA}(1,0) $
	\item NNARMA(1,1): $ y_t = f^\text{NN}(x_t) + u_t $, $u_t \sim \text{ARMA}(1,1) $ \newline
	$\vdots$ 
	\item NNARMA(3,3): $ y_t = f^\text{NN}(x_t) + u_t $, $u_t \sim \text{ARMA}(3,3) $ 
\end{itemize}
In this setting, only the $\text{NNARMA}(1,2)$ specification is correctly specified. 

\begin{figure}[t!]
	\centering
	\begin{subfigure}{\textwidth}
		\begin{subfigure}[c]{0.47\textwidth}
			\centering
			\includegraphics[width=\textwidth]{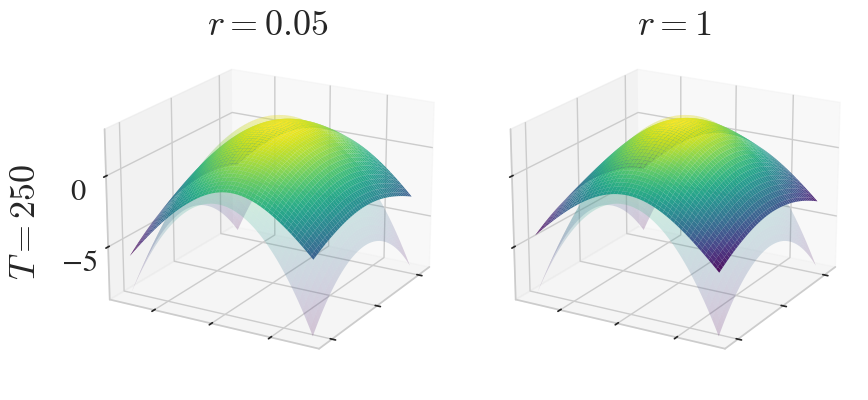}
		\end{subfigure}
		\hfill
		\begin{subfigure}[c]{0.47\textwidth}
			\centering
			\includegraphics[width=\textwidth]{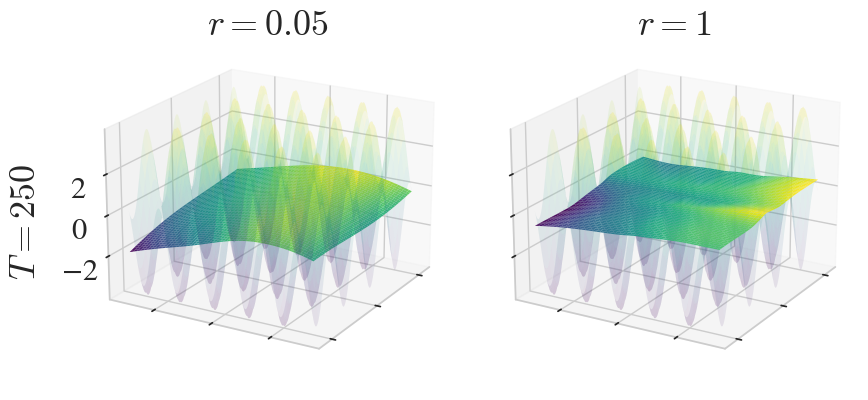}
		\end{subfigure} \\ \vspace*{-0.45cm}
		\begin{subfigure}[c]{0.47\textwidth}
			\centering
			\includegraphics[width=\textwidth]{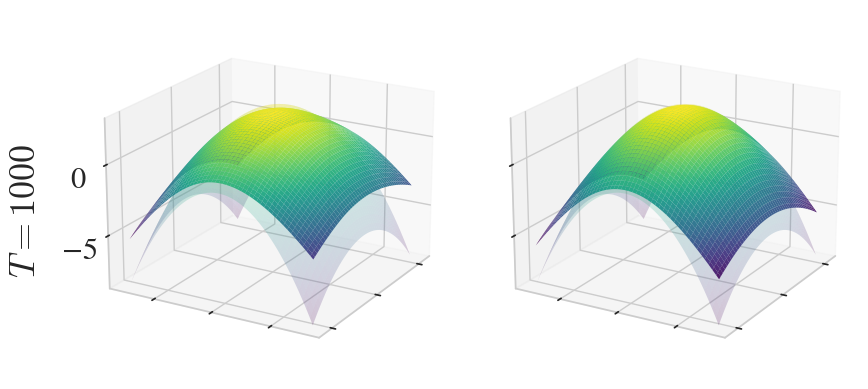}
		\end{subfigure}
		\hfill
		\begin{subfigure}[c]{0.47\textwidth}
			\centering
			\includegraphics[width=\textwidth]{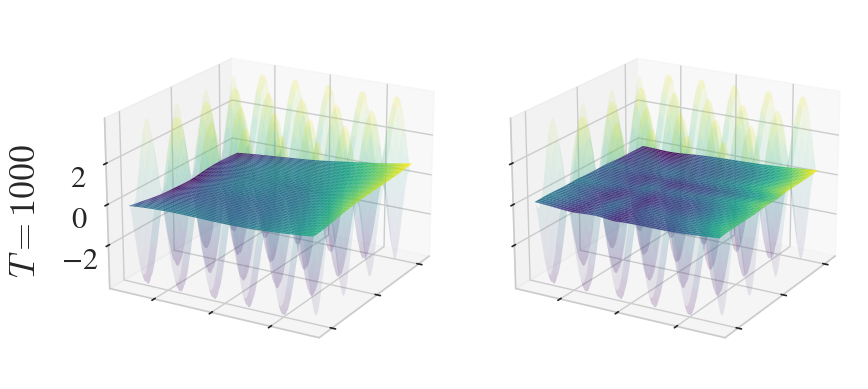}
		\end{subfigure} \\ \vspace*{-0.45cm}
		\begin{subfigure}[c]{0.47\textwidth}
			\centering
			\includegraphics[width=\textwidth]{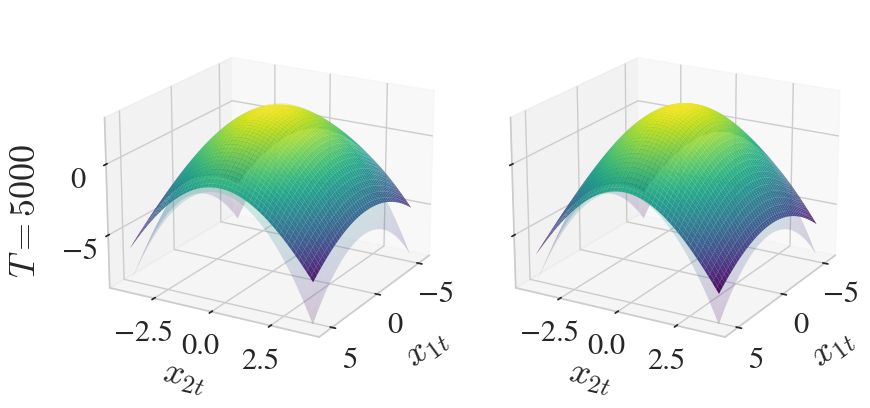}
		\end{subfigure}
		\hfill
		\begin{subfigure}[c]{0.47\textwidth}
			\centering
			\includegraphics[width=\textwidth]{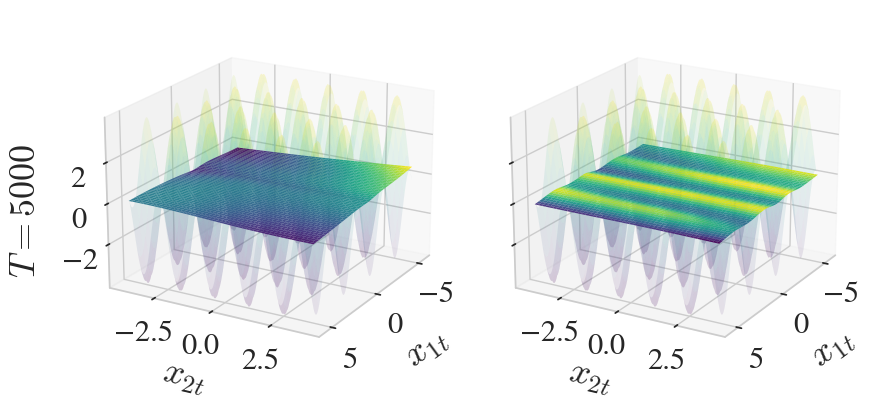}
		\end{subfigure}
		\vspace*{0.1cm}
		\subcaption{NNARMA(0,0)}
	\end{subfigure}
	
	\begin{subfigure}{\textwidth}
		\begin{subfigure}[c]{0.47\textwidth}
			\centering
			\includegraphics[width=\textwidth]{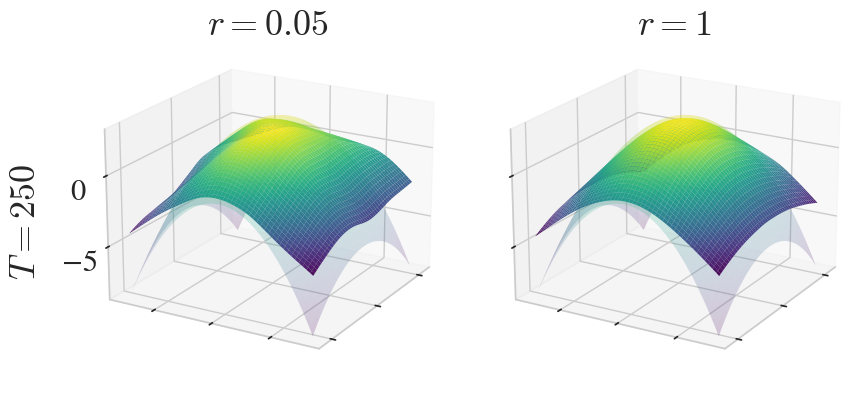}
		\end{subfigure}
		\hfill
		\begin{subfigure}[c]{0.47\textwidth}
			\centering
			\includegraphics[width=\textwidth]{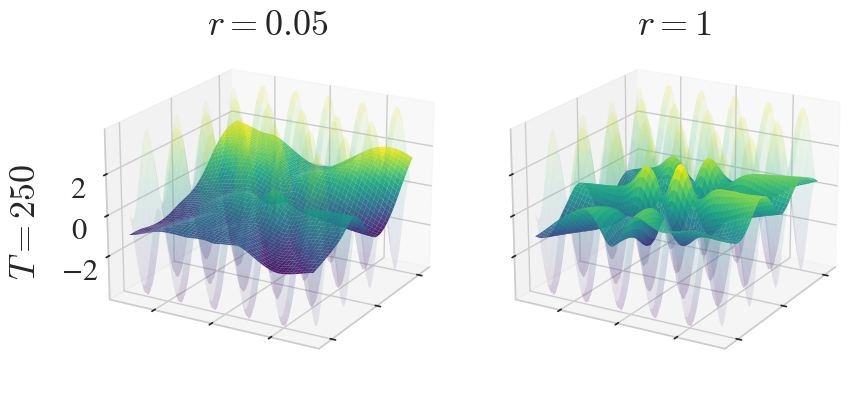}
		\end{subfigure} \\ \vspace*{-0.45cm}
		\begin{subfigure}[c]{0.47\textwidth}
			\centering
			\includegraphics[width=\textwidth]{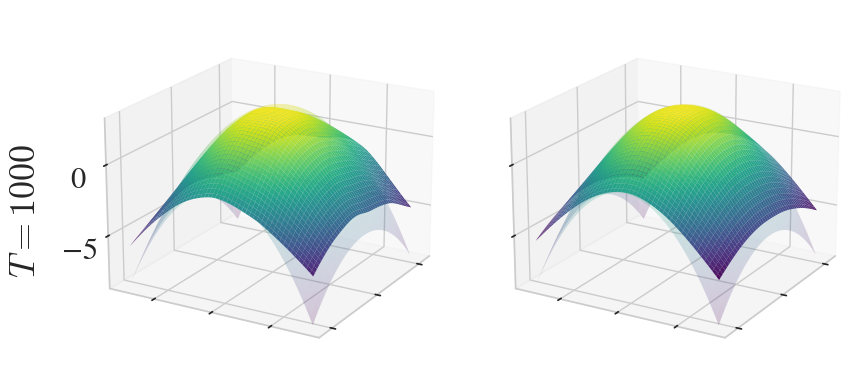}
		\end{subfigure}
		\hfill
		\begin{subfigure}[c]{0.47\textwidth}
			\centering
			\includegraphics[width=\textwidth]{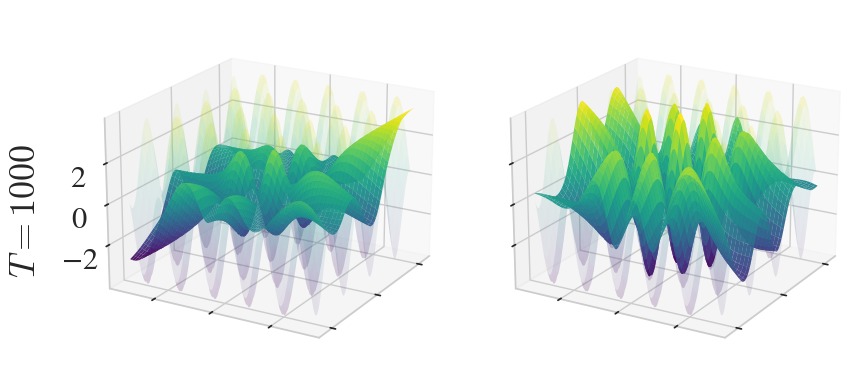}
		\end{subfigure} \\ \vspace*{-0.45cm}
		\begin{subfigure}[c]{0.47\textwidth}
			\centering
			\includegraphics[width=\textwidth]{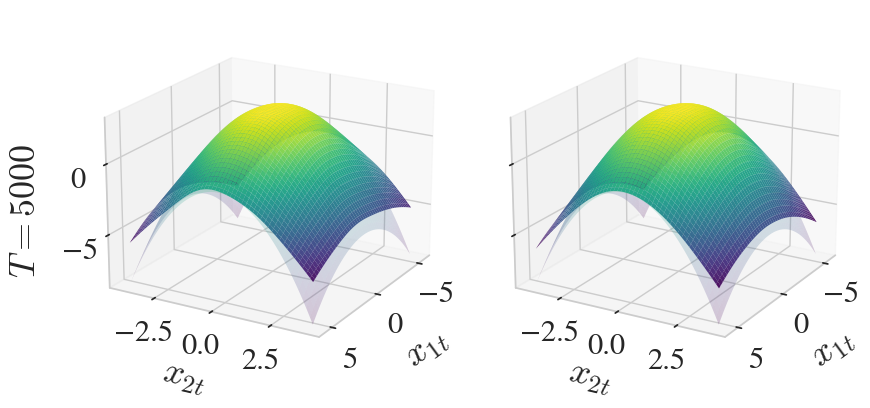}
		\end{subfigure}
		\hfill
		\begin{subfigure}[c]{0.47\textwidth}
			\centering
			\includegraphics[width=\textwidth]{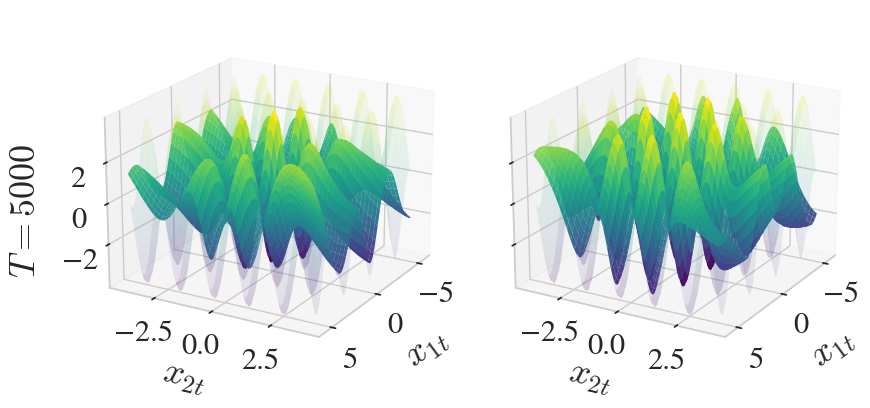}
		\end{subfigure}
		\vspace*{0.1cm}
		\subcaption{NNARMA(0,1)}
	\end{subfigure}
	\caption{Average estimates of the hump-shaped and sinusoidal regression functions under ARMA misspecification for $\text{NNARMA}(0,0)$ (panel a) and $\text{NNARMA}(0,1)$ (panel b), with $\text{NNARMA}(1,2)$ as the correct specification. Subplots are arranged by $r$ (columns) and $T$ (rows). The true function is shown transparently in the background.}
	\label{fig:ARMA mis in-sample 1}
\end{figure}

\begin{figure}[t!]
	\centering
	\begin{subfigure}{\textwidth}
		\begin{subfigure}[c]{0.47\textwidth}
			\centering
			\includegraphics[width=\textwidth]{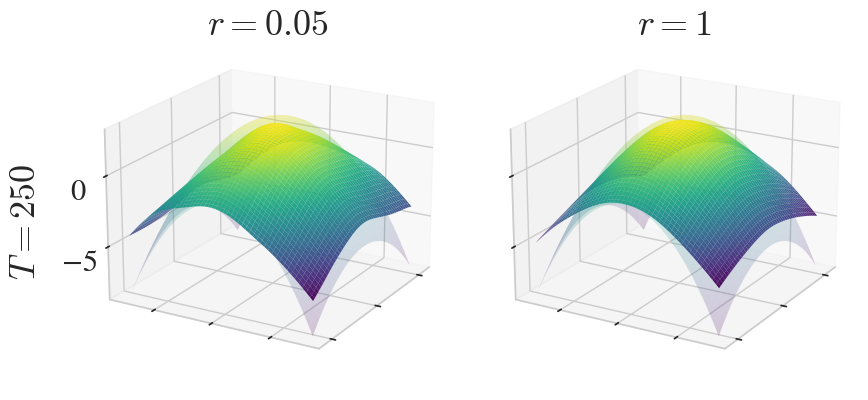}
		\end{subfigure}
		\hfill
		\begin{subfigure}[c]{0.47\textwidth}
			\centering
			\includegraphics[width=\textwidth]{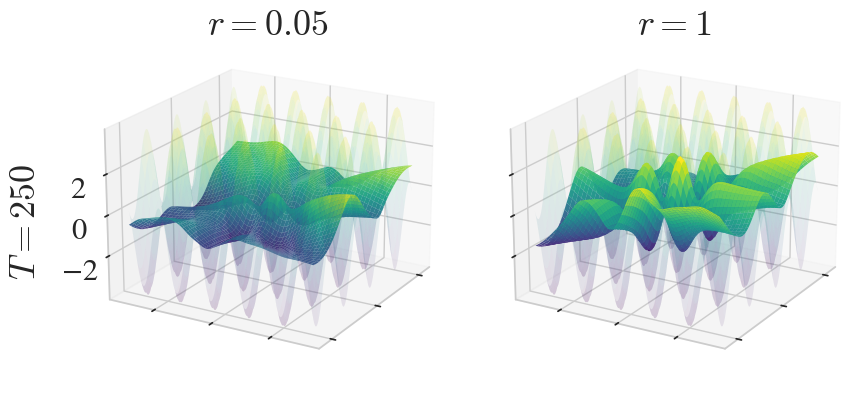}
		\end{subfigure} \\ \vspace*{-0.45cm}
		\begin{subfigure}[c]{0.47\textwidth}
			\centering
			\includegraphics[width=\textwidth]{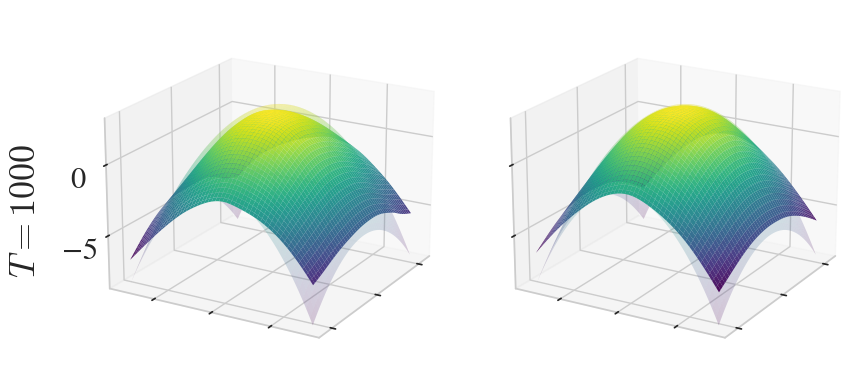}
		\end{subfigure}
		\hfill
		\begin{subfigure}[c]{0.47\textwidth}
			\centering
			\includegraphics[width=\textwidth]{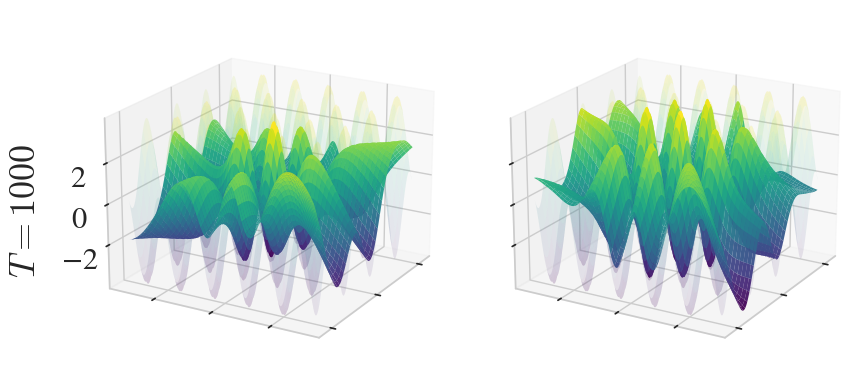}
		\end{subfigure} \\ \vspace*{-0.45cm}
		\begin{subfigure}[c]{0.47\textwidth}
			\centering
			\includegraphics[width=\textwidth]{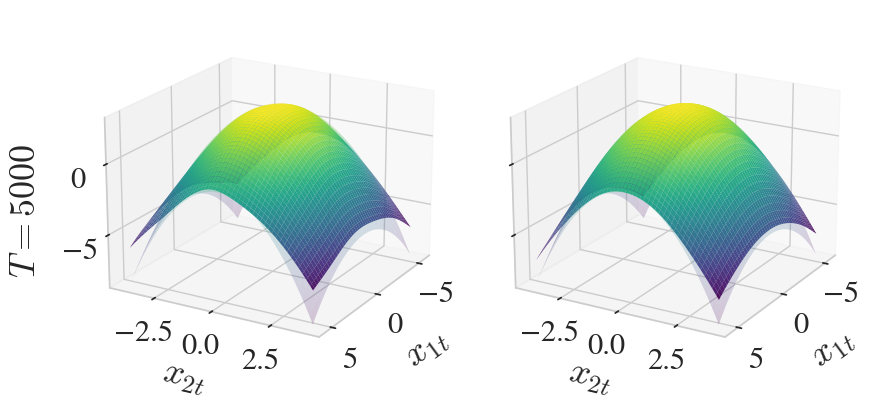}
		\end{subfigure}
		\hfill
		\begin{subfigure}[c]{0.47\textwidth}
			\centering
			\includegraphics[width=\textwidth]{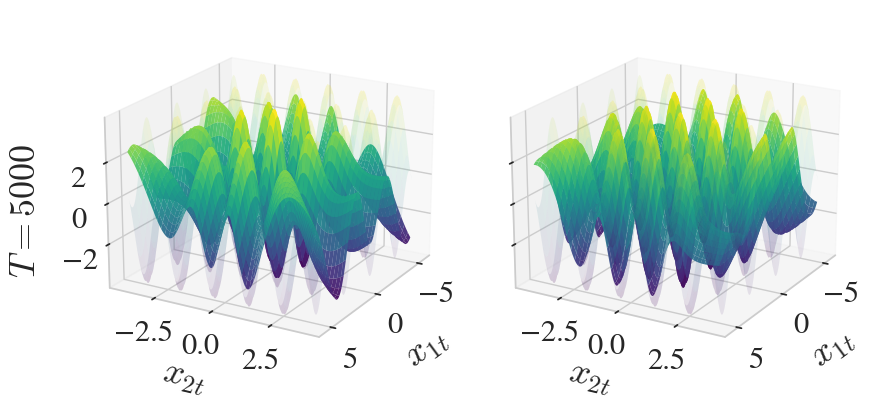}
		\end{subfigure}
		\vspace*{0.1cm}
		\subcaption{NNARMA(1,1)}
	\end{subfigure}
	
	\begin{subfigure}{\textwidth}
		\begin{subfigure}[c]{0.47\textwidth}
			\centering
			\includegraphics[width=\textwidth]{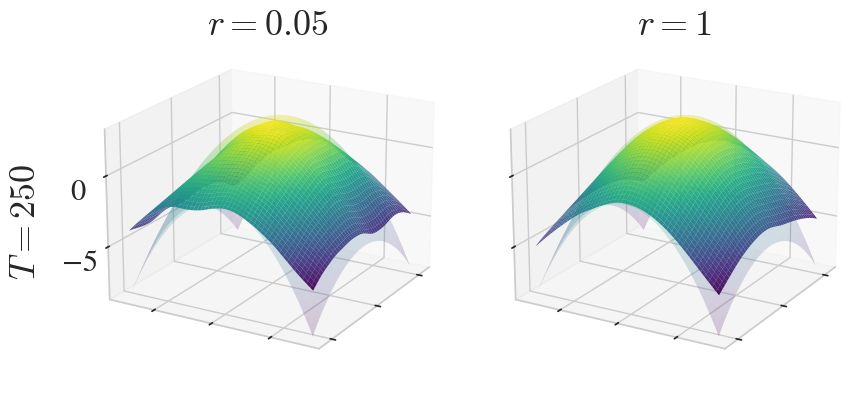}
		\end{subfigure}
		\hfill
		\begin{subfigure}[c]{0.47\textwidth}
			\centering
			\includegraphics[width=\textwidth]{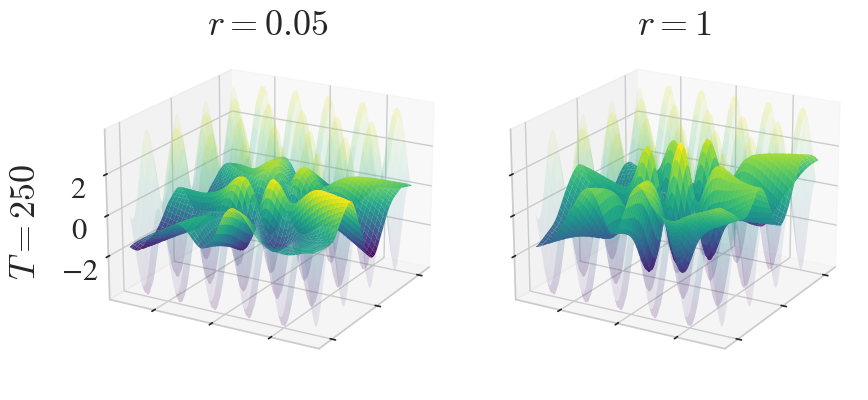}
		\end{subfigure} \\ \vspace*{-0.45cm}
		\begin{subfigure}[c]{0.47\textwidth}
			\centering
			\includegraphics[width=\textwidth]{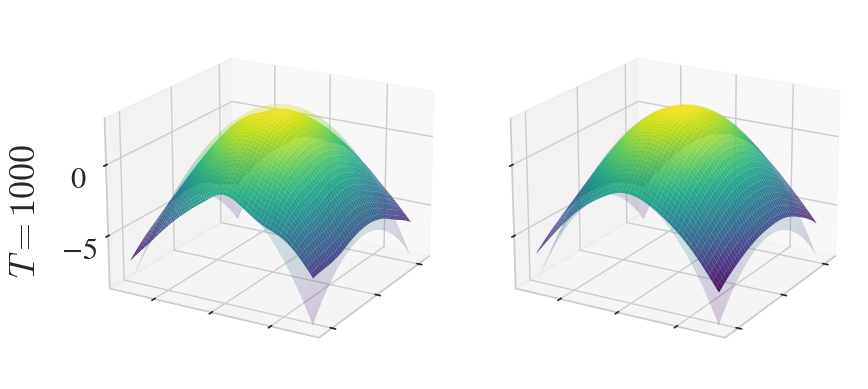}
		\end{subfigure}
		\hfill
		\begin{subfigure}[c]{0.47\textwidth}
			\centering
			\includegraphics[width=\textwidth]{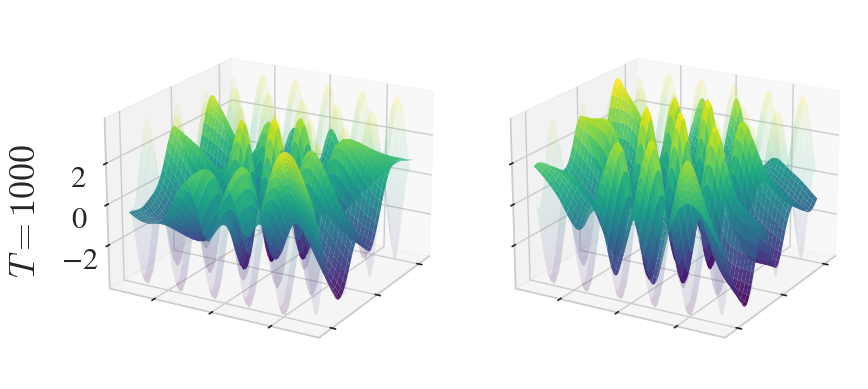}
		\end{subfigure} \\ \vspace*{-0.45cm}
		\begin{subfigure}[c]{0.47\textwidth}
			\centering
			\includegraphics[width=\textwidth]{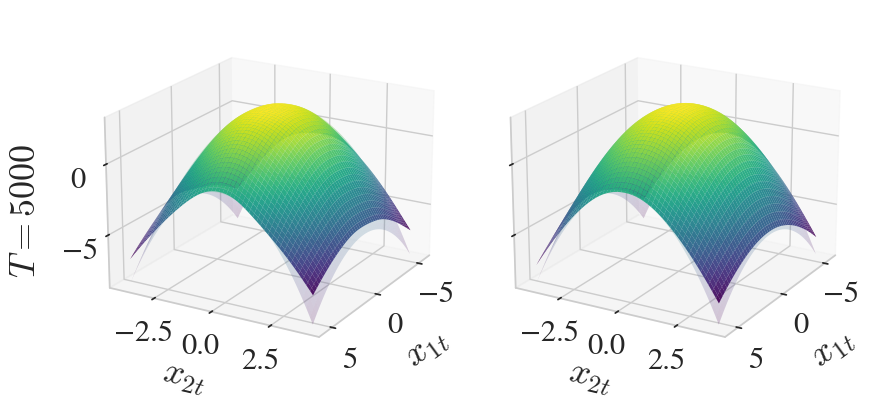}
		\end{subfigure}
		\hfill
		\begin{subfigure}[c]{0.47\textwidth}
			\centering
			\includegraphics[width=\textwidth]{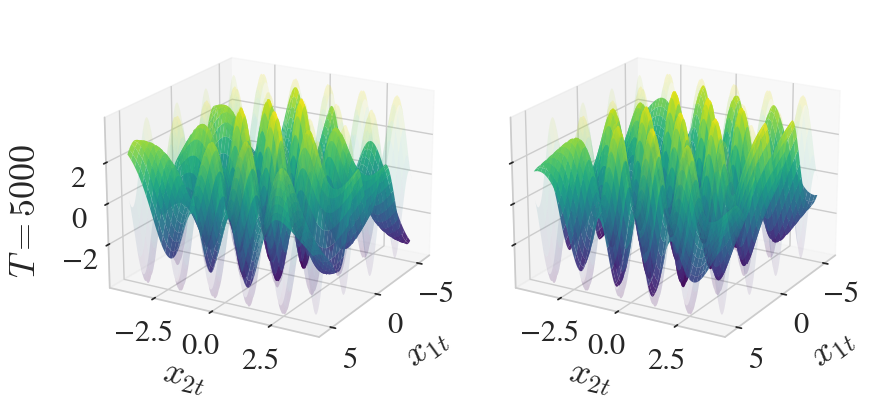}
		\end{subfigure} 
		\vspace*{0.1cm}
		\subcaption{NNARMA(3,3)}
	\end{subfigure}
	\caption{Average estimates of the hump-shaped and sinusoidal regression functions under ARMA misspecification for $\text{NNARMA}(1,1)$ (panel a) and $\text{NNARMA}(3,3)$ (panel b), with $\text{NNARMA}(1,2)$ as the correct specification. Subplots are arranged by $r$ (columns) and $T$ (rows). The true function is shown transparently in the background.}
	\label{fig:ARMA mis in-sample 2}
\end{figure}

\begin{figure}[t!]
	\centering
	\begin{subfigure}[c]{\textwidth}
		\centering
		\includegraphics[width=\textwidth]{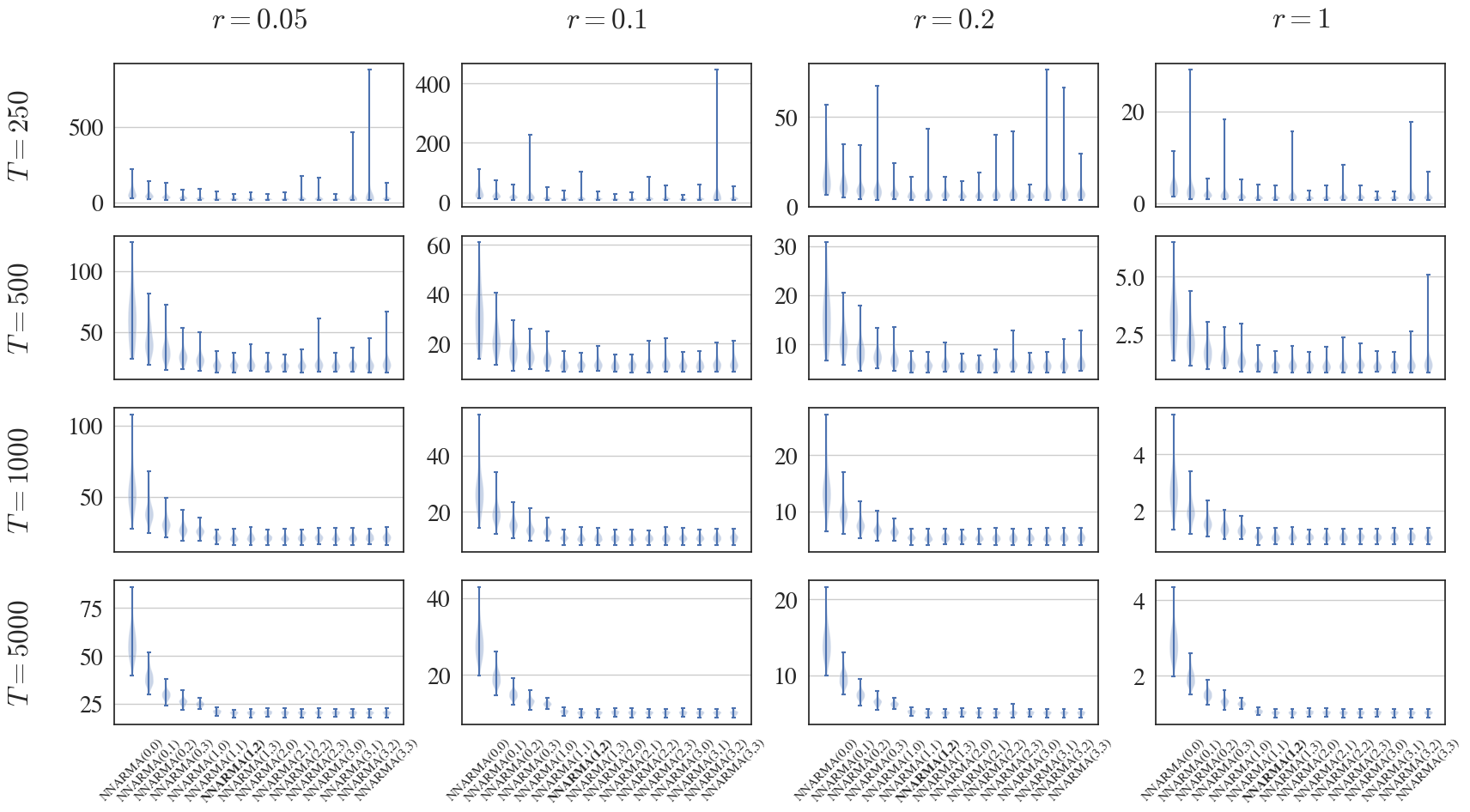}
		\subcaption{MSE sampling distribution}
	\end{subfigure} 	
	\begin{subfigure}[c]{\textwidth}
		\centering
		\includegraphics[width=\textwidth]{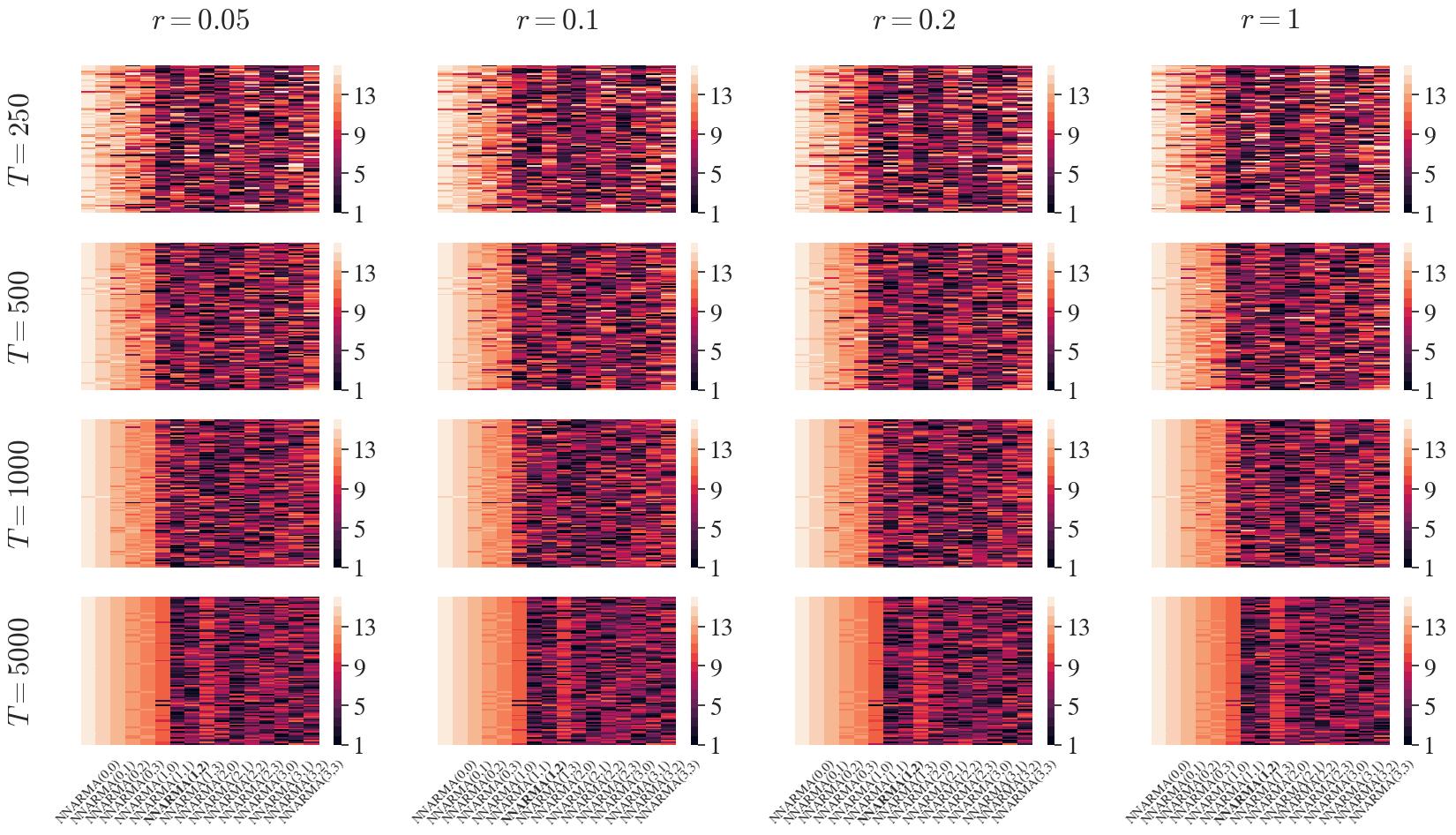}
		\subcaption{Per-sample model ranking by MSE}
	\end{subfigure} 
	\caption{Out-of-sample results under ARMA misspecification, with $\text{NNARMA}(1,2)$ as the correct specification (bold). Hump-shaped regression function: MSE distribution (panel a) and per-sample model rankings (panel b), ordered from lowest MSE (1/dark) to highest MSE (16/light). Subplots are arranged by $r$ (columns) and $T$ (rows).}
	\label{fig: mis ARMA hump}
\end{figure}

\begin{figure}[t!]
	\centering
	\begin{subfigure}[c]{\textwidth}
		\centering
		\includegraphics[width=\textwidth]{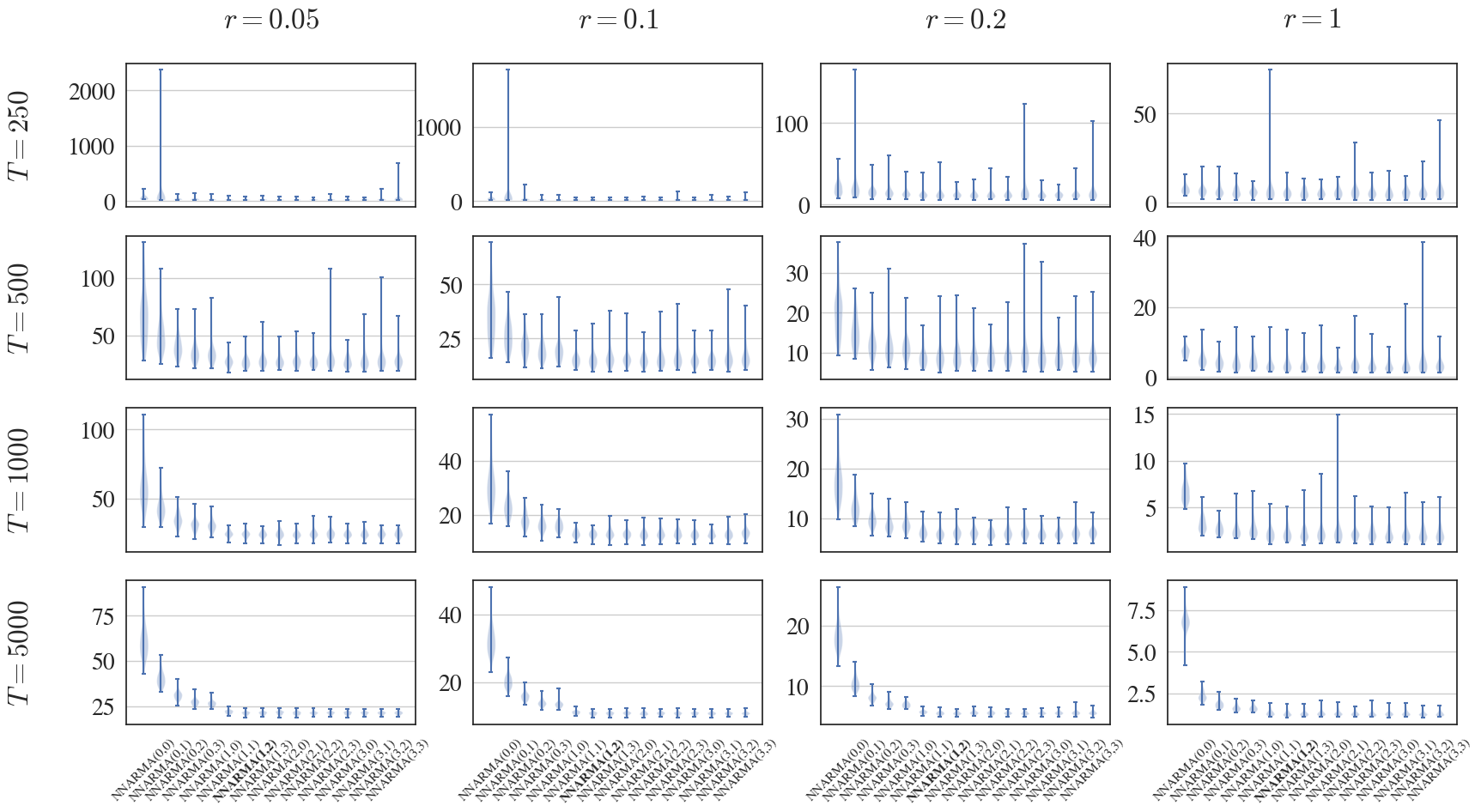}
		\subcaption{MSE sampling distribution}
	\end{subfigure} 
	\begin{subfigure}[c]{\textwidth}
		\centering
		\includegraphics[width=\textwidth]{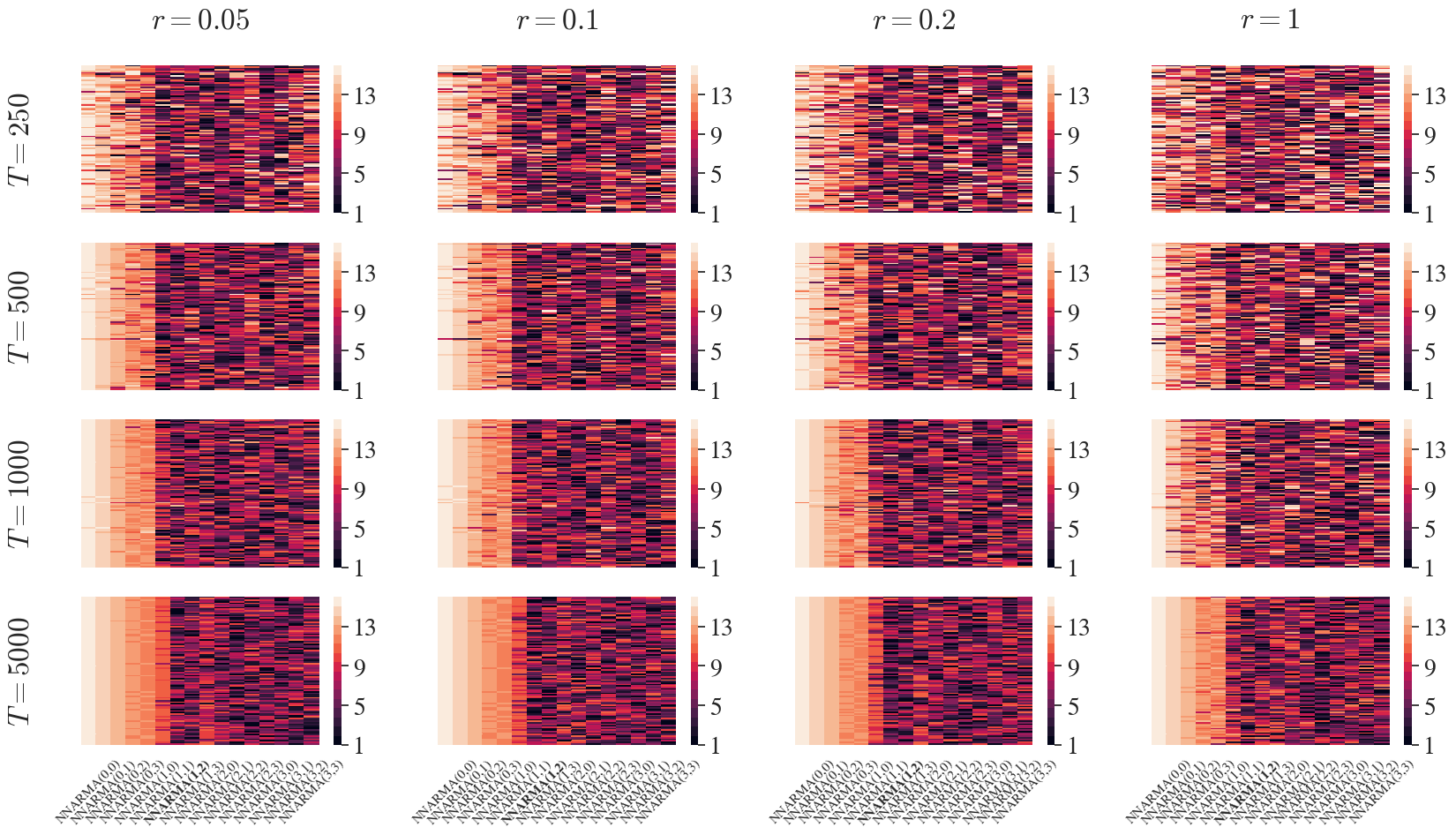}
		\subcaption{Per-sample model ranking by MSE}
	\end{subfigure} 
	\caption{Out-of-sample results under ARMA misspecification, with $\text{NNARMA}(1,2)$ as the correct specification (bold). Sinusoidal regression function: MSE distribution (panel a) and per-sample model rankings (panel b), ordered from lowest MSE (1/dark) to highest MSE (16/light). Subplots are arranged by $r$ (columns) and $T$ (rows).}
	\label{fig: mis ARMA sine waves}
\end{figure}

Figures \ref{fig:ARMA mis in-sample 1} and \ref{fig:ARMA mis in-sample 2} replicate Figure \ref{fig: estimated reg func hump sine} from Section \ref{sec: Monte Carlo (main text)} and display average estimates of the hump-shaped and sinusoidal regression functions for selected misspecifications---$\text{NNARMA}(0,0)$, $\text{NNARMA}(0,1)$, $\text{NNARMA}(1,1)$, and $\text{NNARMA}(3,3)$---over a limited range of $r$ and $T$. The $\text{NNARMA}(0,0)$ specification does not account for serial correlation in the disturbance process. For the hump-shaped regression function, larger sample sizes than for the remaining specifications are required to accurately capture the true function, especially for low $r$. For the sinusoidal function, systematic variation is effectively treated as noise, resulting in nearly flat estimated surfaces. The specification $\text{NNARMA}(0,1)$ is substantially underspecified but already yields a marked improvement in the estimated surfaces. In settings with low $r$, the hump-shaped function is estimated more accurately. For the sinusoidal function, the estimated surfaces capture some systematic variation rather than collapsing to a flat surface. For the mildly underspecified $\text{NNARMA}(1,1)$ and the substantially overspecified $\text{NNARMA}(3,3)$, the surface estimates resemble those from the correct $\text{NNARMA}(1,2)$ specification in Figure \ref{fig: estimated reg func hump sine}.

Figures \ref{fig: mis ARMA hump} and \ref{fig: mis ARMA sine waves} present the out-of-sample results for all NNARMA specifications and corroborates the in-sample results. Panel (a) reveals that prediction accuracy is lower than that of the correct specification when the ARMA structure is substantially underspecified. In particular, the $\text{NNARMA}(0,0)$ specification yields substantially higher MSE-values, especially for the sinusoidal function. Also the specifications $\text{NNARMA}(0,1)$, $\text{NNARMA}(0,2)$, $\text{NNARMA}(0,3)$, and $\text{NNARMA}(1,0)$ produce less accurate predictions than $\text{NNARMA(1,2)}$. The differences are easier to discern as $T$ increases and $r$ decreases. The accuracy of $\text{NNARMA(1,1)}$ and all specifications where the ARMA structure is overspecified fluctuates around the same level as $\text{NNARMA(1,2)}$ for all $T$ and $r$, and no model is systematically preferred, see panel (b).

\FloatBarrier
\subsection{Dynamic misspecification}\label{sec: MC Misspecification Dynamics}
\FloatBarrier

We next analyze how omitting lagged variables affects the out-of-sample predictive performance of the NNARMA model. We modify the simulation setup used in the preceding analysis so that the NNARMA model is dynamically misspecified. First, we omit lagged input variables from the NNARMA model. Next, we omit both lagged input and lagged output variables. In both sets of experiments, the NNARMA model selects $p$ and $q$ automatically for each replication using the practical model selection strategy described in Section \ref{sec: Model selection (methodology)}.

\vspace{0.75em}
\noindent\textbf{Omission of lagged inputs}\label{sec: MC omitting lagged exogenous variables}

\noindent To examine the effect of omitting lagged regressors from the NNARMA model, we generate the dependent variables as
\begin{align}\label{eq: MC dependent variable dynamic misspecification}
	y_t = f(\tilde x_{1t}, \tilde x_{2t}) + u_t, \hspace*{0.2cm} t=1,\ldots,T,
\end{align}
where
\begin{align}
	\tilde x_{jt} = 0.4 x_{jt} + 0.3 x_{jt-1} + 0.2 x_{jt-2}, \hspace*{0.2cm} j = 1,2, \label{eq: MC regressors dynamic misspecification}
\end{align}
and $ u_t \sim NIID(0,1/r)$. The exogenous variables $x_{1t}, x_{2t}$ are generated as in \eqref{eq: MC exogenous variables 1}--\eqref{eq: MC exogenous variables 2}. We also consider the same choices of $r$, $T$, hump-shaped, and sinusoidal regression function as above. While the regression function in \eqref{eq: MC dependent variable dynamic misspecification} includes lagged regressors, we estimate the NNARMA model without lags and compare its out-of-sample prediction accuracy with that of the benchmarks. We include in the comparison the correctly specified benchmark with two lagged regressors, referred to as $\text{NN}(0,2)$.

\begin{figure}[t!]
	\centering
	\begin{subfigure}[c]{\textwidth}
		\centering
		\includegraphics[width=\textwidth]{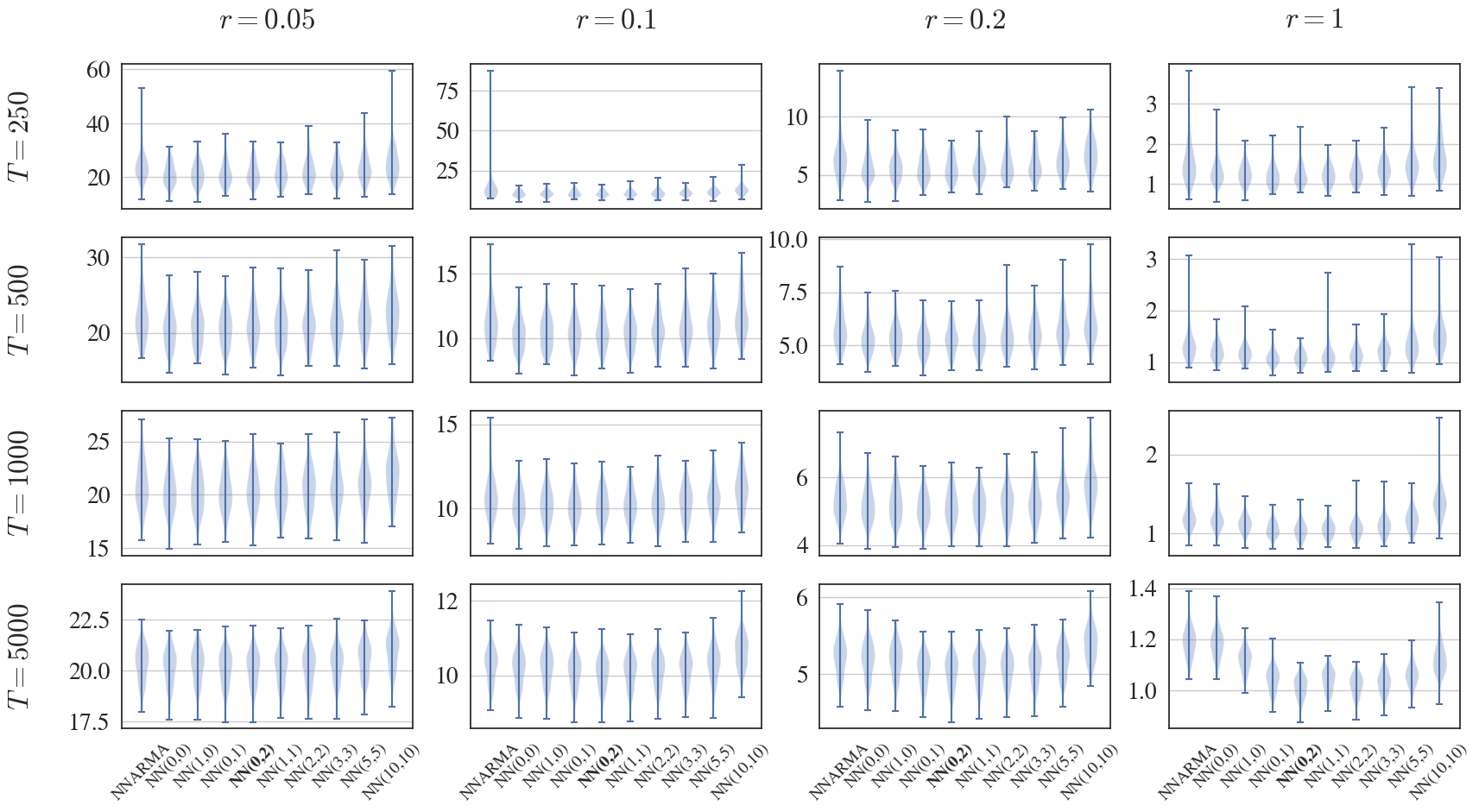}
		\subcaption{MSE sampling distribution}
	\end{subfigure} 
	\begin{subfigure}[c]{\textwidth}
		\centering
		\includegraphics[width=\textwidth]{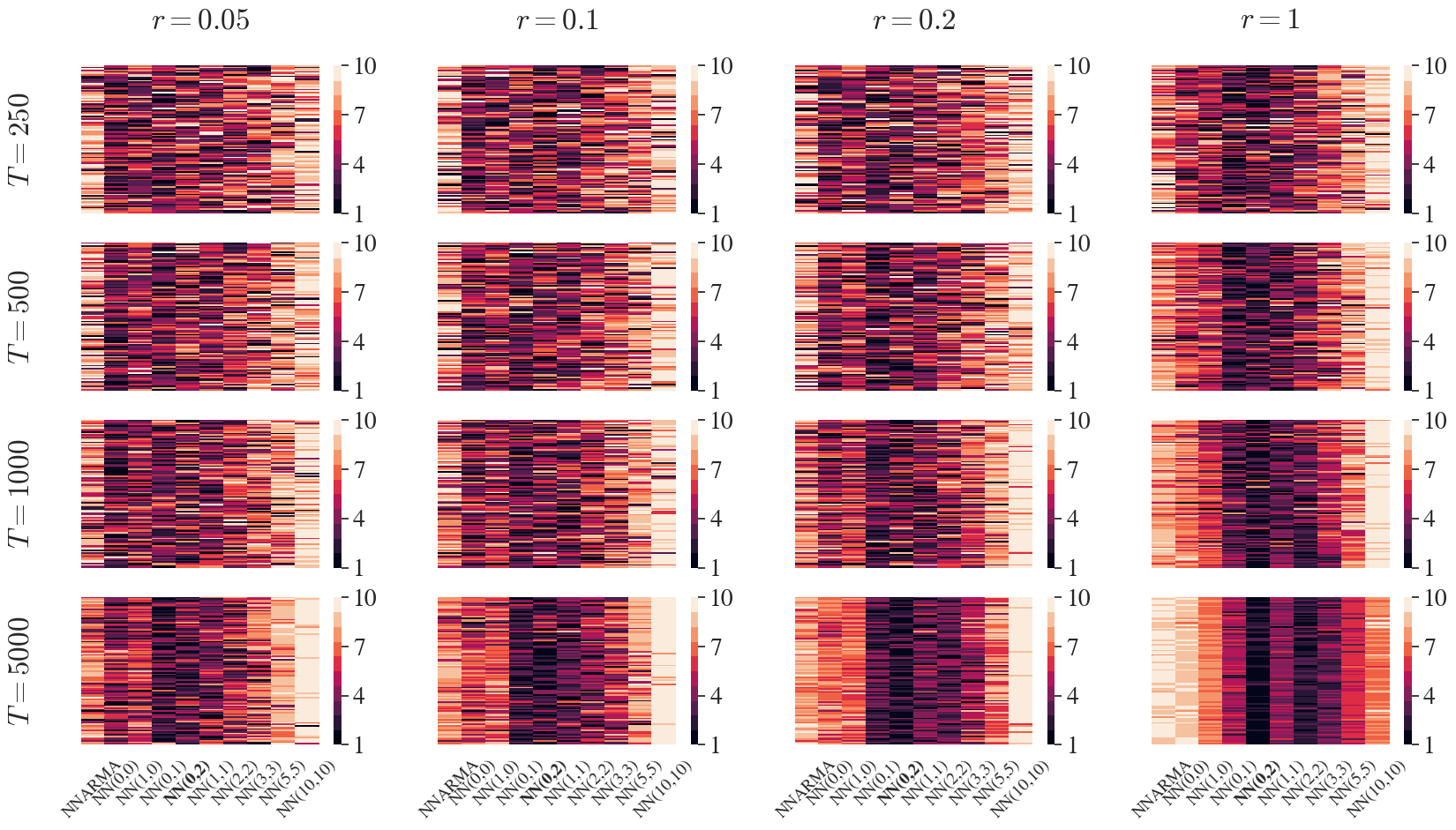}
		\subcaption{Per-sample model ranking by MSE}
	\end{subfigure} 
	\caption{Out-of-sample results under dynamic misspecification of NNARMA, with $\text{NN}(0,2)$ as the correct specification (bold). Hump-shaped regression function: MSE distribution (panel a) and per-sample model rankings (panel b), ordered from lowest MSE (1/dark) to highest MSE (10/light). Subplots are arranged by $r$ (columns) and $T$ (rows). The NNARMA model selects $p$ and $q$ automatically for each replication using the practical model selection strategy described in Section \ref{sec: Model selection (methodology)}.}
	\label{fig: omit input hump}
\end{figure}

\begin{figure}[t!]
	\centering
	\begin{subfigure}[c]{\textwidth}
		\centering
		\includegraphics[width=\textwidth]{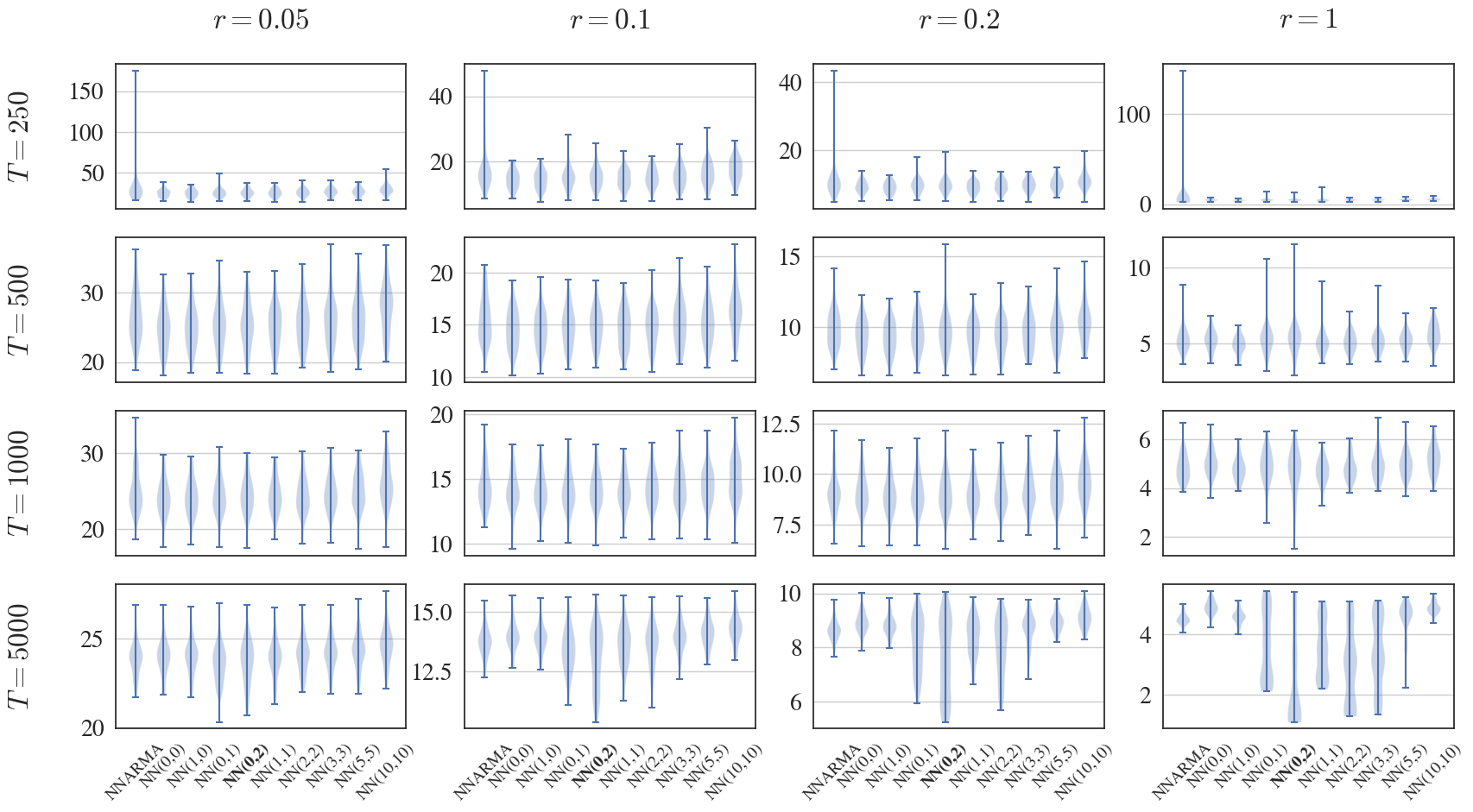}
		\subcaption{MSE sampling distribution}
	\end{subfigure} 
	\begin{subfigure}[c]{\textwidth}
		\centering
		\includegraphics[width=\textwidth]{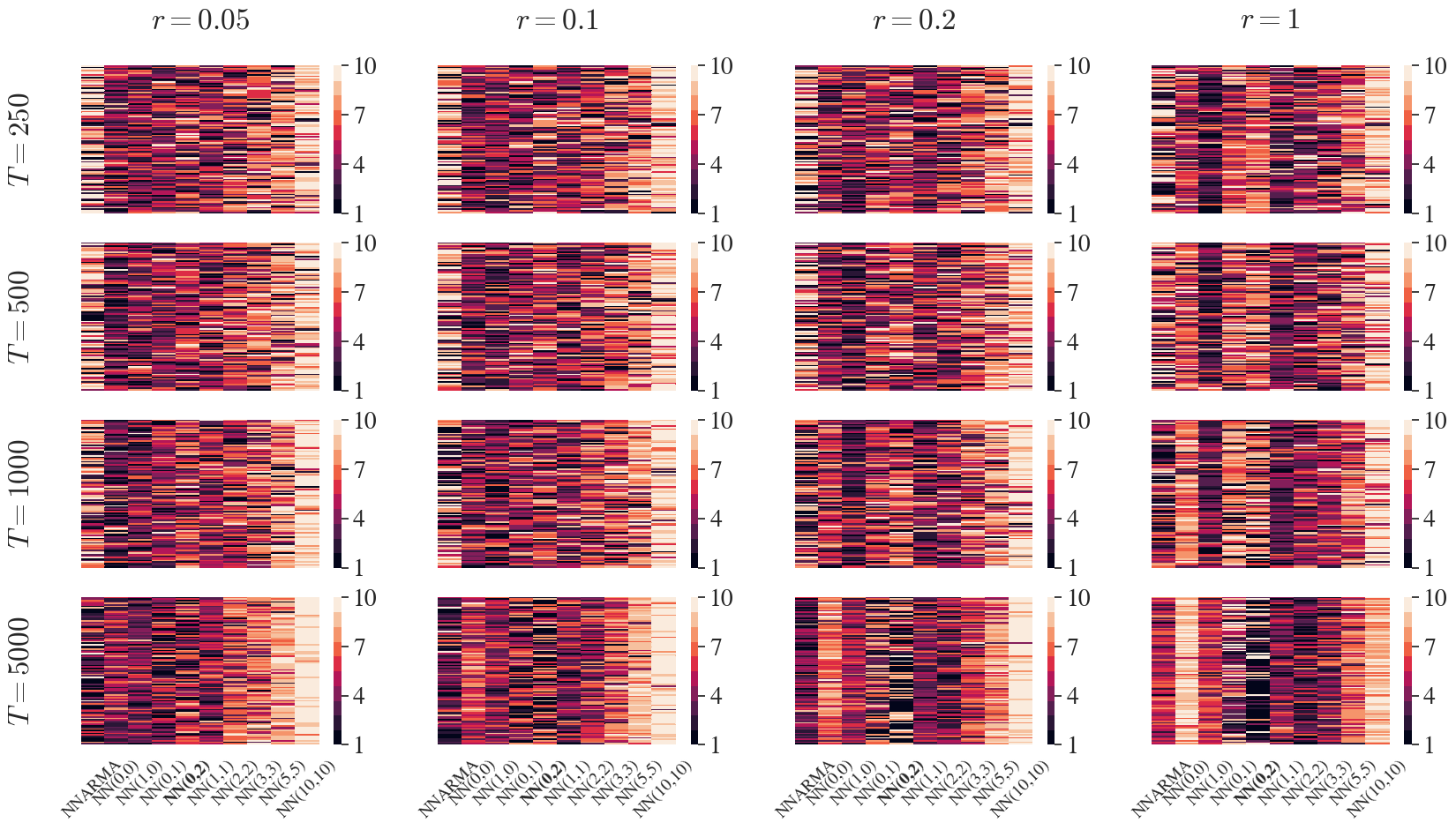}
		\subcaption{Per-sample model ranking by MSE}
	\end{subfigure} 
	\caption{Out-of-sample results under dynamic misspecification of NNARMA, with $\text{NN}(0,2)$ as the correct specification (bold). Sinusoidal regression function: MSE distribution (panel a) and per-sample model rankings (panel b), ordered from lowest MSE (1/dark) to highest MSE (10/light). Subplots are arranged by $r$ (columns) and $T$ (rows). The NNARMA model selects $p$ and $q$ automatically for each replication using the practical model selection strategy described in Section \ref{sec: Model selection (methodology)}.}
	\label{fig: omit input sine waves}
\end{figure}

Results are presented in Figures \ref{fig: omit input hump} and \ref{fig: omit input sine waves}. Differences in predictive accuracy between the correct specification $\text{NN}(0,2) $ and the NNARMA model are generally small, but tend to become more pronounced in favor of $\text{NN}(0,2) $ for high $r$ and large $T$; the largest differences occur for the sinusoidal function with $r=1$ and $T=5000$ (panel a). For the hump-shaped regression function, NNARMA is often the least accurate model for the combination of $r=1$ and $T=5000$ (panel b). In most of the remaining scenarios, the accuracy of the NNARMA model fluctuates around the same level as the benchmarks, including the correct specification $\text{NN}(0,2) $.

\vspace{0.75em}
\noindent\textbf{Omission of lagged inputs and outputs}\label{sec: MC omitting lagged dependent variables and lagged exogenous variables}

\noindent We augment the data generating process in \ref{eq: MC dependent variable dynamic misspecification}--\ref{eq: MC regressors dynamic misspecification} with lagged dependent variables:
\begin{align}\label{eq: MC dependent variable dynamic misspecification lagged dependent variable}
	y_t = f(\tilde x_{1t}, \tilde x_{2t}) + 0.5 y_{t-1} + 0.4 y_{t-2} + u_t, \hspace*{0.2cm} t=1,\ldots,T.
\end{align}
The setup is otherwise as above. Again, we estimate the NNARMA model without lags and compare its out-of-sample prediction accuracy with that of the benchmarks. In this setting, the benchmark specification NN(2,2) is the correct model specification.

\begin{figure}[t!]
	\centering
	\begin{subfigure}[c]{\textwidth}
		\centering
		\includegraphics[width=\textwidth]{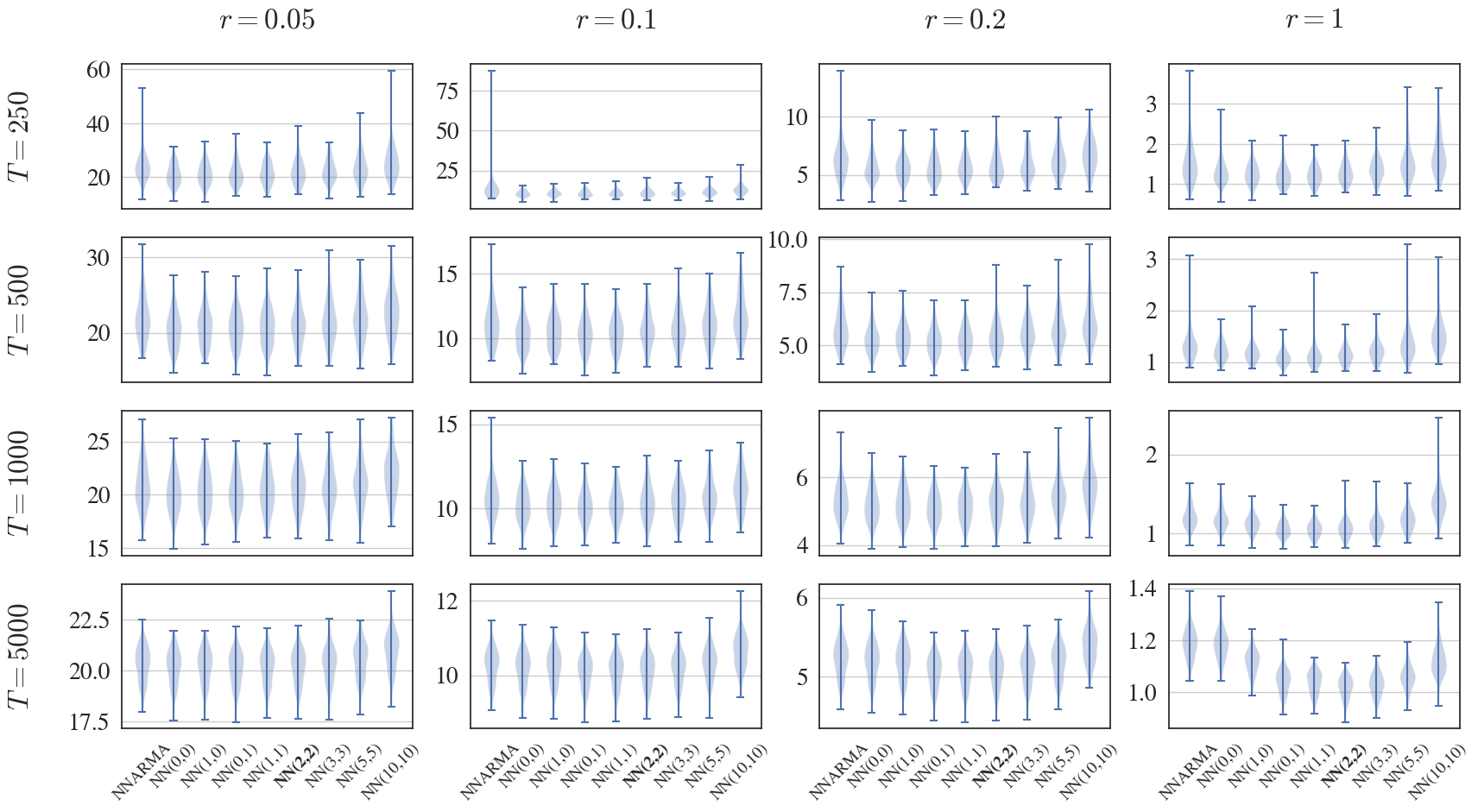}
		\subcaption{MSE sampling distribution}
	\end{subfigure} 
	\begin{subfigure}[c]{\textwidth}
		\centering
		\includegraphics[width=\textwidth]{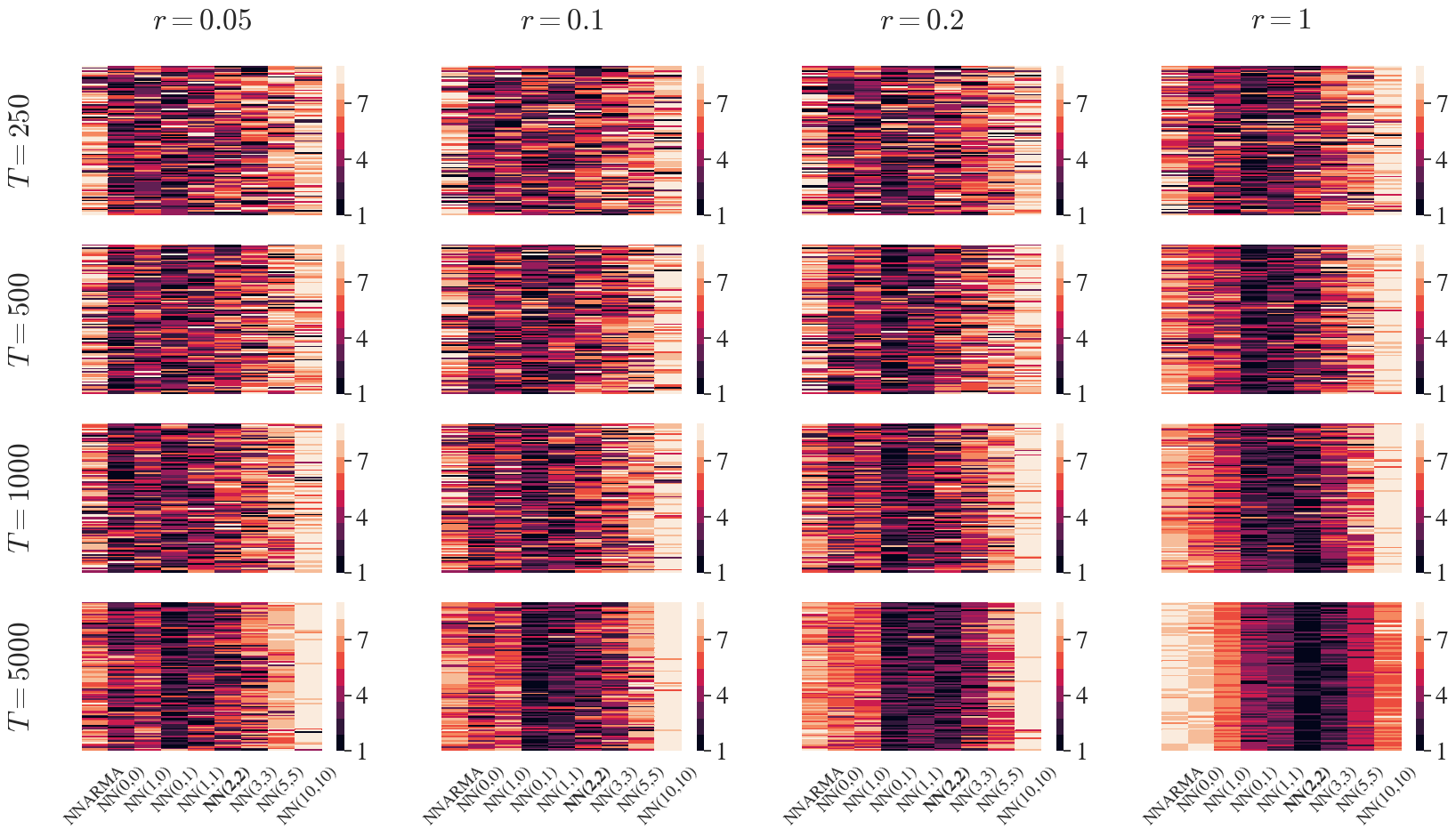}
		\subcaption{Per-sample model ranking by MSE}
	\end{subfigure} 
	\caption{Out-of-sample results under dynamic misspecification of NNARMA, with $\text{NN}(2,2)$ as the correct specification (bold). Hump-shaped regression function: MSE distribution (panel a) and per-sample model rankings (panel b), ordered from lowest MSE (1/dark) to highest MSE (9/light). Subplots are arranged by $r$ (columns) and $T$ (rows). The NNARMA model selects $p$ and $q$ automatically for each replication using the practical model selection strategy described in Section \ref{sec: Model selection (methodology)}.}
	\label{fig: omit input output hump}
\end{figure}

\begin{figure}[t!]
	\centering
	\begin{subfigure}[c]{\textwidth}
		\centering
		\includegraphics[width=\textwidth]{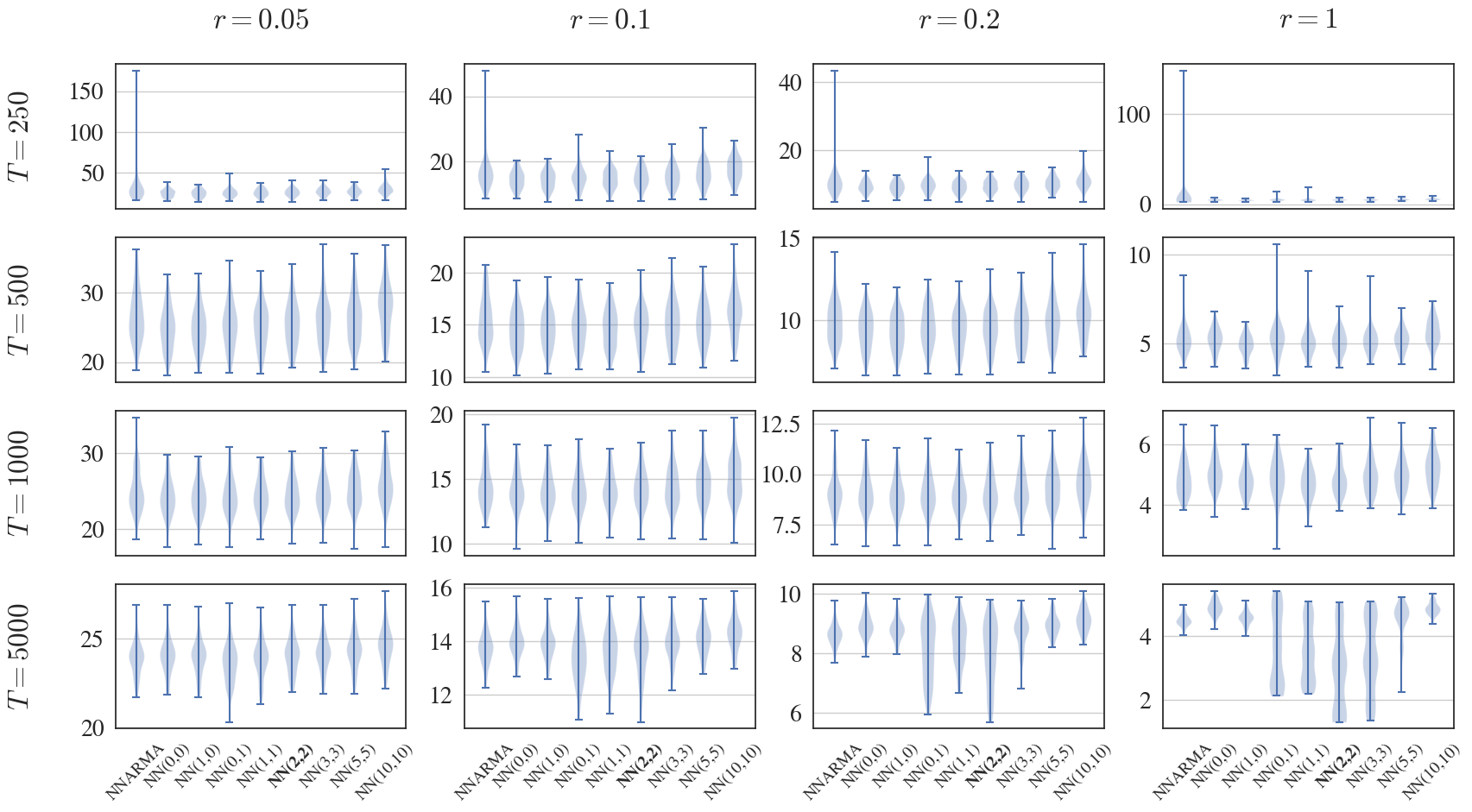}
		\subcaption{MSE sampling distribution}
	\end{subfigure} 
	\begin{subfigure}[c]{\textwidth}
		\centering
		\includegraphics[width=\textwidth]{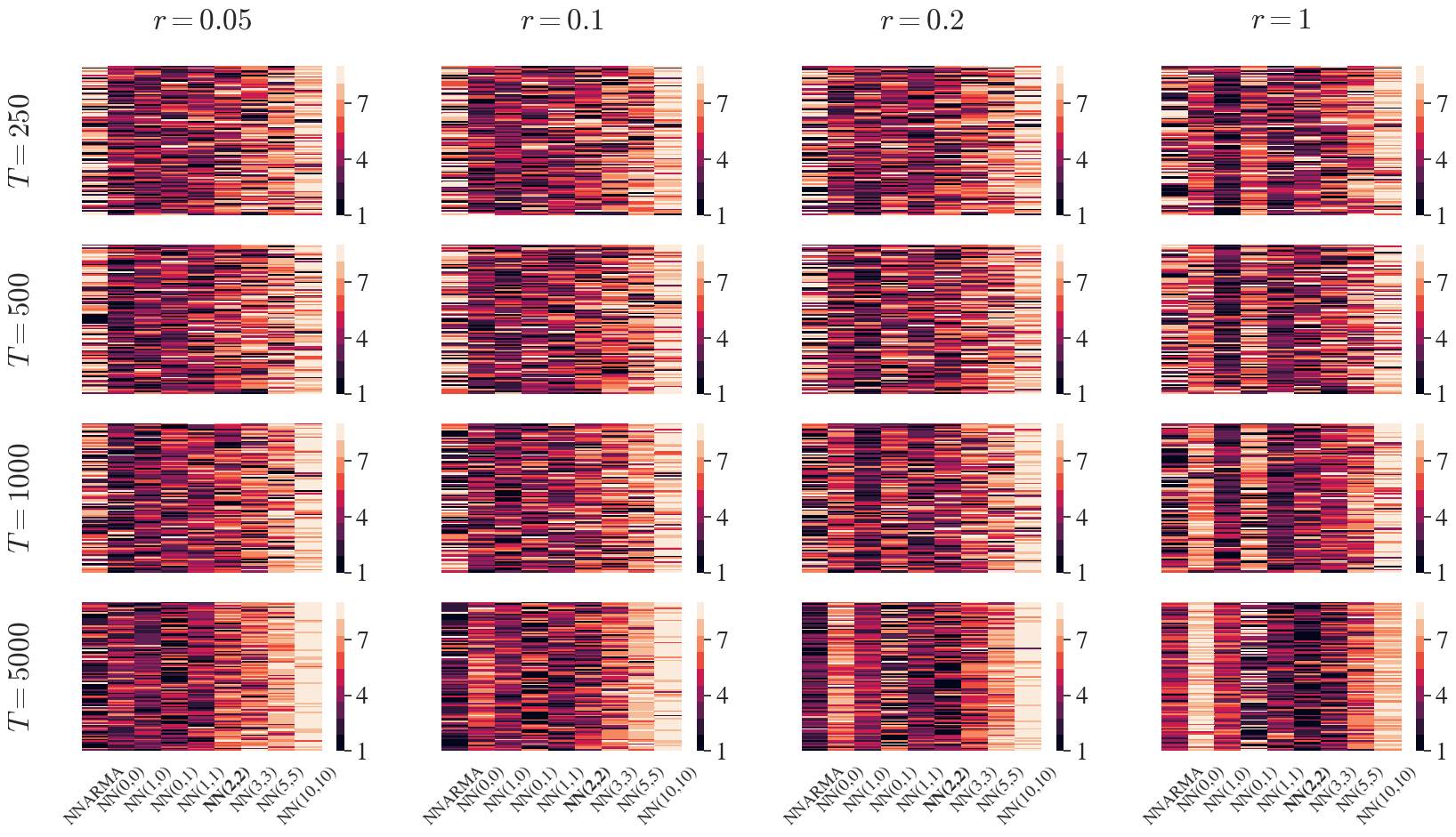}
		\subcaption{Per-sample model ranking by MSE}
	\end{subfigure} 
	\caption{Out-of-sample results under dynamic misspecification of NNARMA, with $\text{NN}(2,2)$ as the correct specification (bold). Sinusoidal regression funciont: MSE distribution (panel a) and per-sample model rankings (panel b), ordered from lowest MSE (1/dark) to highest MSE (9/light). Subplots are arranged by $r$ (columns) and $T$ (rows). The NNARMA model selects $p$ and $q$ automatically for each replication using the practical model selection strategy described in Section \ref{sec: Model selection (methodology)}.}
	\label{fig: omit input output sine waves}
\end{figure}

Figures \ref{fig: omit input output hump} and \ref{fig: omit input output sine waves} present the results, which largely mirror those in Figures \ref{fig: omit input hump} and \ref{fig: omit input sine waves}, but with $\text{NN}(2,2) $ as the correct specification. The largest differences in predictive accuracy between $\text{NN}(2,2) $ and the NNARMA model again occur for the sinusoidal function with $r=1$ and $T=5000$, where the former performs substantially better (panel a). For the sinusoidal function with $r \in \{0.05,0.1,0.2\}$ and $T=5000$, the NNARMA model is more accurate than the correct specification $\text{NN}(2,2) $ in most replications (panel b).

\newpage
\FloatBarrier
\subsection{Model selection}\label{sec:Model selection MC Appendix}
\FloatBarrier
This section motivates the simple and practical strategy for model selection in the NNARMA model discussed in Section \ref{sec: Model selection (methodology)}, using Monte Carlo evidence and the setup described in Appendix \ref{sec: MC setup Appendix}. In turn, we discuss the choice of ARMA specification and neural network architecture.

\vspace{0.75em}
\noindent\textbf{ARMA specification}\label{Model selection Appendix ARMA}

\begin{figure}[t!]
	\centering
	\begin{subfigure}[c]{\textwidth}
		\centering
		\includegraphics[width=\textwidth]{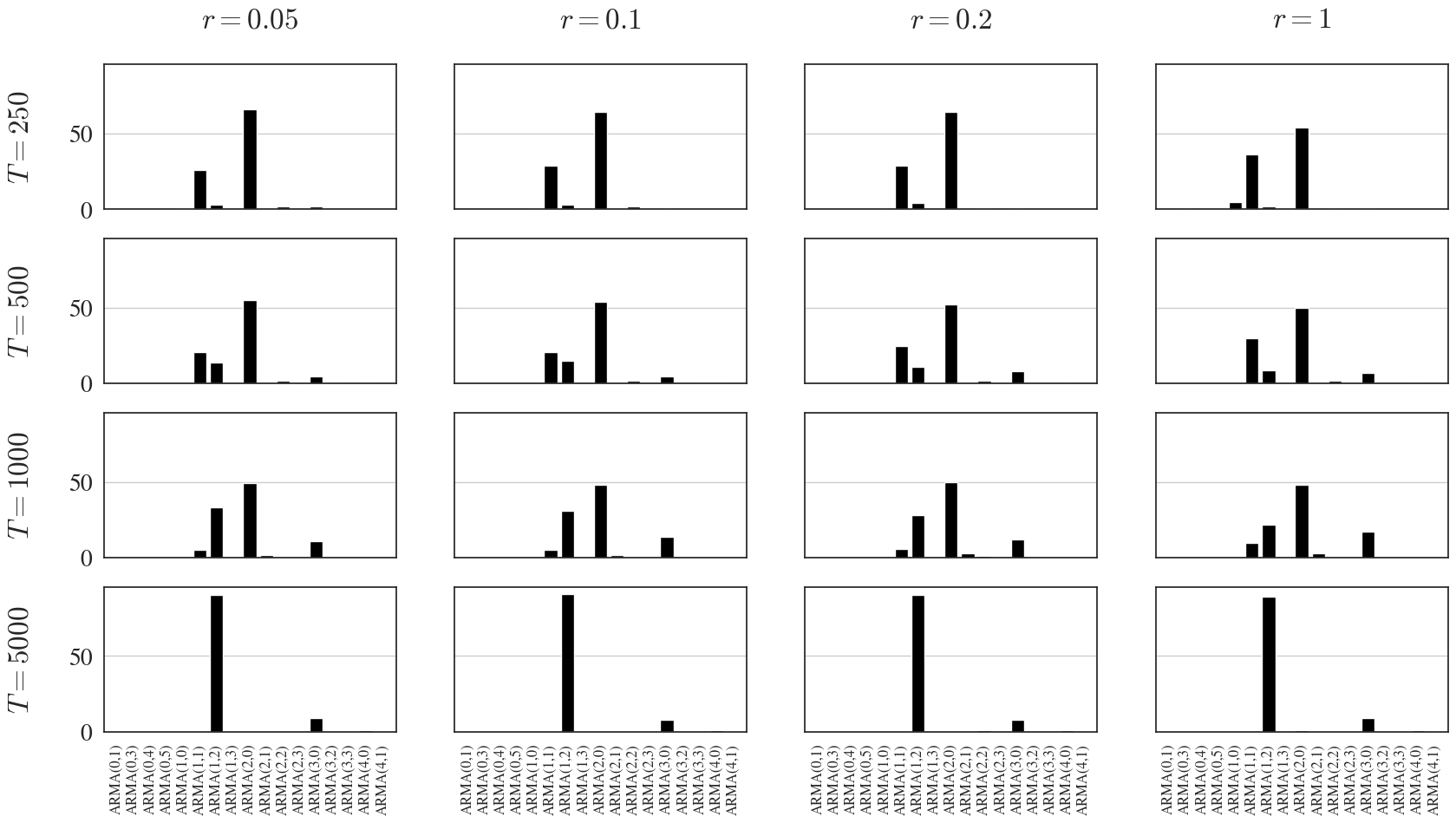}
		\subcaption{Hump-shaped regression function}
	\end{subfigure} 
	\begin{subfigure}[c]{\textwidth}
		\centering
		\includegraphics[width=\textwidth]{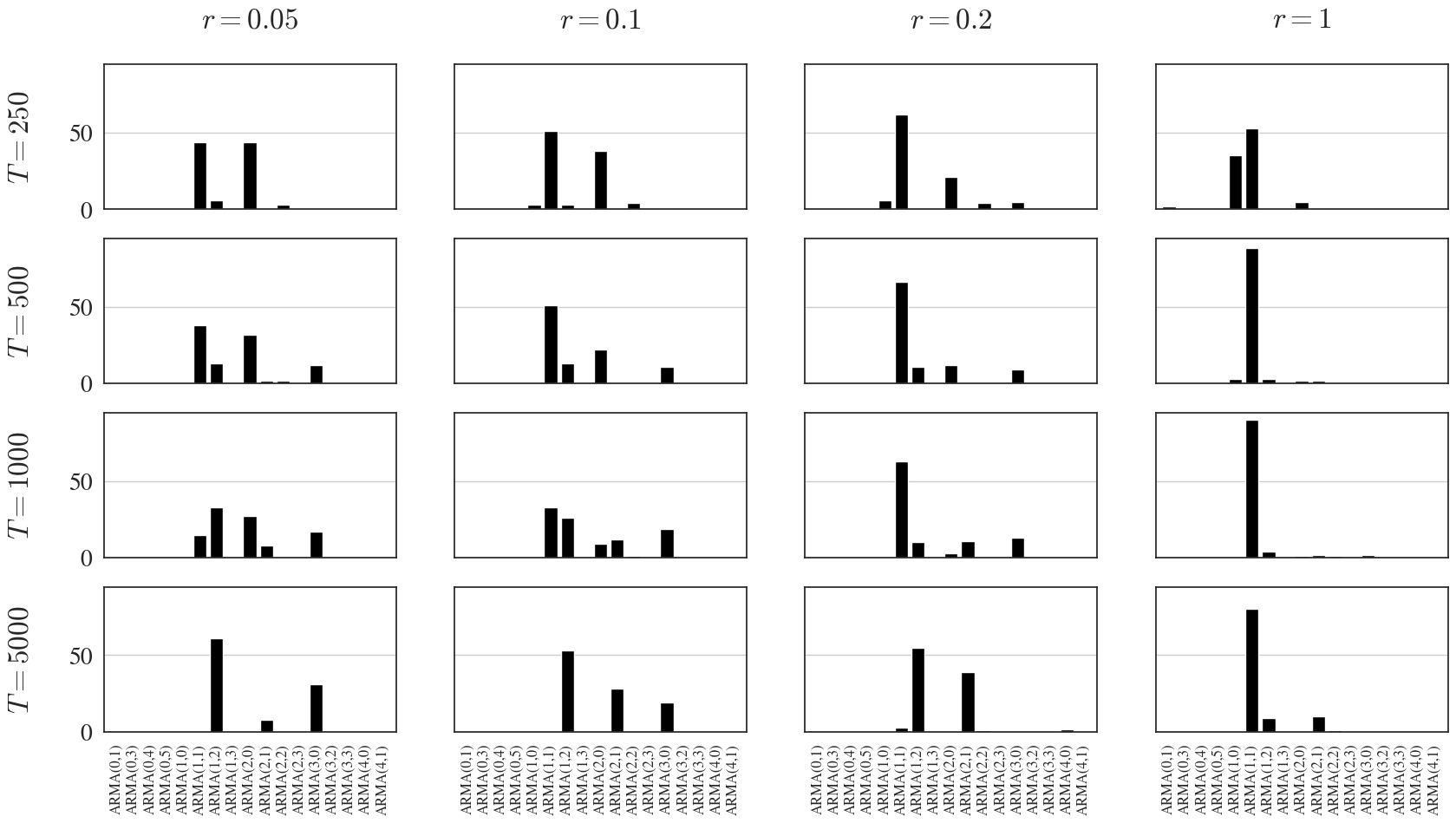}
		\subcaption{Sinusoidal regression function}
	\end{subfigure} 
	\caption{Distribution of selected ARMA specifications across Monte Carlo replications for the hump-shaped (panel a) and sinusoidal (panel b) regression functions. Bars show the percentage of replications selecting each ARMA specification (summing to $100\%$ within each panel), excluding specifications never selected. Subplots are arranged by $r$ (columns) and $T$ (rows).}
	\label{fig: ARMA specifications appendix}
\end{figure}

\noindent Figure \ref{fig: ARMA specifications appendix} illustrates the distribution of ARMA specifications selected across Monte Carlo replications using our practical strategy. We consider 36 $\text{ARMA}(p,q)$ specification with $p,q \in \{0,1,2,3,4,5\}$ as candidates. Results are presented for the hump-shaped and sinusoidal regression functions, using the same simulated data as in Appendices \ref{sec: MC Finite sample properties}--\ref{sec: MC Misspecification}. Across $(r,T)$ scenarios, the distribution of selected ARMA specifications is concentrated around the true $\text{ARMA}(1,2)$ specification. As $T$ increases, the distribution generally collapses to the true specification. For the sinusoidal function with $r=1$, however, convergence appears slower and may require a sample size larger than $5000$. Our strategy for selecting $p$ and $q$ is based on the BIC. We also considered the AIC, but the distribution of selected ARMA specifications was less concentrated and did not converge to the true specification as $T$ increased.

\vspace{0.75em}
\noindent\textbf{Neural network architecture}\label{sec: NN architecture}

\begin{figure}[t!]
	\centering
	\begin{subfigure}[c]{\textwidth}
		\centering
		\includegraphics[width=\textwidth]{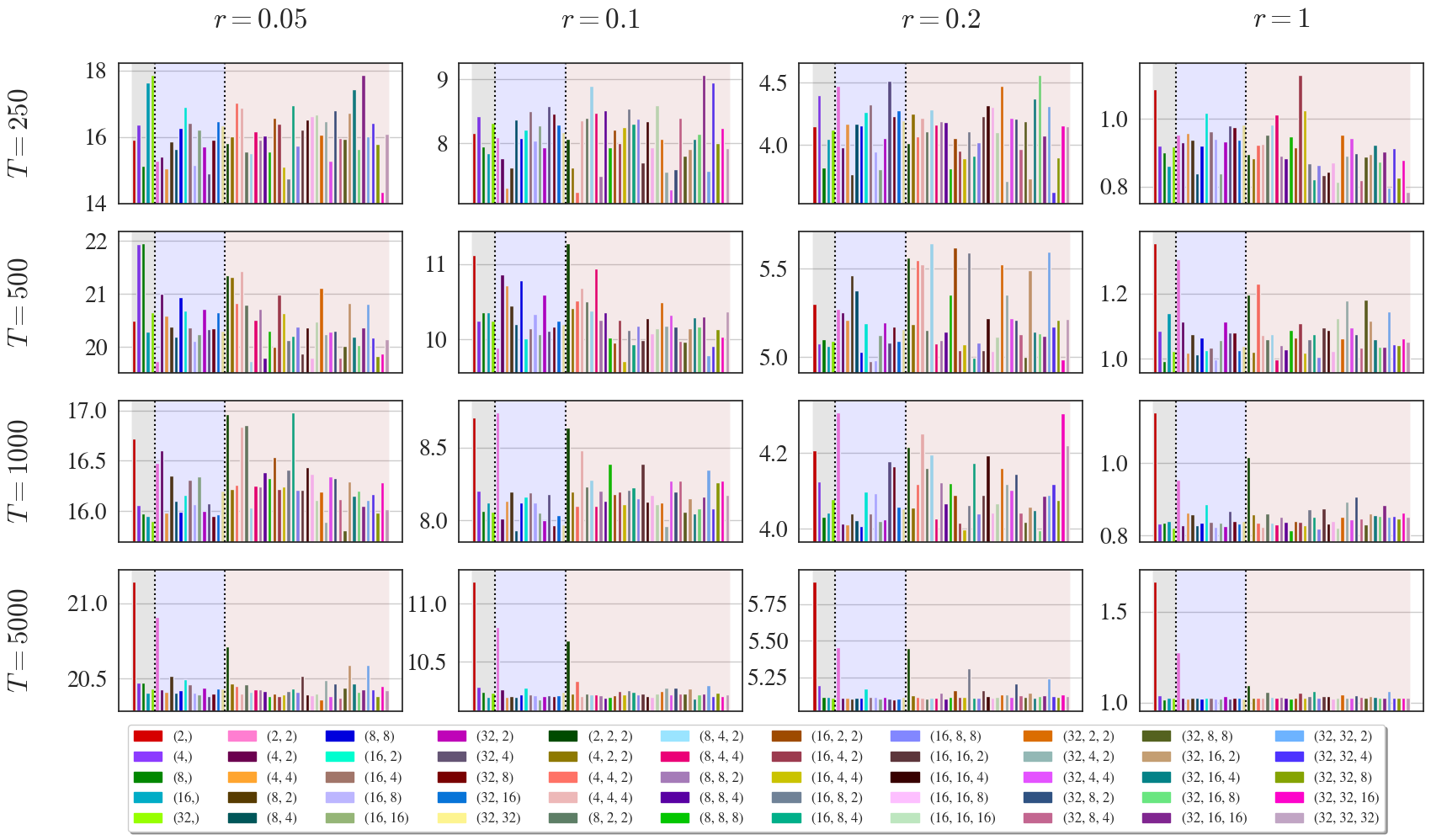}
		\subcaption{Hump-shaped regression function}
	\end{subfigure} \\ \vspace*{0.4cm}
	\begin{subfigure}[c]{\textwidth}
		\centering
		\includegraphics[width=\textwidth]{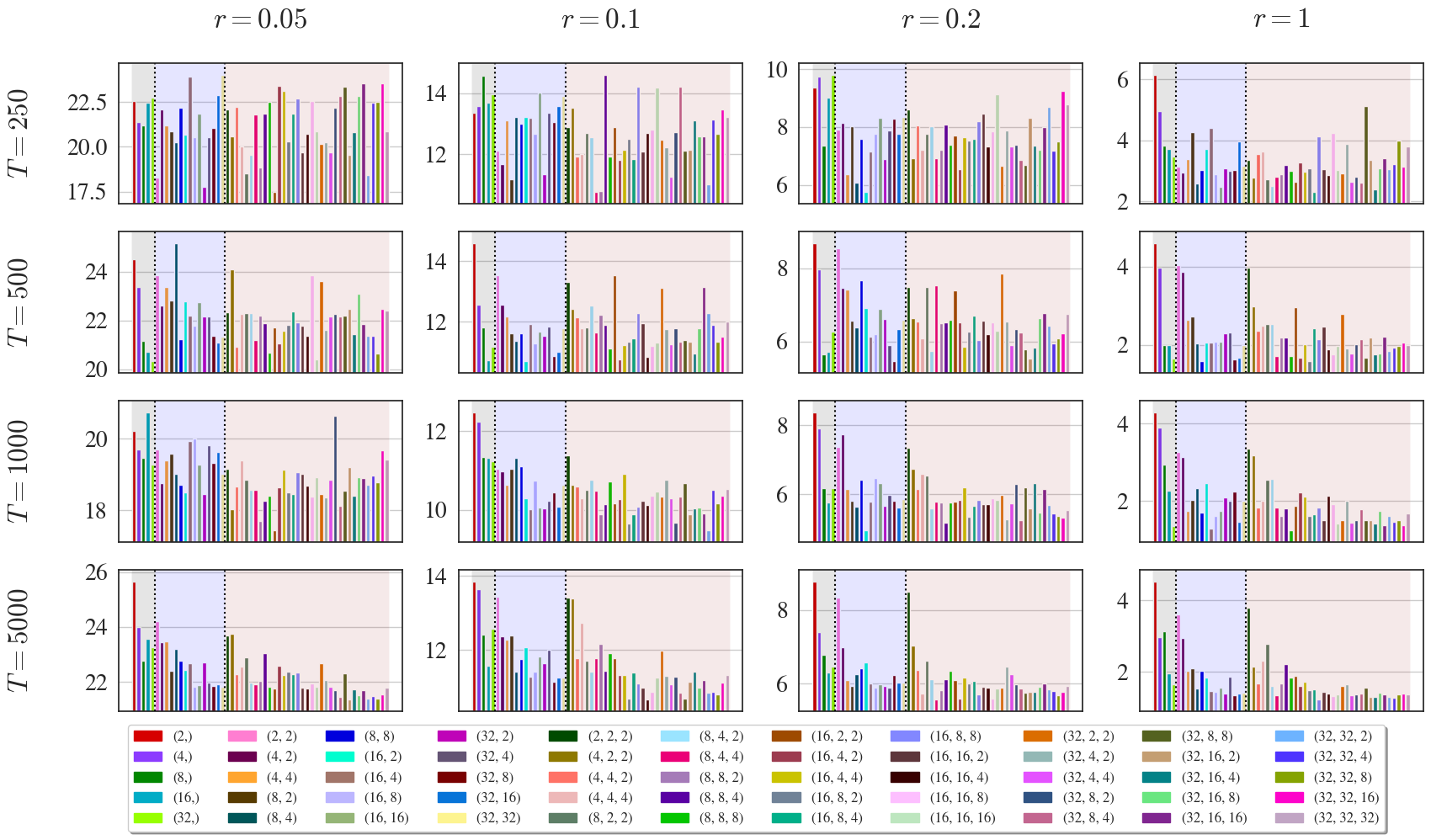}
		\subcaption{Sinusoidal regression function}
	\end{subfigure}
	\caption{Validation loss of NNARMA across network architectures for the hump-shaped (panel a) and sinusoidal (panel b) regression functions. The ``grey", ``blue", and ``brown" blocks indicate one, two, and three hidden layers, respectively. ``$ (a, b, c) $" indicates a network architecture with $ a $ units in the first layer, $ b $ in the second, and $ c $ in the third. Subplots are arranged by $r$ (columns) and $T$ (rows).}
	\label{fig: val loss stand}
\end{figure}

\noindent Figure \ref{fig: val loss stand} shows the validation loss of the NNARMA model for different network architectures, computed from an initial simulated sample. Results are presented for both the hump-shaped and sinusoidal regression functions. In line with \cite{masters1993, Kelly2020, Jensen2023}, we consider 55 different rectangular and pyramid-shaped network architectures. Such architectures are useful for learning gradually more abstract transformations of the input variables and for keeping the number of free parameters at a reasonable level \citep{masters1993}. The set of achitectures is divided into three blocks consisting of one, two, and three hidden layers. The initial architectures in each block are the narrowest and simplest by parameter count. 

The validation loss is largely stable across architectures. It is mostly very narrow networks that sometimes produce higher losses. In particular, for the sinusoidal regression function with $r \in \{0.2, 1\}$ or $T\in \{1000, 5000\}$, the validation loss spikes for the initial set of architectures in each block, see panel (b). These results suggests that, under early stopping, performance is largely insensitive to the specific architecture once the network is sufficiently flexible. The two-layered architecture (32 and 16 units) used in the Monte Carlo experiments and the empirical study provides an illustrative and sufficiently flexible specification without extensive tuning. 

\FloatBarrier
\subsection{Initialization}\label{sec:Initialization MC Appendix}
\FloatBarrier

This section describes the initialization of the Adam and Powell algorithms used to solve the inner and outer minimization problems in \eqref{eq: theta_2} and \eqref{eq: theta_1}, respectively, and presents a simulation-based comparison of cold and warm starts for the Adam algorithm.

For each candidate value of $ \theta_1 $, the Adam algorithm solves the inner minimization problem with respect to $ \theta_2 $. With cold starts, $\theta_2$ is initialized from scratch for each value of $\theta_1$: the slope coefficients in the neural network ($ \beta, \Gamma^{(1)},\ldots,\Gamma^{(h)} $) are initialized using the Gaussian initialization proposed by \cite{He_2015}, and the intercepts ($ \kappa^{(1)},\ldots,\kappa^{(h)} $) are initialized at zero, following standard practice in the neural networks literature \citep{Goodfellow-et-al-2016}. With warm starts, $\theta_2$ is initialized from the solution obtained for the previous candidate value of $\theta_1$. In the Powell algorithm, Initial values of $\theta_1$ are obtained by estimating the ARMA parameters as in \eqref{eq: theta_1} based on residuals from an initial neural network regression as in \eqref{eq: theta_2}, assuming serially uncorrelated disturbances for $u_t$, such that $\theta _1 \in \emptyset$ and $C(\theta_1)=I_T$.

\begin{figure}[t!]
	\centering
	\includegraphics[width=\textwidth]{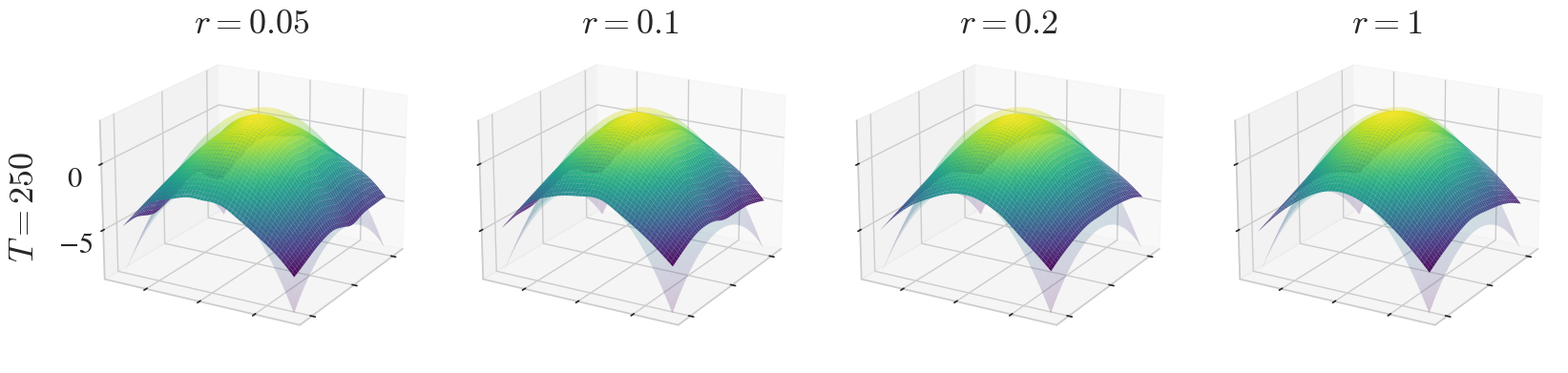} \\ \vspace*{-0.5cm}
	\includegraphics[width=\textwidth]{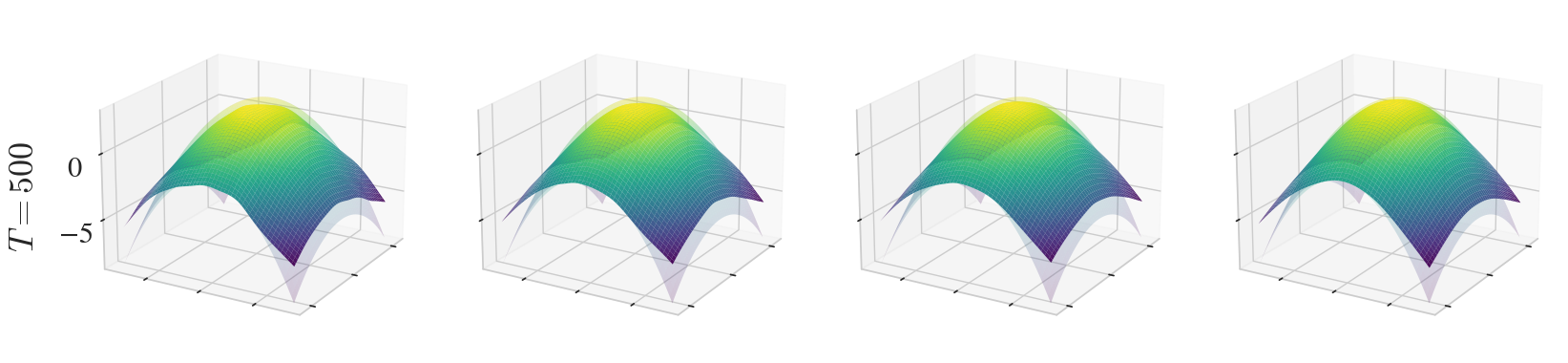} \\ \vspace*{-0.5cm}
	\includegraphics[width=\textwidth]{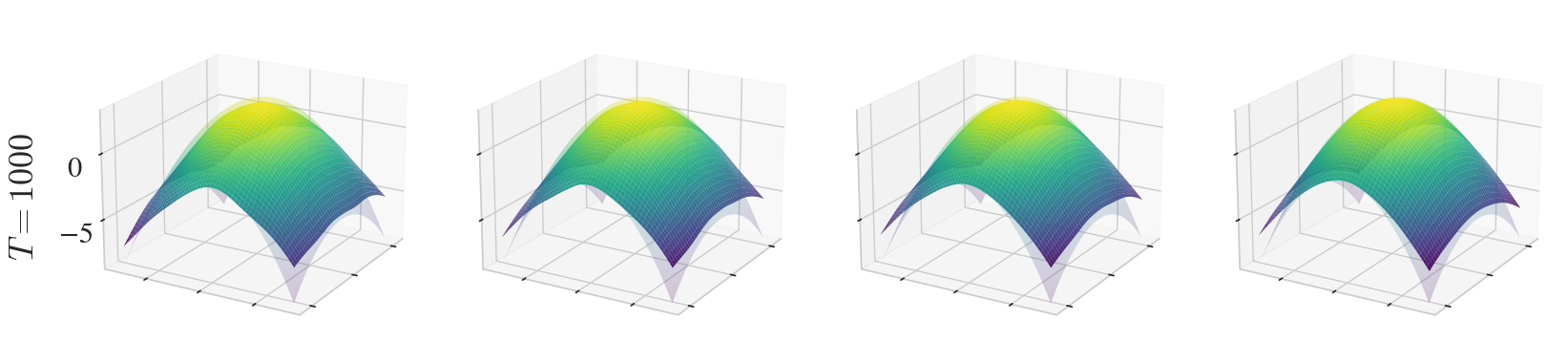} \\ \vspace*{-0.5cm}
	\includegraphics[width=\textwidth]{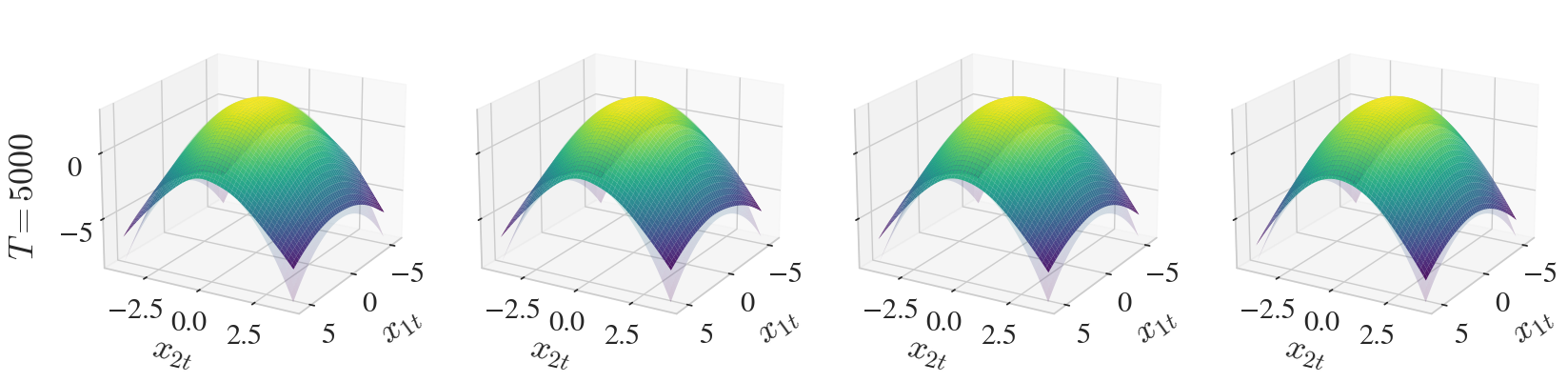} 
	\caption{Average estimate of the hump-shaped regression function with warm starts. Subplots are arranged by $r$ (columns) and $T$ (rows). The true function is shown transparently in the background. Figure \ref{fig: estimated reg func hump} show the corresponding results for cold starts.}
	\label{fig: estimated reg func hump warm}
\end{figure}

\begin{figure}[t!]
	\centering
	\includegraphics[width=\textwidth]{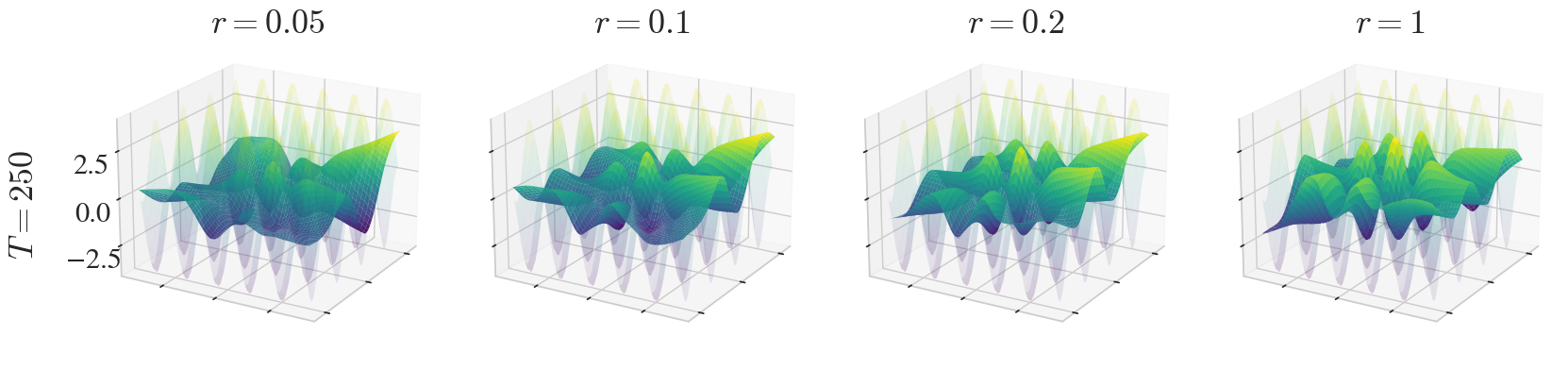} \\ \vspace*{-0.5cm}
	\includegraphics[width=\textwidth]{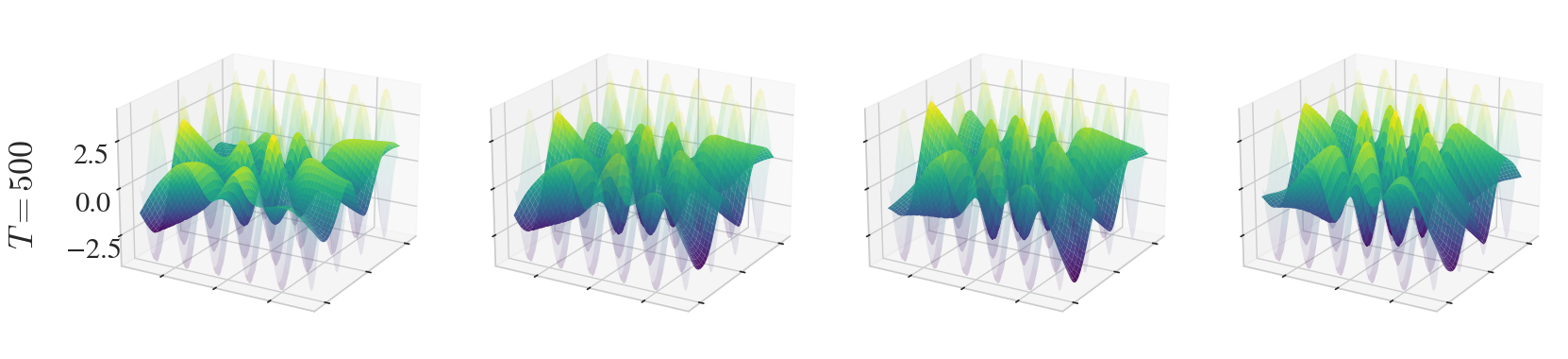} \\ \vspace*{-0.5cm}
	\includegraphics[width=\textwidth]{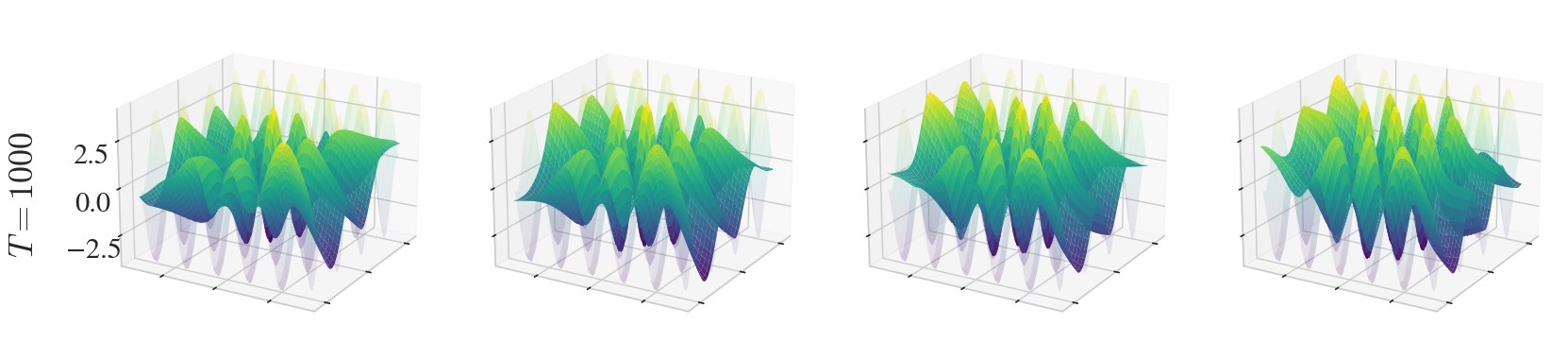} \\ \vspace*{-0.5cm}
	\includegraphics[width=\textwidth]{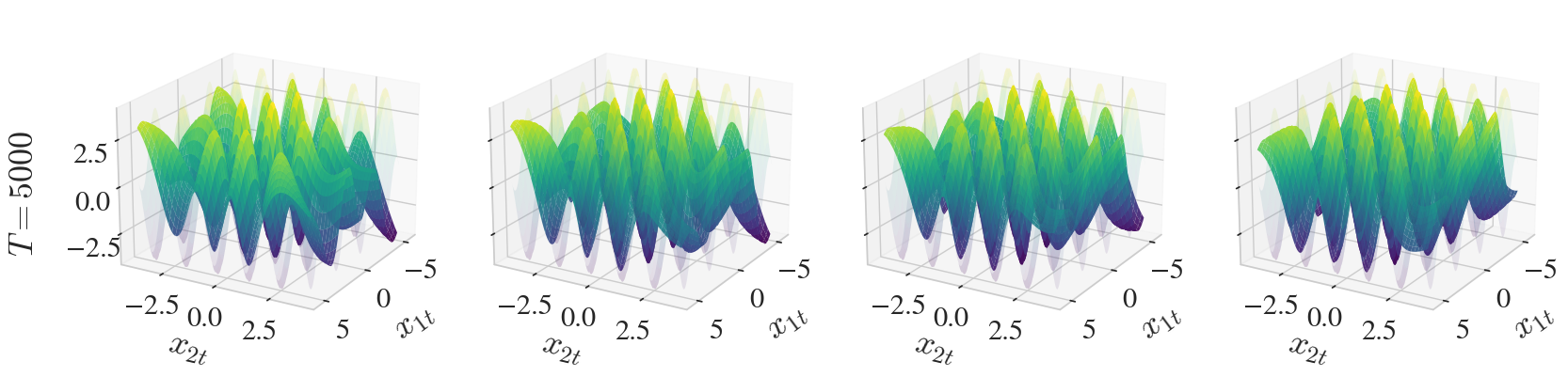} 
	\caption{Average estimate of the sinusoidal regression function with warm starts. Subplots are arranged by $r$ (columns) and $T$ (rows). The true function is shown transparently in the background. Figure \ref{fig: estimated reg func sine} show the corresponding results for cold starts.}
	\label{fig: estimated reg func sine warm}
\end{figure}

To compare cold and warm starts for the Adam algorithm, we reproduce the finite-sample simulations results from Appendix \ref{sec: MC Finite sample properties} using warm starts instead of cold starts, while keeping everything else fixed and using the same simulated data. Figures \ref{fig: estimated reg func hump warm} and \ref{fig: estimated reg func sine warm} display the average estimate of the hump-shaped and sinusoidal regression function using warm starts. They are directly comparable to Figures  \ref{fig: estimated reg func hump} and \ref{fig: estimated reg func sine}, which display the corresponding results obtained using cold starts. The surface estimates and their convergence as $r$ and $T$ increases are similar across cold and warm starts. 

\begin{figure}[t!]
	\centering
	\begin{subfigure}[c]{\textwidth}
		\centering
		\includegraphics[width=\textwidth]{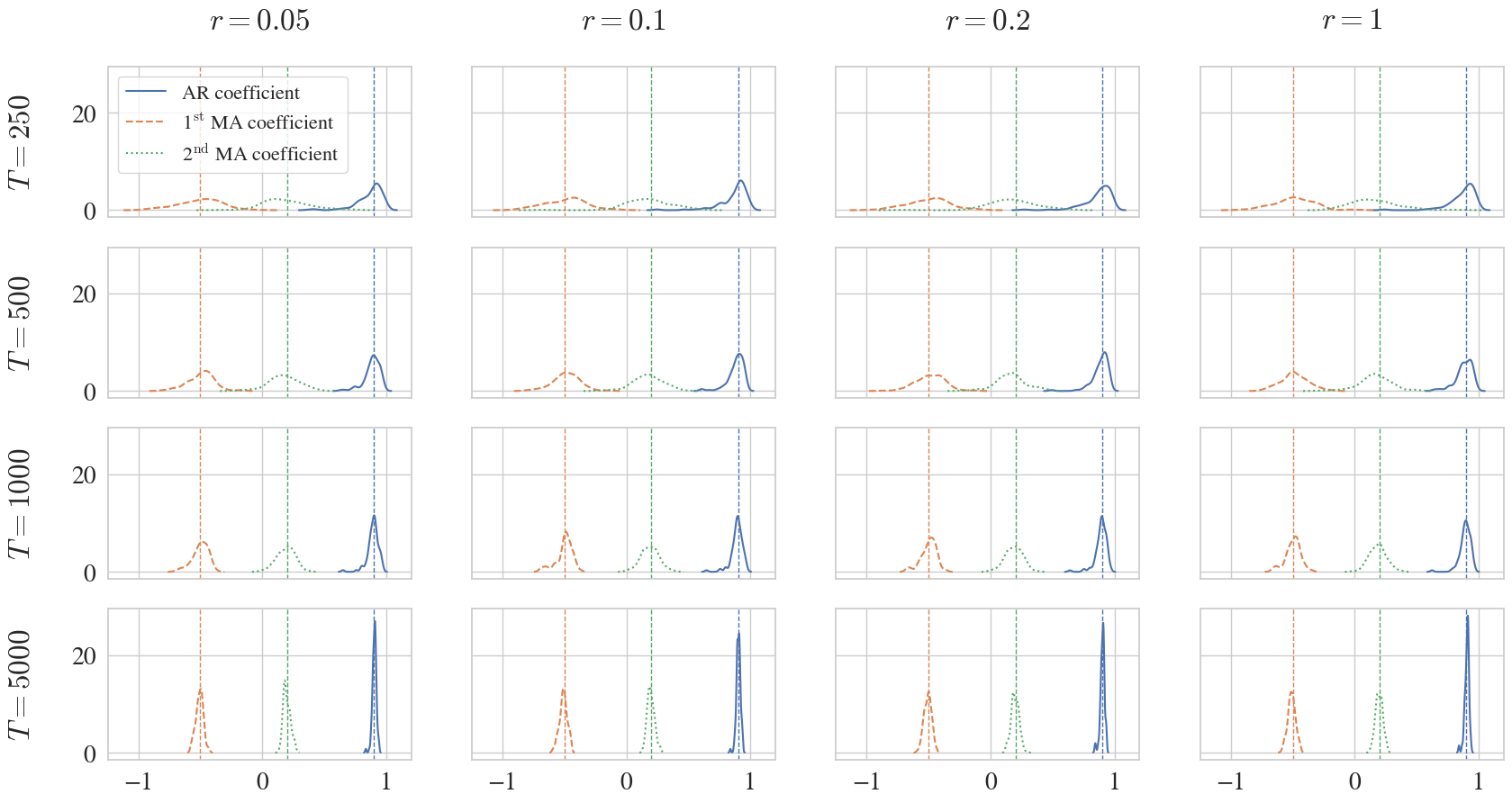}
		\subcaption{Hump-shaped regression function}
	\end{subfigure} 
	
	\begin{subfigure}[c]{\textwidth}
		\centering
		\includegraphics[width=\textwidth]{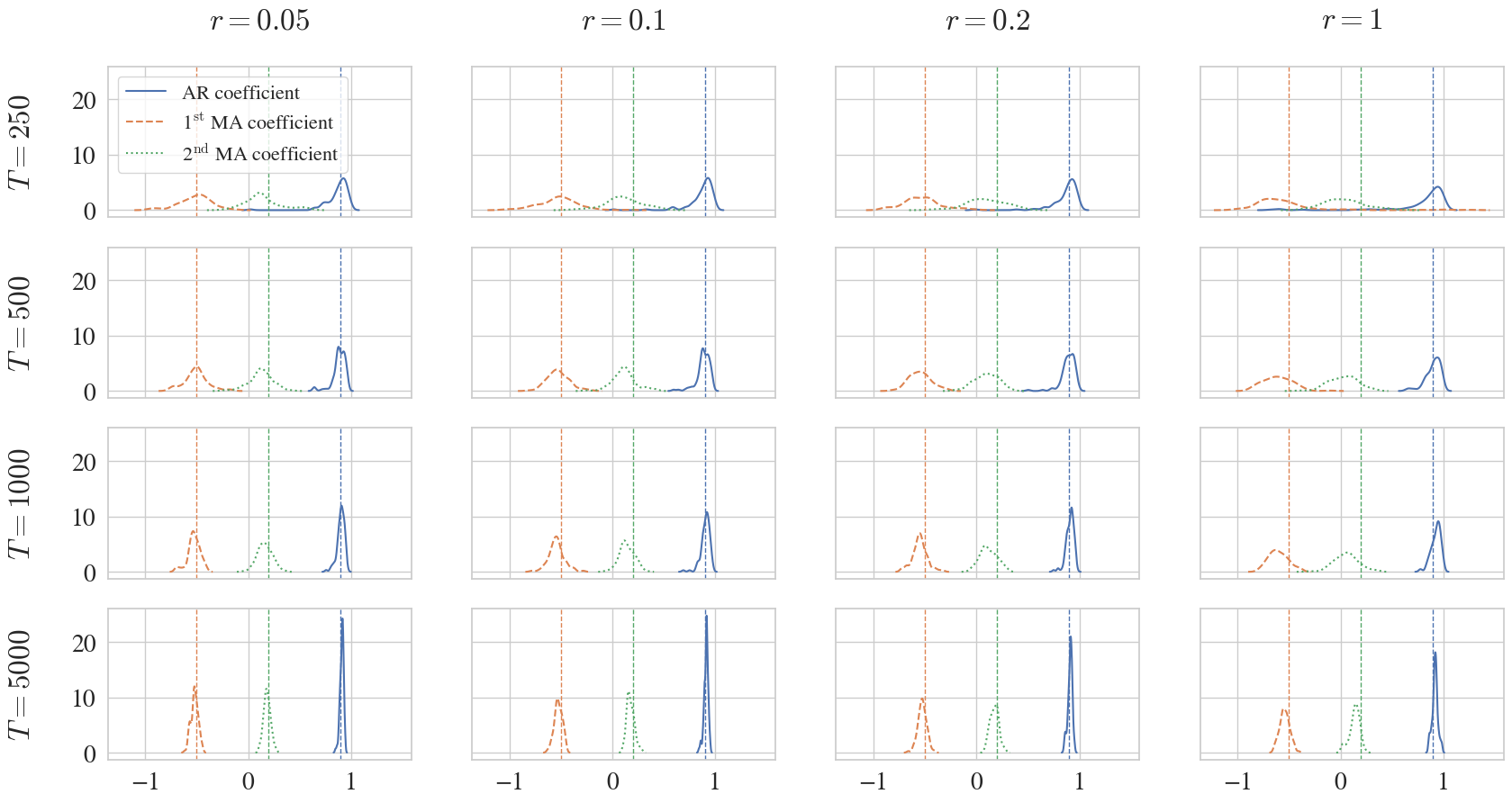}
		\subcaption{Sinusoidal regression function}
	\end{subfigure} 
	
	\caption{ARMA coefficient estimates for the hump-shaped (panel a) and sinusoidal (panel b) regression function with warm starts, both shown as sampling distributions. Subplots are arranged by $r$ (columns) and $T$ (rows). Vertical dashed lines indicate the true coefficient values. Panel (a) of Figures \ref{fig: estimated arma params hump} and \ref{fig: estimated arma params sine} show the corresponding results for cold starts.}
	\label{fig: estimated arma params warm init}
\end{figure}

Corresponding to the surface estimates in Figures \ref{fig: estimated reg func hump warm} and \ref{fig: estimated reg func sine warm}, Figure \ref{fig: estimated arma params warm init} displays the sampling distributions of the ARMA coefficient estimates for the hump-shaped function (panel a) and the sinusoidal function (panel b) using warm starts. The results are directly comparable to those reported in panel (a) of Figures \ref{fig: estimated arma params hump} and \ref{fig: estimated arma params sine}, which display the corresponding results for cold starts. The most noticeable difference is that the sampling distributions obtained with cold start are slightly more spread out across $r$ and $T$, especially for the sinusoidal function.

\newpage
\FloatBarrier
\section{Additional cloud cover results}\label{sec: Appendix supplementary figures}
\FloatBarrier

\begin{figure}[h!]
	\centering
	\includegraphics[width=\textwidth]{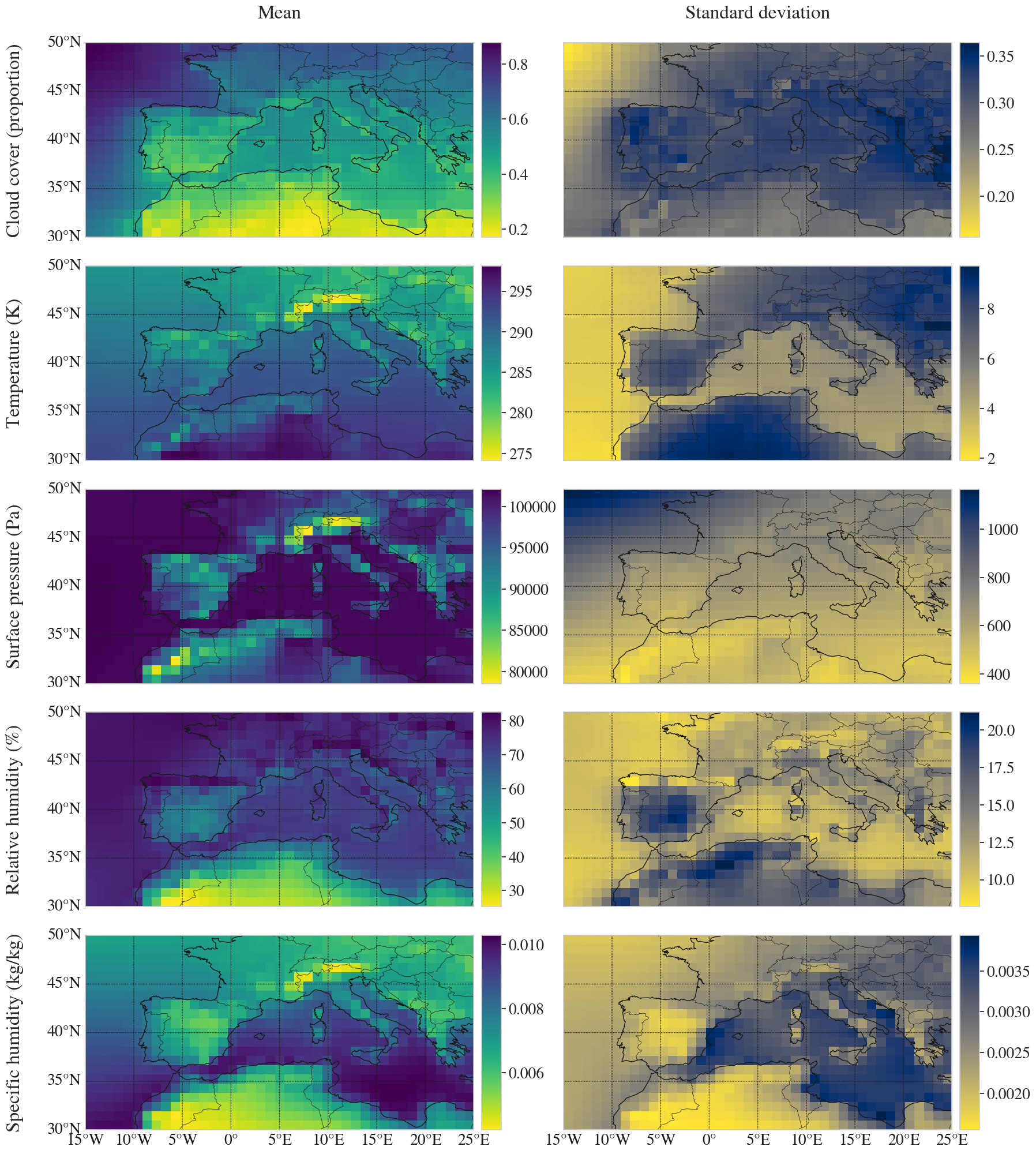}
	\caption{Sample mean (left) and standard deviation (right) of each variable employed in Section \ref{sec: CFC application}, computed over time separately at each geographical location.}
	\label{fig: Sumary plot appendix}
\end{figure}

\begin{figure}[h!]
	\centering
	\includegraphics[width=\textwidth]{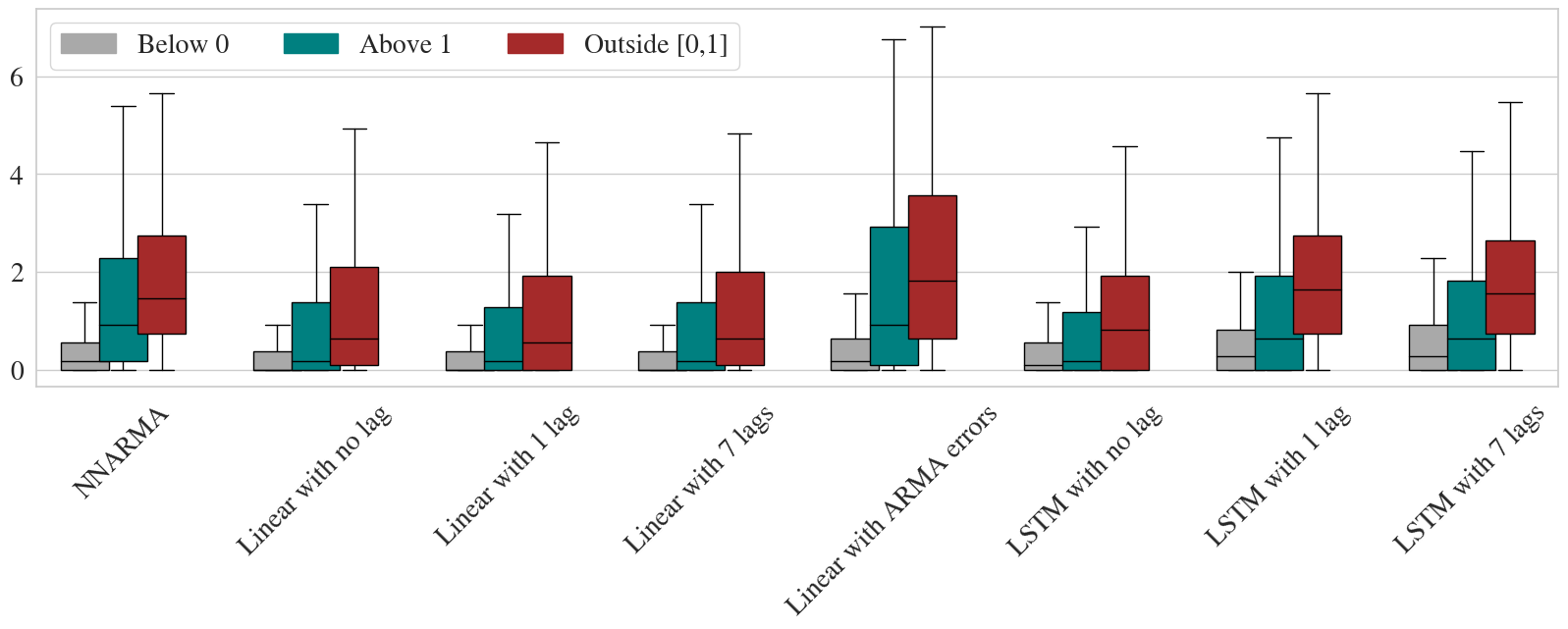}
	\caption{Box plots, across geographical locations, of the percentage of unrestricted out-of-sample predictions falling below 0 (gray), above 1 (teal), and outside the admissible range $[0,1]$ (brown).}
	\label{fig: Truncations box plot}
\end{figure}

\begin{figure}[h!]
	\centering
	\includegraphics[width=\textwidth]{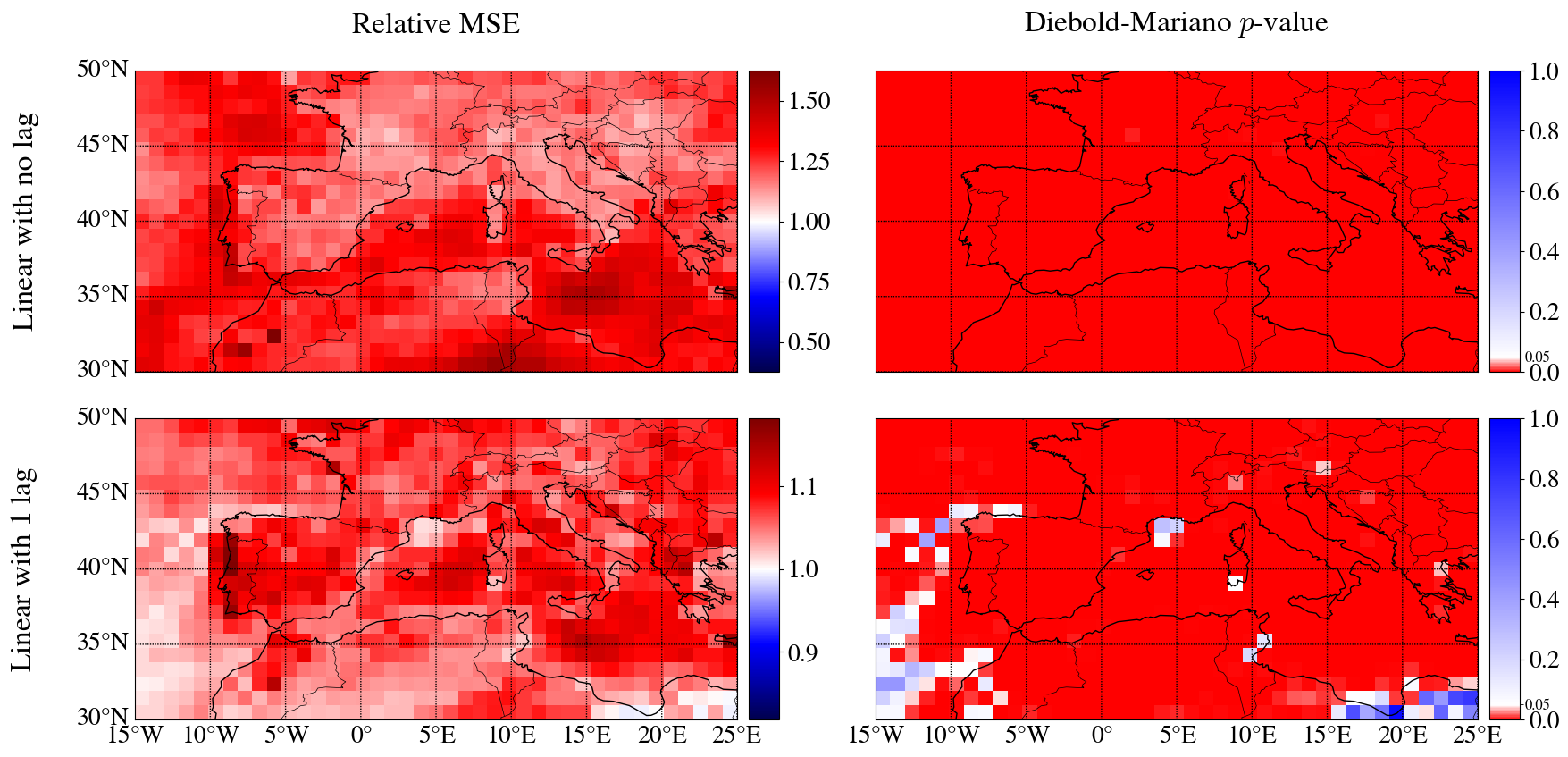}
	\caption{Comparison of the linear benchmarks. Relative MSEs over the test sample normalized to the linear model with ARMA errors (left), where values above one (red) indicate lower error for this model. $p$-values from a two-sided Diebold-Mariano test of equal predictive accuracy to the linear model with ARMA errros (right), where values below $0.05$ (red) indicate significant differences in predictive accuracy at that level.}
	\label{fig: lin vs lin}
\end{figure}

\newpage
\FloatBarrier
\clearpage
\subsection{Results for an expanded set of benchmarks}\label{sec: Appendix expanded set of benchmarks}
\FloatBarrier

\begin{figure}[h!]
	\centering
	\includegraphics[width=\textwidth]{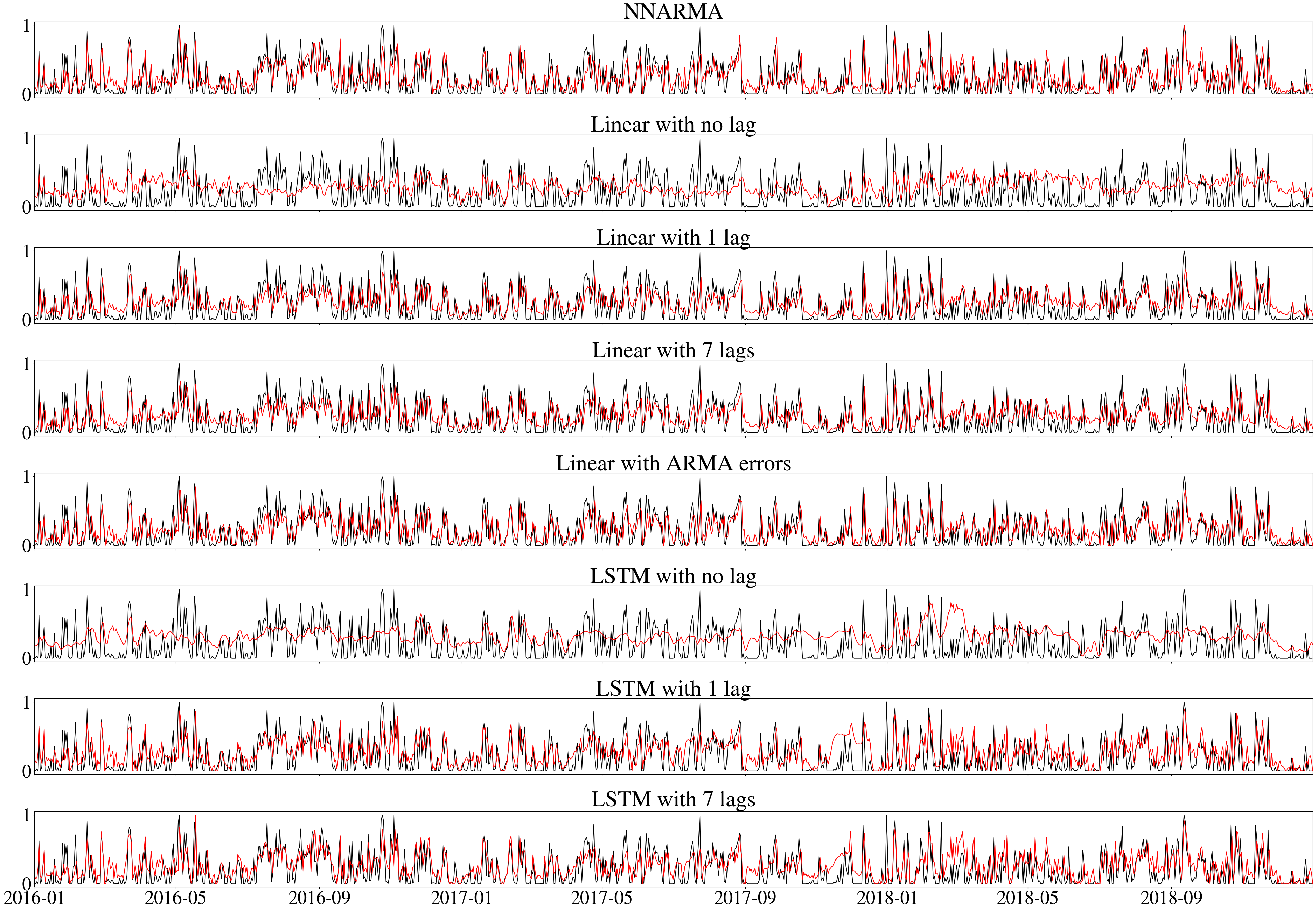}
	\caption{NNARMA and benchmark predictions over the tests sample (red) together observed cloud fractional cover (black) for the location labeled 1 in Figure \ref{fig: topography map}.}
	\label{fig: Appendix TS plot many 1}
\end{figure}

\begin{figure}[h!]
	\centering
	\includegraphics[width=\textwidth]{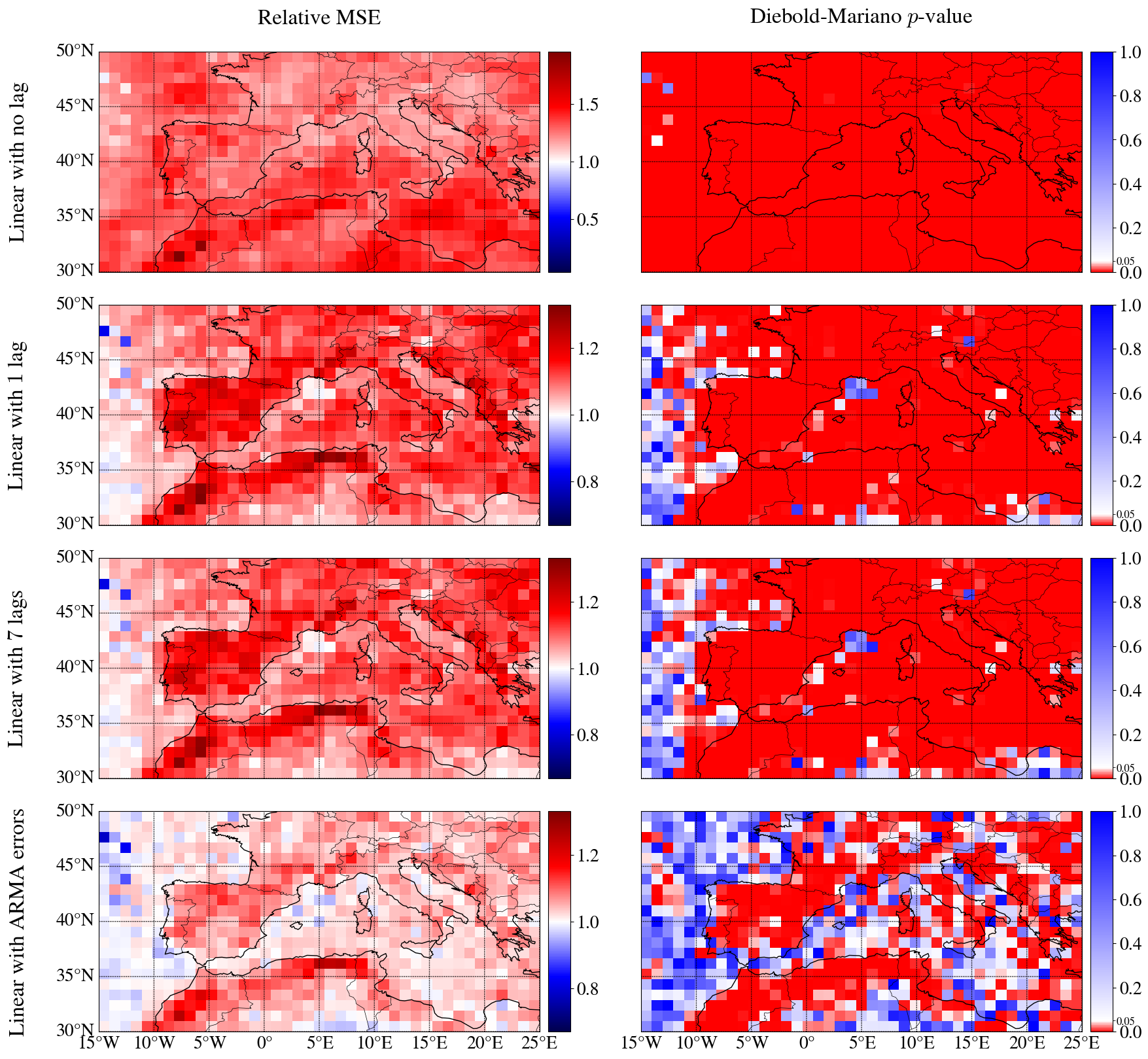}
	\caption{NNARMA and the linear benchmarks. Relative MSEs over the test sample normalized to NNARMA (left), where values above one (red) indicate lower error for NNARMA. $p$-values from a two-sided Diebold-Mariano test of equal predictive accuracy to NNARMA (right), where values below $0.05$ (red) indicate significant differences in predictive accuracy at that level.}
	\label{fig: Appendix linear MSEDM}
\end{figure}

\begin{figure}[t!]
	\centering
	\includegraphics[width=\textwidth]{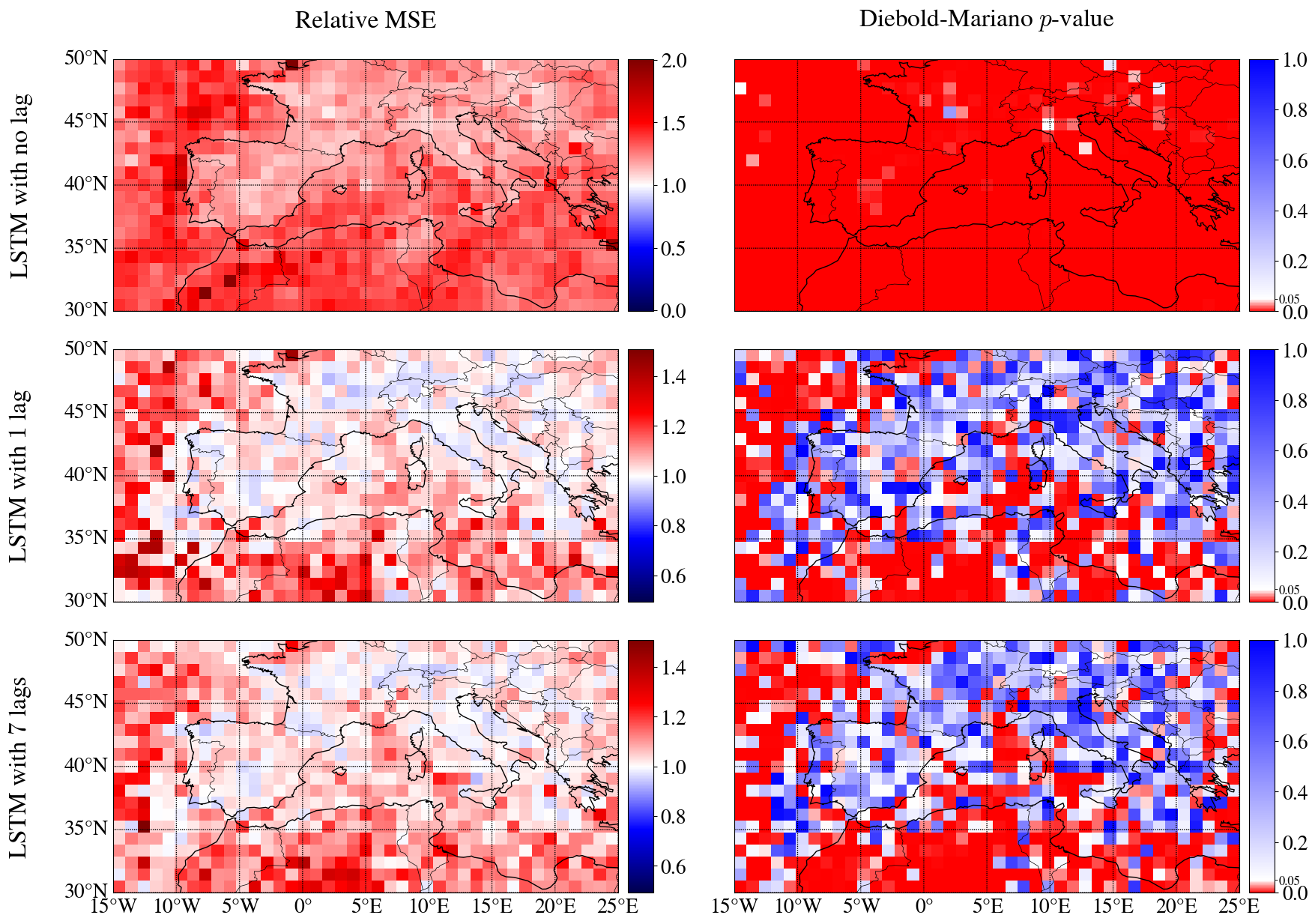}
	\caption{NNARMA and the LSTM network. Relative MSEs over the test sample normalized to NNARMA (left), where values above one (red) indicate lower error for NNARMA. $p$-values from a two-sided Diebold-Mariano test of equal predictive accuracy to NNARMA (right), where values below $0.05$ (red) indicate significant differences in predictive accuracy at that level.}
	\label{fig: Appendix LSTM MSEDM}
\end{figure}

\end{document}